\documentclass[a4paper,10pt]{article}
\pdfoutput=1 
\usepackage{jheppub} 
\usepackage{float}
\usepackage{color}
\usepackage[dvipsnames]{xcolor}
\usepackage{newlfont}
\usepackage{graphicx}
\usepackage{amssymb}
\usepackage{amsmath}
\usepackage{float}
\usepackage{amsfonts}
\usepackage{tablefootnote}

\usepackage{bm}
\usepackage{verbatim}
\usepackage[latin1]{inputenc}
\usepackage{bbm}
\usepackage{multirow}
\usepackage[thinlines]{easytable}
\usepackage{braket}
\usepackage{amsthm}
\usepackage{bbold}
\usepackage{mathrsfs}
\usepackage[varg]{txfonts} 
\usepackage{amsfonts}

\hypersetup{linkcolor=red,          
	citecolor=magenta,        
	filecolor=magenta,      
	urlcolor=teal}    

\newcommand{\ba}{\begin{array}}
\newcommand{\ea}{\end{array}}

\newcommand{\mc}{\mathcal}

\newcommand{\mb}{\mathbb}

\newcommand{\mbf}{\mathbf}

\newcommand{\td}{\text{d}}

\newcommand{\mcK}{\mathcal{K}}
\newcommand{\mcC}{\mathcal{C}}
\newcommand{\mcW}{\mathcal{W}}
\newcommand{\bmal}{\bm{\alpha}}
\newcommand{\bmx}{\bm{x}}
\newcommand{\bmX}{\bm{X}}
\newcommand{\bmxi}{\bm{\xi}}
\newcommand{\op}[2]{\mathbf{#1}_{#2}}
\newcommand{\tiW}{\tilde{W}}

\newcommand{\avgc}[1]{\big<\big<\, #1\,\big>\big>}
\newcommand{\ketc}[1]{\big| \, #1\, \big>\big>}
\newcommand{\brac}[1]{\big<\big<\, #1\, \big|}
\newcommand{\comt}[1]{[ \, #1\, ]}
\newcommand{\braketb}[1]{\Big<\, #1\,\Big>}
\newcommand{\bift}[1]{\Big( #1\Big)}
\newcommand{\bitd}[1]{\Big[ #1\Big]}
\newcommand{\Bitd}[1]{\Bigg[ #1\Bigg]}
\newcommand{\tr}{\text{Tr}}
\newcommand{\bbf}[1]{\tilde{\mathbb{#1}}}
\newcommand{\calch}[1]{\check{\mathcal{#1}}}
\newcommand{\texch}[1]{\check{\text{#1}}}
\newcommand{\bbch}[1]{\check{\mathbb{#1}}}
\newcommand{\suop}[2]{\mathbf{#1}_{\bm{#2}}}
\newcommand{\suopf}[2]{\tilde{\mathbf{#1}}_{\bm{#2}}}
\newcommand{\tilbb}[1]{\tilde{\mathbb{#1}}}
\newcommand{\calchs}[3]{\check{\mathcal{#1}}_{\bm{#2}, \bm{#3}}}

\title{A phase space approach to the wavefunction spreading and operator growth in the Krylov basis}

\author[a]{Kunal Pal,}
\author[b]{Kuntal Pal,}
\author[b,c]{Keun-Young Kim}
\affiliation[a]{Asia Pacific Center for Theoretical Physics, Pohang 37673, Republic of Korea}
\affiliation[b]{Department of Physics and Photon Science, Gwangju Institute of Science and Technology, 123 Cheomdan-gwagiro, Gwangju 61005, Republic of Korea}\affiliation[c]{Research Center for Photon Science Technology, Gwangju Institute of Science and Technology, 123 Cheomdan-gwagiro, Gwangju 61005, Republic of Korea}
\emailAdd{kunal.pal@apctp.org,  kuntalpal@gist.ac.kr, fortoe@gist.ac.kr}

\abstract{
    In the Wigner-Weyl phase space formulation of quantum mechanics, we analyse the problem of the spreading of an initial state or an initial operator under time evolution when described in terms of the Krylov basis. After constructing the phase space functions corresponding to the Krylov basis states generated by a Hamiltonian from a given initial state by using the Weyl transformation,  we subsequently use them to cast the Krylov state complexity as an integral over the phase space in terms of the Wigner function of the time-evolved initial state, so that the contribution of the classical Liouville equation and higher-order quantum corrections to the Wigner function time evolution equation towards the Krylov state complexity can be identified. Next, we construct the double phase space functions associated with the Krylov basis for operators by using a suitable generalisation of the Weyl transformation applicable for superoperators, and use them to rewrite the Krylov operator complexity as an integral over the double phase space in terms of a generalisation of the usual Wigner function. These results, in particular, show that the complexity measures based on the expansion of a time-evolved state (or an operator) in the Krylov basis can be thought to belong to a general class of complexity measures constructed from the expansion coefficients of the time-dependent Wigner function in an orthonormal basis in the phase space, and help us to connect these complexity measures with measures of complexity of time evolved state based on harmonic expansion of the time-dependent Wigner function. }
    
\begin{document}
	\maketitle
\newpage
\section{Introduction}
The phase-space formulation of quantum mechanics, which dates back to the early days of the development of the theory, is an elegant and powerful formalism that not only makes the bridge from classical to quantum dynamics logical and rigorous but also highlights the striking features of quantum mechanics very clearly. The possibility of a phase-space description of quantised systems, with non-commutating dynamical variables, has its root in the work of Weyl on the quantisation of classical dynamical systems \cite{weyl1950theory}, and by Wigner \cite{wigner1932quantum}, who, in the context of quantum mechanical corrections to the classical statistical mechanics, first introduced the quasiprobability distribution defined in terms of the phase space variables, now known as the Wigner function. The underlying connection between Wigner and Weyl's approaches, on the other hand, was understood rigorously after the seminal works of Groenewold \cite{Groenewold} and Moyal \cite{moyal1949quantum}.  Roughly, in the phase space formulation, if we consider a set of classical observable functions $f(q,p)$ of the phase space canonical coordinates $(q,p)$, then the Weyl quantisation can be performed within this set of observables, and essentially amounts to replacing the classical Poisson bracket with the so-called Moyal bracket defined in terms of the Moyal star product (which, in turn, is defined as the conjugation between the Weyl symbol associated with a quantum mechanical operator), and replacing the classical probability distribution functions by their quantum counterparts.  
One of the most important advantageous features of this phase space formulation of quantum mechanics is that, since it provides a way of computing quantum mechanical expectation values as a classical-like average over scalar variables, the quantum corrections become more transparent in this formulation, allowing a smooth transition from quantum to classical mechanics \cite{Hillery:1983ms, Hirshfeld:2002yjb, marino2021advanced}. 

The phase-space description of quantum dynamics similarly maps the von Neumann equation for the time evolution of density matrices to the quantum Liouville equation on phase space, which determines the time evolution of the Wigner function, and is strongly reminiscent of the classical Liouville equation in terms of the phase-space distribution of classical dynamics \cite{wigner1932quantum, Hillery:1983ms}. In other words, in this approach, one 
can formulate the dynamics in both quantum and classical mechanics in terms of unitary flows on the Hilbert space of square-integrable functions defined on the phase space. In classical dynamics, these flows are generated by the observables $f(q,p)$ acting through the Poisson bracket in the usual manner, while for quantum dynamics, these flows are generated by the Moyal bracket \cite{Hillery:1983ms, Osborn:1994sf, blaszak2019quantum}. 
The phase space formulation of quantum mechanics has been utilised to address a wide range of problems, including the understanding of quantum signatures of chaos in quantum systems \cite{berry1977semi, TAKAHASHI}, quantum mechanical scattering and collision problems \cite{remler1975use, RevModPhys.55.245}, and probing sub-Planckian structures in phase space decoherence \cite{zurek2001sub}.\footnote{See \cite{kim1991phase, kim2005physics, Hillery:1983ms} for various other applications of the phase space formulation of quantum mechanics.}  

This approach to the Hamiltonian dynamics of the quantum mechanical system provides, in particular, a natural way to describe the complexity of time evolution of a quantum system, with a well-defined classical limit, in parallel with the classical counterpart, a topic we will be interested in the current work.\footnote{For an extensive discussion of chaos and Lyapunov exponent in the context of classical and quantum distribution functions, we refer to \cite{PhysRevE.56.5174, gong2003chaos, gu1985group, Wang:2019juv, Sokolev, balachandran2010phase}.} The complexity of a quantum mechanical system, a manifestation of the underlying irreversibility of the dynamics, has gained a lot of interest in the recent literature, in particular for lattice systems, where the complexity is often quantified as how an initial `simple' operator becomes `complex' at later times under the Heisenberg evolution generated by the Hamiltonian. The complexity of operator growth is then on a similar footing to the scrambling of information throughout the degrees of freedom of the quantum systems under the evolution.

Building upon the concept of operator growth for generic Hamiltonian systems, the \textit{universal operator growth hypothesis} was advanced in \cite{Parker:2018yvk}, which was subsequently verified by numerous numerical and, in some cases, analytical examples. At the heart of this hypothesis is the Lanczos recursion procedure, which generates the so-called Krylov basis, which is the orthonormal set of (operator) basis constructed from the set of (non-normalised) bases generated by the Hamiltonian evolution of the initial `simple' operator. Furthermore, as a quantifier of the complexity of a Heisenberg time-evolved operator, subsequently known as the Krylov operator complexity, was introduced in this work, and it was shown that it provides an upper bound to a family of generalised complexities up to a multiplicative constant factor. 

The analogue of this approach of quantifying the complexity of the time evolution of a Heisenberg operator based on Krylov subspace, for Schrodinger time evolution of an initial state by the governing Hamiltonian, by employing the Krylov subspace generated by the Hamiltonian was constructed in \cite{Balasubramanian:2022tpr}, where the fact that at any instant of time after the start of the time evolution, the evolved state is contained in this subspace of the full Hilbert space was utilised to show the special role played by the corresponding Krylov basis in the time evolution.

These two elegant ways of quantifying the complexity evolution of an initial operator or state have received a flurry of further exploration in the recent literature, which not only shed light on fundamental issues of quantum dynamics of classical integrable and chaotic systems, but also proved to be an unifying theme between a wide range of novel quantum phenomena. For example, the long-time behaviour of Krylov operator complexity were elucidated in \cite{Barbon:2019wsy}, operator growth in nonintegrable quantum systems were explored in  \cite{Dymarsky:2019elm}, Krylov complexity in two dimensional conformal field theory in various contexts were studied in \cite{Dymarsky:2021bjq, Banerjee:2022ime, Kundu:2023hbk, Malvimat:2024vhr, Chattopadhyay:2024pdj, Caputa:2025local}, suppression of Krylov complexity in integrable models were established in \cite{Rabinovici:2021qqt}, universality of Krylov complexity and the entropy of operator growth was pointed out in \cite{Fan:2022xaa}, Krylov subspace construction was extended for open quantum system in \cite{Bhattacharyaopen, Liu:2022open}, it was used to quantify the properties of autocorrelation function in \cite{Zhang:2023wtr}, Lanczos spectrum for random matrix theory was studied in \cite{Loc:2024oen}. 
 
Furthermore, Krylov complexity for Calabi-Yau quantum mechanics was studied in \cite{Du:2022ocp}, for quantum field theory in \cite{Avdoshkin:2022xuw, Camargo:2022rnt}, for Lifshitz scalar field theories in \cite{Vasli:2023syq}, for matrix models in \cite{Iizuka:2023fba}, for Gaussian states in \cite{Adhikari:2023evu}, for density matrix operators in \cite{Caputa:2024density}, for finite temperature scalar field theory in \cite{He:2024xjp}, for Schrodinger field theory in \cite{He:2025guu, He:2024hkw}. Various generalisations of Krylov complexity have also appeared in \cite{Fan:2023ohh, He:2024pox, FarajiAstaneh2, Patramanis:2023cwz}.

On the other hand,  the Krylov state complexity was used to probe quantum phase transition in the Lipkin-Meshkov-Glick model in \cite{Afrasiar:2022efk, Bento:2023bjn};  
to quantify the scar states in \cite{Hu:2025scar, PhysRevB.106.205150, Nandy:2023brt}; to study dynamical phase transition in multi-mode Bose-Einstein condensate in \cite{Zhou:2024rtg}; to quantify integrability to chaotic transition in spin systems in \cite{Camargo:2024deu, FarajiAstaneh1}; in the context of quantum maps and qubit dynamics in \cite{Seetharaman:2024ket}, mixed phase space in \cite{Huh:2024mixedphase}, analogues representing gravitational systems in \cite{Jeong:2024brick, Aguilar-Gutierrez:2025hbf}, as well as for dynamical phase transition in \cite{Takahashi:2025mdt}, driven systems in \cite{Nizami:2024ltk, PhysRevE.108.054222} and free Fermion models in \cite{Gautam:2023pny}.  
 
A very generic relationship between the moments of the work done on a quantum system under a sudden quench of the Hamiltonian parameters and the Lanczos coefficients for a time evolution under the post-quench Hamiltonian was established in \cite{Palstatistics,Gill:2023umm}, providing a physical interpretation of the Lanczos coefficients in this setting.  Different features of the time evolution of Krylov state complexity in interacting quantum systems after a sudden quantum quench, as dictated by the survival probability of the initial state, were elucidated in \cite{Gautam:2023bcm}.\footnote{This set of references represents only a very small sample of the extensive works that have appeared in the recent literature. Explorations of Krylov complexity when the evolution is generated by a non-Hermitian Hamiltonian, 

were carried out in \cite{Bhattacharya:2023yec, Bhattacharya:2024hto}, 
whereas, it was explored for a non-Hermitian Ising chain in \cite{Medina-Guerra, Medina-Guerra2}. 
A possible way to measure Krylov state complexity by utilising a measurement-efficient basis was proposed in \cite{Cindrak:2024exj}. Information theoretic quantities motivated by Krylov state complexity in the context of reservoir computing were explored in \cite{Cindrak:2024reservoirs}, it was used as a probe of confinement in \cite{Jiang:2025wpj}. The relationship between Krylov complexity and operator quantum speed limit was exemplified in \cite{Gill:2024acg}, whereas in \cite{Aguilar-Gutierrez:2023nyk} it was argued that it does not measure any `distance' between quantum states. For an overview of different  applications of this approach to quantum dynamics, as well as more complete references of the Krylov subspace methods in quantum dynamics of operators and states, we refer to the recent reviews \cite{Nandy:2024evd, Rabinovici:2025otw}.}

A central point of this work, which connects the above discussions on the phase space picture of time evolution and its complexity, is that  \textit{the time evolution of the complexity of a quantum state should have an equivalent understanding as the evolution of the complexity of the Wigner function of the time-evolved state}. Building upon this philosophy, in the first part of this work, we establish several properties of the Wigner functions associated with the Krylov basis generated by the Hamiltonian of a quantum system with continuous phase space variables, starting from an arbitrary initial state on the Hilbert space (see Sections \ref{Ph_sp_Krylov} and \ref{Krylov_and_Moyal}).\footnote{Here we note that the phase space Wigner functions corresponding to the Krylov basis vectors was studied previously for systems with finite phase space variables in  \cite{Basu:2024tgg}.} These functions, in particular, help us to write a version of the Lanczos algorithm in terms of functions defined in the phase space, and rewrite the Krylov state complexity as a phase space average of the Wigner function of the time-evolved state and a function constructed from these phase space Krylov functions (see Section \ref{SC_phase}). As a direct consequence of this, using the well-known time evolution equation for the pure state Wigner functions - the so-called quantum Liouville equation, we can separate the classical and quantum contributions of the quantum Liouville equation to the time derivative of the Krylov state complexity (Section \ref{sec_WF_dot}).

Importantly, this interpretation of the Krylov state complexity as a phase space average over the Wigner function makes it possible to connect the Krylov state complexity with the notion of complexity of time evolution, based on the growth of the harmonics of the time-evolved Wigner functions, which have been studied extensively in the literature \cite{Beneti, Sokolev, balachandran2010phase}. Specifically, we show that the Krylov state complexity and the complexity based on the harmonic expansion of the Wigner function can be considered to belong to the same general class of quantities one can construct from the time-dependent coefficients of an expansion of the Wigner function with respect to a complete orthonormal basis for the phase space functions (Section \ref{SC_phase}). 

A natural question one can ask at this point is how to extend this phase-space-based approach to describe the growth of Heisenberg time-evolved operators and the Krylov supervector basis generated by the Liouvillian. To address this question, a significant portion of the present work (Section \ref{sec_op_phase_sp}) focuses on the phase space representation of superoperators. Here, our goal is two-fold. First, our aim is to provide a unified description of the method to determine the canonical set of superoperators (and their bases) in the operator Hilbert space (the Liouville space), which are essential for defining a generalisation of the Weyl transformation of a general superoperator. To this end, we provide a systematic derivation of these canonical superoperators and their associated bases (along the lines of \cite{Royer2}), and then interpret them in terms of an extended version of phase space. Second, we then use this phase space representation of Liouville space superoperators in the special case where these supervectors are elements of the Krylov superbasis. This enables us to develop a phase space version of the Lanczos algorithm for generating the Krylov subspace associated with an initial operator, and to interpret the Krylov operator complexity as an average over an extended version of the usual phase space. 
As a byproduct of our definition of the Weyl transformations of superoperators, we can obtain explicit expressions for the Weyl transforms of large classes of superoperators, important for this purpose, directly in terms of the Weyl symbols of the underlying operators. Furthermore, for the purpose of illustration, some specific examples of the computation of Wigner functions associated with superoperators constructed from Weyl-ordered operators have been computed explicitly in this section. 

Important findings of this work are summarised in Section \ref{sec_dis_conc}, where we also provide some indications along the directions this work can be extended. This paper also contains several Appendices, where we have provided some important, but mostly technical details, that were left out of the main text. 

\textbf{Note added.} In a recent ongoing work \cite{Inprep2}, the authors have studied complexity measures based on the Krylov subspace method from the perspective of classical phase space and used them to address the so-called saddle-dominated scrambling. It would be interesting to see how the results of \cite{Inprep2} are related to those of ours. We thank the authors of \cite{Inprep2} for related discussions. 

\section{Weyl transformation, Wigner function and the Krylov basis }

In this section, we first discuss some elementary relations connecting the Wigner function, Weyl transform, and their Fourier transforms. Then we shall utilise these to define the Wigner functions for a specific set of orthonormal complete basis, the Krylov basis generated by the Hamiltonian starting from a given initial state. The system we shall consider in this paper is a quantum system where a particle moves in a potential, so that the corresponding classical phase space can be represented by the position and momentum of the particle, $q$ and $p$, respectively.\footnote{In this paper, we consider one-dimensional systems only. However, the following discussions can be straightforwardly generalised for a particle moving in a higher-dimensional potential as well \cite{Hillery:1983ms}.} 
Our goal here is to write down some important relations that appear when one considers the spreading of an initial wavefunction with time in the Krylov basis in terms of the Weyl transformations of some related operators, such as the density matrix of the state under consideration and the spreading operator in the Krylov basis for such systems having a two-dimensional phase space. Specifically, as we shall show in a later section, from these definitions, it is possible to understand the notion of complexity of a time-evolved state defined with respect to the Krylov basis generated by the Hamiltonian as an integral over the phase space of the system under consideration.

\textbf{Notations.}
We denote by $\bm{x}$ the vector $\bm{x}=(q,p)^T$ and with $\bm{\alpha}$ the vector, $\bm{\alpha}=(\alpha_1, \alpha_2)^T$ with real components. Meanwhile, $\bmX=(\mathbf{Q, \mathbf{P}})^T$ is the vector containing the position and the momentum operators.  We will use these compact notations alternatively, depending on which one is convenient. In addition, we shall denote $\int \td \bm{x} = \int \td q ~\td p$, $\delta(\bmx-\bmx^\prime)=\delta(q-q^\prime)\delta(p-p^\prime)$, and a function of $q$ and $p$ by $f(\bmx) = f(q,p)$. Other notations used will be specified in the text as they are introduced. 

\subsection{Weyl transformation and Wigner function}\label{Weyl_Wigner}

We start by defining the \textit{Weyl symbol} of an operator $\mathbf{O}$\footnote{In this paper, we shall consider the Weyl symbol of self-adjoint operators only. Quantum mechanical operators are denoted in this paper by bold-faced capital letters, such as $\mbf{A}$ (with some standard exceptions, such as the density matrix operator, or the creation and annihilation operators). The Weyl transformation of an operator is denoted by the blackboard version, e.g., $\mb{A}$. }, here denoted by $\mathbb{O}(q,p)$, through the \textit{Weyl transform}\footnote{This procedure is sometimes refereed to as the \textit{Wigner transform} or \textit{Weyl dequantisation} map, i.e., $\mathbb{O}(q,p)=\mc{Q}_W^{-1}(\mathbf{O})$ \cite{Ali:2004ft}.} \cite{Hillery:1983ms,Case, leaf1968weyl} 
\begin{equation}\label{Weyl-tr}
	\mathbb{O}(q,p) := 2 \int e^{2i p y/\hbar}  \bra{q - y} \mathbf{O} \ket{q+ y}   ~\td y~,
\end{equation}
which maps operators in the Hilbert space $L^2 (\mb{R})$ to functions in the phase space ($\mb{R}^2 \simeq T^*\mb{R}$). 
Equivalently, in terms of the momentum space eigenfunctions, the Weyl symbol can be defined as 
\begin{equation}\label{Weyl-tr2}
	\mathbb{O}(q,p) := 2 \int e^{-2i q k/\hbar} \bra{p - k} \mathbf{O} \ket{p + k} ~\td k~.
\end{equation}

From these definitions, it can be seen that the trace of the product of two operators is proportional to the integral of the product of their Weyl  transforms over the phase space, i.e., 
\begin{equation}\label{transitions}
	2\pi \hbar ~\text{Tr} [\mathbf{O}_1 \mathbf{O}_2] =  \int \mathbb{O}_1 (q,p)~ \mathbb{O}_2 (q,p) ~ \td p ~\td q~,
\end{equation}
and the Weyl transformations of Hermitian operators are real functions of phase space coordinates. The Weyl transformation of the density matrix ($\rho$) is the \textit{Wigner function}\footnote{In this paper, we mainly follow conventions of \cite{Hillery:1983ms}. 
The Wigner function is usually defined with an extra factor of $2 \pi\hbar$ in the denominator compared to the Weyl transformation. From now on, in most parts of this work, we shall set $\hbar=1$, and it will be restored when necessary. We shall not specify the limits of the integrals, which, unless specified otherwise, are from $-\infty$ to $+\infty$, at least for the case of continuous (and unbounded)  phase space variables we are considering. }
\begin{equation}\label{Wignerfunction}
	W (q,p) = \frac{1}{\pi \hbar} \int e^{2i p y/\hbar}  \bra{q - y} \rho \ket{q+ y} ~  \td y~.
\end{equation}
The integral of the Wigner function, defined through this relation, over the entire phase space is unity, while a partial integral over momentum and position coordinates (the marginal distributions) gives the position and momentum space probability distribution, respectively \cite{Hillery:1983ms}. 
For a pure state with $\rho=\ket{\psi}\bra{\psi}$, the Wigner function is therefore given in terms of the position space wavefunctions by the following relation\footnote{To simplify the notions, we shall often use the convention that, when the argument of a position-space wavefunction $q-y$, we denote it as $\psi$, while if the argument is $q+y$, we denote it as $\chi= \psi(q+y)$. } \cite{wigner1932quantum, moyal1949quantum, kim1991phase}
\begin{equation}
	 W (q,p) = \frac{1}{\pi} \int e^{2i p y}  \psi(q-y) ~ \psi^{*}(q+y) ~\td y~.
\end{equation}
Since its introduction by Wigner in 1932 to study quantum corrections to thermodynamic equilibrium \cite{wigner1932quantum}, the Wigner function has been assigned different interpretations. E.g., we refer to \cite{Royer}, where the Wigner function has been interpreted as the expectation value of a parity operator at the phase space point $(q,p)$, and \cite{bondar}, where it was argued that the Wigner quasiprobability distribution is the probability amplitude for a quantum particle to be at some position in the corresponding classical phase space.

One important relation that can be directly obtained from the above definition of the Wigner function is the following. Consider two normalised pure states, $\ket{\psi_1}$, and $\ket{\psi_2}$. If we denote the corresponding Wigner functions by $W_1(q,p)$, and $W_2(q,p)$, respectively, then the transition probability between the states can be written as  
\begin{equation}\label{transition_state}
		|\braket{\psi_1|\psi_2}|^2 = 2\pi \int W_1(q,p) ~W_2 (q,p) ~\td q~ \td p~. 
\end{equation}
Furthermore, the expectation value of an operator $\mathbf{O}$ in a state described by the density matrix $\rho$ can be written as
\begin{equation}\label{expecttaion}
	\braket{\mathbf{O}} = \text{Tr} [ \rho \mathbf{O}] = \int \mathbb{O} (q,p)~ W(q,p) ~\td q~ \td p~.
\end{equation}
Unless stated otherwise, for the purpose of this manuscript, we shall assume that the state of the system under consideration is a pure state.

\paragraph{\textbf{Moyal star product and the phase space functions.}}
The Weyl transform of the product of two (properly ordered) operators corresponds to the star product of the individual Weyl transforms \cite{Groenewold, Hirshfeld:2002yjb}, i.e., 
\begin{equation}\label{star_def}
    \mathbb{A}(q,p) \star \mathbb{B}(q,p) =2 \int e^{2ipy} ~\bra{q - y} \mathbf{A} \mathbf{B} \ket{q+ y}   ~\td y~. 
\end{equation}
This relation corresponds to the statement that two generic functions $\mathbb{A}(q,p)$ and $\mathbb{B}(q,p)$ defined in the phase space compose through the star product non-commutatively. More precisely, if we consider the Weyl transform of two Weyl-ordered operators, then it corresponds to the star product of their Weyl transforms (sometimes also called the classical kernel of the operator). 
Now, substituting the relation\footnote{This rule for associating an operator with a phase space function is known as the \textit{Weyl correspondence formula}. For other rules, such as the standard and anti-standard rules for position and momentum operators or the normal and anti-normal ordering rules with the creation and annihilation operators, we refer to \cite{agarwal1970calculus, Agarwal2, gadella1995moyal, Hillery:1983ms} and references therein.} (which is an operator integral) \cite{Hillery:1983ms}
\begin{equation}\label{Weyl_to_op}
    \mathbf{A}(\bm{X}) = \frac{1}{(2 \pi)^2} \int \td \bm{\alpha} \td \bm{x} ~\mathbb{A}(\bm{x})~ \exp{\Big[i \bm{\alpha}. (\bm{X-\bm{x)}}\Big]}
\end{equation}
between an operator $\mathbf{A}(\bm{X})$ and its Weyl transform (equivalent to \eqref{Weyl-tr}) in \eqref{star_def}, we have the following conventional integral representation of the star product\footnote{There are several other different representation of the $\star$-product, specifically differentiation based formulas involving the so-called Bopp operators.  In general, these formulas provide an asymptotic expansion in $\hbar$ of the Weyl symbol of the product $\mbf{AB}$. See, e.g.,  \cite{Hillery:1983ms, Hirshfeld:2002yjb} for these expressions of $\star$-product. }  \cite{baker, Hirshfeld:2002yjb, marino2021advanced}
\begin{equation}\label{star_Fourier}
\begin{aligned}
    \mathbb{A}(\bm{x}) \star \mathbb{B}(\bm{x}) &= \frac{1}{\pi^2} \int \td \bm{x}_1 ~\td \bm{x}_2 ~ \mathbb{A}(\bm{x}_1) ~\mathbb{B}(\bm{x}_2)~\exp{\big[ 4 i \Delta(\bm{x}, \bm{x}_1,\bm{x_2})\big]}~\\
    &=\frac{1}{\pi^2} \int \td \bm{x}_1 ~\td \bm{x}_2 ~ \mathbb{A}(\bmx+\bm{x}_1) ~\mathbb{B}(\bmx+\bm{x}_2)~\exp{\big[ 2 i \braket{\bmx_1,\bmx_2}_{s}\big]}~,
        \end{aligned}
\end{equation}
where $\Delta(\bm{x}, \bm{x}_1,\bm{x_2})$ denotes the area of the triangle made by the points $(\bm{x, \bm{x}_1, \bm{x_2})}$ in the phase space (see e.g., \cite{Zachos:2000zh}), and is given by the expression,
\begin{equation}\label{area_triang}
    \Delta(\bm{x}, \bm{x}_1,\bm{x_2}) = \frac{1}{2} \Big[p(q_2-q_1)+ p_1(q-q_2)+p_2(q_1-q)\Big]~=\frac{1}{2} \Big[ \braket{\bmx, \bm{x}_1}_s +\braket{\bmx_1, \bm{x}_2}_s+\braket{\bmx_2, \bm{x}}_s\Big]~.
\end{equation}
In the second expression, we have rewritten the area in terms of the symplectic product, $\braket{\bmx_1, \bmx_2}_s=q_1p_2-p_1q_2$. 

Using the star product formula above, one can write the trace relation of two operators, such as eq. \eqref{transitions}, as the phase space integrals over the corresponding star products \cite{fairlie1964formulation, gadella1995moyal}
\begin{equation}\label{star_to_pq}
\begin{split}
    2\pi \hbar ~\text{Tr} [\mathbf{O}_1 \mathbf{O}_2] =    \int \mathbb{O}_1 (q,p)~ \mathbb{O}_2 (q,p) ~ \td p ~\td q~=\int \mathbb{O}_1 (q,p)\star \mathbb{O}_2 (q,p) ~ \td p ~\td q~\\
        =\int \mathbb{O}_2 (q,p)\star \mathbb{O}_1 (q,p) ~ \td p ~\td q~.
\end{split}
\end{equation}

We will use the above relations for the Krylov basis later in this paper. However, before proceeding further, it is important to note that, since we have defined the Wigner function \eqref{Wignerfunction} with a different normalisation compared to the Weyl transform of a generic operator (eq. \eqref{Weyl-tr}), when the relations in \eqref{star_def} and \eqref{star_Fourier} are used for a density matrix or its Wigner function, we need to be careful with the overall normalisation. Specifically, in \eqref{star_def} we need to divide by a factor of $2 \pi$ in the right-hand side, which would correspond to a multiplication by this factor in the right-hand side of \eqref{Weyl_to_op}, thereby leaving the overall constant factor in \eqref{star_Fourier} intact. 

As an example of the $\star$-product composition, consider the eigenvalue equation for the Hamiltonian $\mbf{H}(\bm{X})\ket{E_n}=E_n\ket{E_n}$, then the Wigner function corresponding to the eigenstate density matrix $\rho_n=\ket{E_n}\bra{E_n}$ satisfy the star product equations \cite{fairlie1964formulation, Hirshfeld:2002yjb}: $\mb{H}(\bm{x)}\star W_{nn}=E_n W_{nn}$ and $W_{nn} \star \mb{H}(\bm{x)}=E_n W_{nn}$, for $E_{n} \in \mb{R}$. More generally, for the operator $\rho_{nm}=\ket{n}\bra{m}$, one has (these are known as the \textit{star-genvalue equation} in the literature \cite{Hirshfeld:2002yjb}),
\begin{equation}\label{star_eigen}
    \mb{H}(\bm{x})\star W_{nm}=E_n W_{nm},~~ \text{and}~~W_{nm} \star \mb{H}(\bm{x})=E_m W_{nm}~.
\end{equation} 
Defining a commutator through the $\star$-product \cite{moyal1949quantum, baker, Hirshfeld:2002yjb}
\begin{equation}
    [\mb{A}(\bm{x}), \mb{B}(\bm{x})]_\star=\mb{A}(\bm{x}) \star \mb{B}(\bm{x}) - \mb{B}(\bm{x}) \star \mb{A}(\bm{x})~, 
\end{equation}
we have, $[\mb{H}(\bm{x}), W_{nm}(\bm{x)}]_\star=(E_n-E_m)W_{nm}(\bm{x})$. The $\star$-product commutator is also known as the Moyal bracket, since this is the image of the usual quantum commutator under Weyl transformation.\footnote{Some authors also define the Moyal bracket with an additional factor of (restoring $\hbar) $ $(i\hbar)^{-1}$ with the commutator, see, e.g., \cite{Osborn:1994sf, prosser1983correspondence}. }
\subsection{Phase space Krylov functions}\label{Ph_sp_Krylov}
As pointed out first by Moyal \cite{moyal1949quantum}, an orthonormal complete basis in the Hilbert space, under Weyl transform, maps to a set complete orthonormal functions in the phase space, so that with respect to this basis, any well-behaved phase space distribution function can be expanded (for the purposes of this paper, these functions are $L^2(\mb{R}^2)$). 
We now consider the Krylov basis, $\ket{K_n}$, generated by the Hamiltonian of the quantum system under consideration, acting on a given initial state $\ket{\psi_0}$, and find the complete orthonormal set of functions, and subsequently study various properties of them. 
The action of the Hamiltonian on the $n$th Krylov basis element is given by
\begin{equation}\label{H_Krylov}
     \mbf{H} \ket{K_n}=a_n \ket{K_n} + b_{n+1}\ket{K_{n+1}}+b_n \ket{K_{n-1}}~,
\end{equation}
where $a_n$ and $b_n$ are two sets of Lanczos coefficients \cite{viswanath1994recursion, Dehesa, golub2009matrices}. The first element of the Krylov subspace is the initial state $\ket{K_0}=\ket{\psi_0}$, and the $n$-th element is related to the initial state as $\ket{K_n}=\mc{P}_n(\mbf{H})\ket{\psi_0}$, where $\mc{P}_n(\mbf{H})$ denote a set of normalised orthogonal polynomials (usually known as the Krylov polynomials, with $n=0 \cdots D_K-1$, $D_K$ being the dimension of the Krylov subspace) which satisfy the recurrence relation, $\mbf{H} \mc{P}_n(\mbf{H}) = a_n \mc{P}_n(\mbf{H}) + b_{n+1} \mc{P}_{n+1}(\mbf{H}) +b_n \mc{P}_{n-1}(\mbf{H})$, with $\mc{P}_0(\mbf{H})=1$, and $b_0=0$ \cite{Dehesa}. Furthermore, these polynomials form a complete basis in the space of the polynomials of degree $D_K-1$ over the spectrum of the restriction of the Hamiltonian over the Krylov subspace\footnote{From now on, we shall assume that the shape of the potential energy function is such that there are no degeneracies in the energy spectrum.} (from the initial vector $\ket{\psi_0}$), i.e., 
\begin{equation}\label{polyn_comt}
    \sum_{n=0}^{D_K-1} \mc{P}_n(E_a)\mc{P}_n(E_b) =|\braket{E_a|\psi_0}|^{-2} \delta_{ab}~.
\end{equation}

Now we write down the expansion of the Wigner function at an instant of time $t$, in terms of the Krylov wavefunctions  $\phi_n(t)=\braket{K_n|\Psi(t)}$, which are commonly utilised to define a notion of complexity associated with the spreading of the time-evolved wavefunction $\ket{\Psi(t)}=\exp{(-i \mbf{H}t) \ket{\psi_0}}$. Using the definition of the Wigner function in \eqref{Wignerfunction} and the completeness of the Krylov basis, one can derive the following expression for the Wigner function for the time-evolved state,
\begin{equation}\label{WFexpanK}
    \begin{aligned}
		W(q,p,t)&=  \frac{1}{\pi} \sum_{n,m}   \int e^{2i p y} ~\psi^K_n(q-y)~ \psi_m^{K*} (q+y)~ \phi_n(t) ~\phi_m^{*}(t) ~\td y ~\\
		&=\sum_{n,m}  \mathcal{W}^K_{nm} (p,q)~ \phi_n(t)~ \phi_m^{*}(t) ~. \qquad \qquad
\end{aligned}
\end{equation}

Here we have denoted the position space wavefunctions in the Krylov basis as $\psi^K_n(q)=  \braket{q| K_n}$, and in the second expression, we have defined  the functions $ \mathcal{W}^K_{nm} (p,q)$ by separating the   time-independent part,  which is the following integral, 
\begin{equation}\label{W_nm_K}
	\mathcal{W}^K_{nm} (p,q)  = \frac{1}{\pi}  \int e^{2i p y} ~\psi^K_n(q-y) \psi_m^{K*} (q+y) ~dy ~.
\end{equation} 
These are just the Weyl transformation of the operators $\mathbf{P}_{nm}= \ket{K_n}\bra{K_m}$ defined in terms of the Krylov basis elements. As we have mentioned previously, in a recent work \cite{Basu:2024tgg}, the Wigner functions corresponding to the Krylov basis vectors were defined and studied for systems with finite phase space variables, specifically in the context of random matrices.

\paragraph{\textbf{Properties of $\mathcal{W}^K_{nm} (\bm{x})$.}}
We now discuss some properties and explicit expressions of the functions $\mathcal{W}^K_{nm} (p,q) $, some of which will play important roles in the latter part of the paper.

1. First note that the elements $\mathcal{W}^K_{nm} (p,q)$ form a Hermitian matrix, since, $\mathcal{W}_{mn}^{K*} (p,q)=\mathcal{W}^K_{nm} (p,q)$.  Also, using the definition of the Krylov basis, one can check that these functions satisfy the following useful relation
\begin{equation}\label{antisW}
	\begin{split}
		\mathcal{W}^K_{[n(n+1)]}(q,p)=  \frac{i b^{-1}_{n+1}}{2\pi} \int dy ~ e^{2i p y}  \bra{q - y} [\mathbf{P}_n, H] 
		- 2 b_n  \mathbf{P}_{[n(n-1)]} 	\ket{q+ y}~,
	\end{split}
\end{equation}
where we have defined the operators, $\mathbf{P}_n= \ket{K_n}\bra{K_n}$ which is the projector onto the $n$-th Krylov basis, $\mathbf{P}_{nm}= \ket{K_n}\bra{K_m}$
(with $m \neq n$), and the square brackets around the indices denote anti-symmetrisation. When the Wigner function is expanded in an arbitrary orthonormal complete basis, a similar set of quantities as those of $\mcW_{nm}(q,p)$ arise, which were referred to as the phase-space eigenfunctions in \cite{moyal1949quantum, leaf1968weyl}, hence we can refer  $\mcW_{nm}(q,p)$  as the \textit{phase space Krylov functions}. The matrix elements of an operator $\mbf{O}$ in the Krylov basis can then be written in terms of these functions as 
\begin{equation}\label{matrix_krylov}
    \braket{K_m|\mbf{O}|K_n} = \int \td \bm{x} ~ \mb{O} (\bm{x}) ~\mcW_{nm}(\bm{x})~.
\end{equation}

2. These functions form a complete orthonormal set, since, using the orthonormality and completeness relation of the Krylov basis, one can show that they satisfy the following relations, 
\begin{equation}\label{W_k_product}
    \int \mathcal{W}^K_{nm} (\bmx) ~ \mathcal{W}_{ij}^{K*}(\bmx) ~\td \bmx = \frac{1}{2 \pi} \delta_{mi} \delta_{nj}~,
\end{equation}
and 
\begin{equation}
    \sum_{nm} \mcW^K_{nm} (q,p) \mcW_{nm}^{K*} (q^\prime,p^\prime) =\frac{1}{2\pi} \delta(q-q^\prime) \delta(p-p^\prime)~. 
\end{equation}
Therefore, $\mcW^K_{nm}(q,p)$ is a complete orthonormal basis of the phase space functions $L^2(\mb{R}^2)$. 
The relation in \eqref{WFexpanK} is just the expansion of the Wigner function of the time-evolved state in terms of this complete orthonormal set of functions. 

3. It is also easy to write down the relationship between the phase space Krylov functions and the corresponding quantities for the energy eigenfunctions ($W_{ab}$). This is given, in terms of the orthogonal polynomials $\mc{P}_n(E)$ by 
\begin{equation}
    \mcW^K_{nm} (\bmx) = \sum_{ab} \braket{E_a|\rho_{0}|E_b}\mc{P}_n(E_a)\mc{P}_m(E_b) W_{ab} (\bmx)~,
\end{equation}
where $\rho_{0}=\ket{\psi_0}\bra{\psi_0}$ is the initial state density matrix, and $W_{ab}(\bmx)$ is the Weyl transform of the operator $\ket{E_{a}}\bra{E_{b}}$ with appropriate normalisation constant \cite{moyal1949quantum}. Using the completeness relation for the Krylov polynomials \eqref{polyn_comt}, or by a direct computation, it can be seen that the above relation between $W_{nm}^{K}(\bmx)$ and $W_{ab}(\bmx)$ can be inverted as
\begin{equation}
  W_{ab} (\bmx)  = \braket{E_b|\rho_{0}|E_a} \sum_{nm} \mc{P}_n(E_a)\mc{P}_m(E_b) \mcW^K_{nm} (\bmx)~,
\end{equation}

4. Using the relation $\ket{K_n}=\mc{P}_n(\mbf{H})\ket{\psi_0}$, we see that these functions have the following simple expressions in terms of the Weyl symbol of the Hamiltonian and the Wigner function of the initial state density matrix, $\rho_{0}=\ket{\psi_0}\bra{\psi_0}$
\begin{equation}
    \mcW_{nm}^K(\bmx)=  \mb{P}_{n}(\bmx)\star W_{0}(\bmx)\star \mb{P}_{m}(\bmx)=\mc{P}_{n}(\star_{n}\mb{H}(\bmx))\star W_0(\bmx) \star \mc{P}_{m}(\star_{m}\mb{H}(\bmx))~,
\end{equation}
where $\mb{P}_{n}(\bmx)$ denotes the Weyl transform of the operator $\mc{P}_n(\mbf{H})$, and in the second expression the compact symbol $\star_{n}\mb{H}$ denotes the star product of $n$ Hamiltonian symbols for each value of $n$ in an expansion of the polynomial $\mc{P}_n(\lambda)$ in powers of $\lambda$. Explicitly computing the expressions for $\mb{P}_{n}(\bmx)$ in terms of $W_{ab}(\bmx)$, the Weyl transform of $\ket{E_{a}}\bra{E_{b}}$, we have,
\begin{equation}\label{WKnm_star_def}
    \mcW_{nm}^K(\bmx) = \sum_{a,b} \mc{P}_{n}(E_{a})\mc{P}_{m}(E_{b}) \bift{W_{aa}(\bmx) \star W_{0}(\bmx) \star W_{bb}(\bmx)} ~.
\end{equation}
For generic initial states, to use this expression, we need to know the Weyl transforms of $\ket{E_{a}}\bra{E_{b}}$ for
diagonal elements of this operator, as well as those of the non-diagonal ones. As an example, 
consider an initial state of the form, which is a generic superposition of the energy eigenstates, 
\begin{equation*}
    \ket{\psi_0} = \sum_{a}c_a \ket{E_{a}}~,
\end{equation*}
so that one needs to evaluate to the $\star$-products of the form $W_{aa}(\bmx) \star W_{ij}(\bmx) \star W_{bb}(\bmx)$ to obtain the phase space Krylov functions, so that $ \mcW_{nm}^K(\bmx)= \sum_{ab}c_ac_b^*\mc{P}_n(E_a)\mc{P}_m(E_b) W_{ab} (\bmx)$.

5. For later use, we note here that the diagonal components of $\mcW_{nm}^K(\bmx)$ are given by  
\begin{equation}\label{Wndef}
	\begin{split}
		W^K_{nn}(q,p) = \frac{1}{\pi} \int e^{2i p y} ~ \braket{q - y| K_n}\braket{K_n | q+ y}  ~ \td y~\\
		=  \frac{1}{\pi} \int e^{2i p y} ~ \psi^K_n(q-y) ~\psi^{K*}_n(q+y)~ \td y~.
	\end{split}
\end{equation}

6. We also note a few other relations in this context: From the orthonormality  of the Krylov basis, we have,
\begin{equation}\label{orthpsi}
	\int \mcW^K_{nm}(q,p) ~ \td q \td p =\int  \psi^{K*}_m(q) \psi^K_n (q) ~ \td q = \delta_{mn}~,
\end{equation}
while the completeness of the Krylov basis indicates that 
\begin{equation}\label{sWn=1}
	2\pi \sum_{n} W^K_{nn} (q,p)=1~.
\end{equation}
Furthermore, these functions satisfy the following relation under the star product,
\begin{equation}\label{WstarW}
    \mcW^K_{nm}(\bm{x})\star \mcW^K_{ij}(\bm{x}) = \frac{1}{2\pi} \delta_{mi} \mcW^K_{nj}(\bm{x})~,
\end{equation}
which can be easily checked to be compatible with eqs. \eqref{W_k_product} and \eqref{orthpsi} by using the relation \eqref{star_to_pq}.

\paragraph{\textbf{Generating functions for the phase space Krylov functions.}}
A compact way of packaging all the phase space Krylov functions is to define the following generating function, in a manner that is analogous to the generating function of the Wigner functions
associated with the energy eigenfunctions (see e.g., \cite{Curtright:2000ux}),
\begin{equation}\label{gen_function}
    G^K(\bm{\mu};\bm{x}) = \sum_{m,n} \frac{\mu_1^m}{\sqrt{m!}} \frac{\mu_2^n}{\sqrt{n!}}~ \mcW_{nm}^K(\bm{x})~,
\end{equation}
where $\bm{\mu}=(\mu_1,\mu_2)$\footnote{Note that $\mu_1$ and $\mu_2$ can, in general, be complex numbers.}, and the sums over the indices $m$ and $n$ are from $0$ to the dimension of the Krylov subspace $D_K$. Later, we will discuss an interpretation of this generating function (after first performing a Fourier transformation) as an expectation value with respect to a specific operator. This expression for the generating function can be rewritten as 
\begin{equation}\label{gen_fun_star}
    G^K(\bm{\mu};\bm{x}) = \sum_{a,b} G_{1}(\mu_1,E_a) G_{2}(\mu_2,E_b) \bift{W_{aa}(\bmx) \star W_{0}(\bmx) \star W_{bb}(\bmx)} ~,
\end{equation}
where we have defined the generating function for the orthogonal polynomials through the relation,
\begin{equation}
    G_{i}(\mu_i,E) = \sum_{n} \frac{\mu_i^n}{\sqrt{n!}} \mc{P}_{n}(E)~.
\end{equation}

Taking different orders of derivatives of the generating function with respect to the parameters $\mu_1$ and $\mu_2$ at $\bm{\mu}=0$ one gets back all the functions $\mcW_{nm}^K(\bm{x})$. 
The average of this function over the phase space is given by 
\begin{equation}
    \int \td \bm{x} ~G^K(\bm{\mu};\bm{x}) = e^{\mu_1\mu_2}~,
\end{equation}
whereas the star composition of two such generating functions can be computed (using \eqref{WstarW}) to be 
\begin{equation}
    G^K(\bm{\mu};\bm{x}) \star G^K(\bm{\nu};\bm{x}) = \frac{1}{2 \pi} e^{\mu_1\nu_2} ~G^K(\mu_2, \nu_1;\bm{x})~. 
\end{equation}
In many situations, the generating functions simplify the computations in phase space. Further relations associated with the generating functions, specifically the star composition of the Hamiltonian kernel and kernel associated with the number operator on the Krylov basis on the generating function, are of considerable interest (see Section \ref{Krylov_and_Moyal}). 

It can be seen that to compute the phase space Krylov functions, either from the expansion in \eqref{WKnm_star_def}, or from the generating function in \eqref{gen_fun_star}, we need to evaluate the $\star$-product of three phase space functions. From the integral representation of the $\star$-product in \eqref{star_Fourier} (and its associativity), one can derive the following integral formula for the required $\star$-product of three generic phase space functions $f_i(\bmx), i=1,2,3$\footnote{The integral representation formula for the $\star$-product of $n$ phase space functions is known, see, e.g., \cite{berezin1974quantization, Osborn:2002yoz, Zachos:2000zh}. There are also other representations for the same (see \cite{Osborn:1994sf, de1998weyl}).}
\begin{equation}\label{star_three}
\begin{aligned}
    f_1(\bmx) \star f_2(\bmx)\star f_3(\bmx) &=\frac{1}{\pi^2} \int \td \bmx_1 \td\bmx_2 \td\bmx_3~ f_1(\bmx_1) f_2(\bmx_2) f_3(\bmx_3) ~\delta\bift{\bmx-\bmx_1+\bmx_2-\bmx_3}\exp{\big[ 4 i \Delta(\bm{x}_1, \bm{x}_2,\bm{x_3})\big]}\\
    &= \frac{1}{\pi^2} \int \td \bmx_1 \td\bmx_2 ~ f_1(\bmx_1) f_2(\bmx_2) f_3(\bmx+\bmx_2-\bmx_1) \exp{\big[ 4 i \Delta(\bm{x}, \bm{x}_1,\bm{x_2})\big]}
    ~,
    \end{aligned}
\end{equation}
where the function $\Delta(\bm{x}, \bm{x}_1,\bm{x_2})$ has been defined in eq. \eqref{area_triang}. It is clear, therefore, that to obtain the required $\star$-product, one needs to know the Wigner function associated with the initial states, as well as the diagonal phase space eigenfunctions. For quantum systems whose classical counterpart is chaotic, one can employ Berry's conjecture \cite{Berry:1977wpp, voros1976semi, voros1977asymptotic} for the averaged Wigner functions corresponding to the energy eigenstates to evaluate the above $\star$-product (see the discussion in Section \ref{sec_dis_conc}). We note that it is also possible that the quantum system has an `effective' Planck constant $\hbar_{eff}$, so that the classical system can be thought to be obtained by taking the limit $\hbar_{eff} \rightarrow 0$.

\paragraph{\textbf{Long-time average.}}
It is also interesting to consider the expression of the Wigner function $W(q,p,t)$ in \eqref{WFexpanK} in terms of the energy basis of the Hamiltonian (denoted as  $\ket{E_a}$). The desired expansion of the Wigner function of the time-evolved state can be obtained to be 
\begin{equation}
		W(q,p,t)=\sum_{n,m} \mathcal{W}^K_{nm} (p,q) \sum_{a,b}  P_{mn} (E_b,E_a)~ \rho_0 (E_a, E_b) ~ e^{-it (E_a-E_b)}~,
\end{equation}
where $\rho_0 (E_a, E_b)  = \braket{E_a | \psi_0}\braket{\psi_0|E_b} = \braket{\psi_0|\rho_{ba}|\psi_0}$ and $ P_{nm} ( E_b,E_a) = \braket{E_b|K_n}\braket{K_m|E_a}= \braket{E_b|\mathbf{P}_{nm}|E_a}$
denotes respectively, the matrix elements of the initial density matrix and the operator $\mathbf{P}_{nm}$ in the energy basis. From this expression, we get the long-time average of the Wigner function to be 
\begin{equation}\label{WFavg}
\begin{split}
	\bar{W}(q,p)= \lim_{T \rightarrow \infty} \frac{1}{T} \int_{0}^{T} W(q,p,t)~\td t
	= \sum_{n,m}  \mathcal{W}^K_{nm} (p,q) ~\sum_{a}  P_{mn} (E_a) ~\rho_0 (E_a) ~.
\end{split}
\end{equation}
This is the quantity that determines the long-time average of the Krylov state complexity, which we discuss later on. 

\subsection{Krylov basis and the Lanczos algorithm for states in the phase space representation}\label{Krylov_and_Moyal}
The action of the Hamiltonian operator on the elements of the Krylov basis is given by the relation in \eqref{H_Krylov}, which reduces the Hamiltonian, in the Krylov basis, to a tridiagonal matrix. A natural question one can ask in the context of the phase-space analysis is what is the analogue of this equation in terms of the phase space Krylov functions i.e., how the classical Hamiltonian kernel $\mb{H}(\bm{x})$ acts on the functions $\mcW^K_{nm}(\bm{x})$?  The star-product equation corresponding to eq. \eqref{H_Krylov} can be evaluated to be given by the following relation
\begin{equation}\label{krylov_star}
    \mathbb{H}(\bm{x}) \star W_{nn}^K (\bm{x}) = a_n W_{nn}^K (\bm{x}) + b_n W_{(n-1)n}^K (\bm{x}) + b_{n+1} W_{(n+1)n}^K (\bm{x})~.
\end{equation}

For a given initial state $\ket{\psi_0}$ and the associated Wigner function $W^K_{00}(\bmx)$, we can obtain the higher order phase space Krylov functions $W^K_{nn}(\bmx)$ associated with $\ket{K_n}$s using \eqref{krylov_star} by computing the relevant star products. The formulas for the Lanczos coefficients to be used are derived in the following. 

Several interesting relations follow directly from \eqref{krylov_star}, including alternative representations for the Lanczos coefficients. Firstly, by performing an average over the phase space variables in \eqref{krylov_star}, and using \eqref{star_to_pq} and \eqref{orthpsi}, we have an expression for the Lanczos coefficients $a_n$,
\begin{equation}
    a_n= \int \td \bm{x}~ \mb{H}(\bm{x})~ W^K_{nn}(\bm{x})~,
\end{equation}
which can be easily checked to be consistent with the usual expression $a_n=\braket{K_n|\mbf{H}|K_n}$. This expression can be further modified towards a more meaningful form by noting that, for the Hamiltonian we are considering in this section, i.e., $\mbf{H}(\bm{X})=\frac{\mbf{P}^2}{2m}+V(\mbf{Q})$, we have $\mb{H}(q,p)= \frac{p^2}{2m}+V(q)$, hence, 
\begin{equation}
    a_n = \frac{1}{2m}\int \td p~ p^2~ |\psi^K_{n}(p)|^2 +  \int \td q~V(q)~ |\psi^K_{n}(q)|^2~,
\end{equation}
where $\psi^K_{n}(p)$, and $\psi^K_{n}(q)$, respectively denote the momentum and position space representation of the 
$n$th Krylov basis element, so that their modulus squares have the usual probabilistic interpretations. Similarly, by performing an average over the phase space in \eqref{krylov_star}, and subsequently using \eqref{star_to_pq} and \eqref{orthpsi}, we have an expression for the Lanczos coefficients $b_n$,
\begin{equation}
    b_n= \int \td \bm{x}~ \mb{H}(\bm{x})~ \mcW^K_{n(n-1)}(\bm{x})~,
\end{equation}

Secondly, using \eqref{WstarW}, we one can derive
\begin{equation}
\begin{split}
    \mcW_{mm}^K (\bm{x}) \star \mathbb{H}(\bm{x}) \star \mcW_{nn}^K (q,p) = \frac{\delta_{nm}}{2\pi} \Big(a_n \mcW_{mn}^K (\bm{x}) + b_n \mcW_{m(n-1)}^K (\bm{x}) + b_{n+1} \mcW_{m(n+1)}^K (\bm{x})\Big)~.
\end{split}
\end{equation}
Once again, we can use this relation to obtain expressions for the Lanczos coefficients by considering cases with $m=n$ and $m=n-1$. 

As a generalisation of \eqref{krylov_star}, we also have the following two relations  
\begin{equation}\label{krylov_star_nm}
    \mathbb{H}(\bm{x}) \star \mcW_{nm}^K (\bm{x}) = a_n \mcW_{nm}^K (\bm{x}) + b_n \mcW_{(n-1)m}^K (\bm{x}) + b_{n+1} \mcW_{(n+1)m}^K (\bm{x})~.
\end{equation}
and 
\begin{equation}\label{krylov_star_mn}
    \mcW_{nm}^K (\bm{x}) \star  \mathbb{H}(\bm{x}) = a_m \mcW_{nm}^K (\bm{x}) + b_m \mcW_{n(m-1)}^K (\bm{x}) + b_{m+1} \mcW_{n(m+1)}^K (\bm{x})~.
\end{equation}

Next, consider the star-action of the classical kernel associated with the Hamiltonian $\mbf{H}(\bm{X})=\frac{\mbf{P}^2}{2m}+V(\mbf{Q})$ on the generating function defined in \eqref{gen_function}. This can be easily obtained to be given by ($\bm{\mu}=(\mu_1,\mu_2)$)
\begin{equation}
\begin{split}
        \mb{H}(\bm{x}) \star G^K(\bm{\mu};\bm{x})= \sum_{m,n}\frac{\mu_1^m}{\sqrt{m!}} \frac{\mu_2^n}{\sqrt{n!}}  \Bigg(a_n \mcW_{nm}^K (\bm{x}) + b_n \mcW_{(n-1)m}^K (\bm{x}) + b_{n+1} \mcW_{(n+1)m}^K (\bm{x})\Bigg)~.
\end{split}
\end{equation}
Now performing an integral over the phase space, we have the following relations  for the Lanczos coefficients in terms of the generating function,
\begin{align}
    a_n&= \frac{1}{n!} \frac{\partial^n}{\partial \mu_1^n}\frac{\partial^n}{\partial \mu_2^n} \int \td \bm{x}~ ~\mb{H}(\bm{x}) \star G^K(\bm{\mu};\bm{x})\Big|_{\bm{\mu}=0}~,\\
    b_n&=\frac{1}{\sqrt{n! (n-1)!}} \frac{\partial^{n-1}}{\partial \mu_1^{n-1}}\frac{\partial^n}{\partial \mu_2^n}\int \td \bm{x}~ ~\mb{H}(\bm{x}) \star G^K(\bm{\mu};\bm{x})\Big|_{\bm{\mu}=0}~.
\end{align}

\subsection{Characteristic function of the Wigner function}
In many applications, it is easier to use a Fourier transform representation of the phase space variables, in particular, that of the Wigner function. 
This will help us to rewrite the expression for the Krylov state complexity discussed later from 
a different perspective, and more importantly, this description will be essential in section \ref{sec_op_phase_sp} where we shall construct a phase space representation of superoperators. Let us first consider the Fourier transform of the Wigner function ($W(\bm{x})$) associated with a generic state,
\begin{equation}
   \tiW(\bm{\alpha})= \int \td \bm{x}~ W(\bm{x})~ e^{i \bm{\alpha} \cdot \bm{x}} ~.
\end{equation}

The quantity $W(\bm{\alpha})$ can be thought of as the \textit{(quantum) characteristic function }of the Wigner distribution function. Strictly speaking, calling this the characteristic function is improper, since the Wigner function, in general, is not a probability distribution function. For this reason, it is sometimes also referred to as the generalised characteristic function \cite{moyal1949quantum, agarwal1970calculus}. 

Next, we note that the function $e^{i \bm{\alpha} \cdot \bm{x}} $ is just the Weyl transformation \eqref{Weyl-tr} of the operator $\mbf{S}_{\bmal}=\exp\big[i \bmal \cdot \bm{X}\big]$, i.e., $\mathbb{S}_{\bmal}(\bm{x})=\exp\big[i \bmal \cdot \bm{x}\big]$. Hence eq. \eqref{expecttaion} implies that $W(\bmal)$ is just the expectation value of the operator $\mbf{S}_{\bmal}$ in the state $\ket{\psi}$, i.e., (with $\rho_\psi=\ket{\psi}\bra{\psi}$) 
\begin{equation}\label{W_FT}
    \tiW(\bmal)=\int W(\bmx) ~ \mathbb{S}_{\bmal}(\bmx) ~\td \bmx = \braket{\psi|\mbf{S}_{\bmal}|\psi} = \text{Tr}\bitd{\rho \mbf{S}_{\bmal}}~.
\end{equation}
Note that the operator $\mbf{S}_\alpha$ is nothing but the (hermitian conjugate of) the \textit{displacement operator} \cite{Glauber}. To see this, consider the standard from the displacement operator, $\mbf{D}(z)=\exp\Big[z a^\dagger - z^*a\Big]$, for a pair of operators $a, a^\dagger$ obeying the commutation relation $[a, a^\dagger]=1$. By choosing, $z=(\alpha_2-i\alpha_1)/\sqrt{2}$ we see that the displacement operator is, $\mbf{D}_{\bm{\alpha}}=\mbf{D}\Big(z=(\alpha_2-i\alpha_1)/\sqrt{2}\Big)=\mbf{S}^\dagger_{\alpha_1, \alpha_2}$, since we have, $a=(\mbf{Q}+i \mbf{P)/\sqrt{2}}$ and $a^\dagger=(\mbf{Q}-i \mbf{P)/\sqrt{2}}$.

Performing a similar computation as above, we can also see that the Fourier transforms of the functions $W^K_{nm}(q,p)$ are given by the matrix elements of the operator $\mbf{S}_{\bmal}$ in the Krylov basis, i.e., 
\begin{equation}\label{wnm_Fourier}
     \tiW^K_{nm}(\bmal)=\int ~\td \bmx ~\mcW_{nm}(\bmx) ~ \mathbb{S}_{\bmal}(\bmx) = \braket{K_m|\mbf{S}_{\bmal}|K_n}~.
\end{equation}
Note that here we have defined the Fourier transforms of only the Wigner functions; however, one can similarly obtain the Fourier transforms of the Weyl symbols of other quantum mechanical operators as well. For an operator $\mbf{O}(\mbf{Q}, \mbf{P})$, the Fourier space representation of its Weyl transform is given by 
\begin{equation}
    \tilde{\mb{O}} (\bmal) = \int \td \bmx~\mb{O} (\bmx) ~\mathbb{S}_{\bmal}(\bmx)~=\text{Tr}\bitd{\mbf{O}(\mbf{Q}, \mbf{P})\mbf{S}_{\bmal} }~.
\end{equation}

For our purposes in Section \ref{sec_op_phase_sp}, here we mention a slight change of notation which will be used from the next subsection: Consider changing $\alpha_1=-\xi_p$ and $\alpha_2=\xi_q$, so that, compared to \eqref{W_FT}, 
$ W(\bmx)$ and  $\tiW(\bmal)$ are related by \textit{symplectic Fourier transformation} in terms of the new variable $\bm{\xi}=(\xi_q,\xi_p)$, rather than the usual one,
\begin{equation}
     \tiW(\bm{\xi})= \int \td \bm{x}~ W(\bm{x})~ e^{-i \braket{ \bm{x}, \bm{\xi}}_s} ~,~~\braket{ \bm{x},\bm{\xi}}_s= q \xi_p -  \xi_qp =\bmx^T J \bm{\xi}~,~~\text{with}~~~J =\left(
		\begin{array}{ccc}
			0 & 1\\
			-1  & 0
		\end{array}
		\right)~.
\end{equation}
Here $J$ is the symplectic matrix. 
In the expressions derived in the following, we will use both these conventions for the Fourier space variables, and any one of them can be converted to the other by using $\alpha_1=-\xi_p$ and $\alpha_2=\xi_q$. As an example,  in terms of the variables $\bm{\xi}$, the expression for the displacement operator and its  Weyl transforms are given by  \cite{Glauber}
\begin{equation}\label{displace_Fourier}
    \mbf{D}_{\bm{\xi}}=\mbf{D}_{\xi_q, \xi_p}=\exp\Big[i(\xi_p \mbf{Q} - \xi_q \mbf{P})\Big] = \exp\Big[i \braket{\mbf{X}, \bm{\xi}}_s\Big]~,~~\text{and}~~\mb{D}_{\bm{\xi}}(\bmx)=\exp\Big[i(\xi_p q - \xi_q p)\Big] = \exp\Big[i \braket{\bmx, \bm{\xi}}_s\Big]~.
\end{equation}
As mentioned above, in terms of the variables $\{\xi_q, \xi_p\}$, the double Fourier transform would become a symplectic Fourier transform. 

\paragraph{\textbf{Fourier transform of the generating function.}}
Let us now consider the Fourier transform of the generating function of the phase space Krylov functions, given in \eqref{gen_function}. Using the relation \eqref{wnm_Fourier}, this can be rewritten as  
\begin{equation}
    G^K(\bm{\mu};\bm{\alpha}) = \int G^K(\bm{\mu};\bm{x}) ~ \mathbb{S}_{\bmal}(\bmx) ~\td \bmx= \sum_{m,n}\frac{\mu_1^m}{\sqrt{m!}} \frac{\mu_2^n}{\sqrt{n!}} \braket{K_m|\mbf{S}_{\bmal}|K_n}~.
\end{equation}
Defining the class of states,\footnote{These states are quite similar in form to the canonical coherent states of harmonic oscillators \cite{gazeau2009coherent, Perelomov:1986tf}. However, we note that $\ket{\mu}^K$, in general, does not have any such interpretation. } 
\begin{equation}
    \ket{\mu}^K = \sum_{n=0}^{D_K} \frac{\mu^n}{\sqrt{n!}} \ket{K_n}~,
\end{equation}
we see that $G(\bm{\mu};\bm{\alpha})$ can be written  as 
the expectation value of the operator $\mbf{S}_{\bmal}$ with respect to  these states,
\begin{equation}
    G(\bm{\mu};\bm{\alpha}) = ~^K\braket{\mu_1|\mbf{S}_{\bmal}|\mu_2}^K~.
\end{equation}
When the initial state and the Hamiltonian are such that the dimension of the  Krylov subspace is countable and infinite, then formally one can define two operators, $a^K$ and $a^{K \dagger}$, with commutation relation $[a^K, a^{K\dagger}]=1$ which act on the elements of the Krylov basis in the way usual raising and lowering operators for the harmonic oscillator operators do. Then we can define an operator 
\begin{equation}
    \mbf{D}^K(z) = \exp \Big[{z a^{K\dagger} - \bar{z}a^K}\Big]~,
\end{equation}
which acts on the lowest element of the Krylov basis, to produce the state $\ket{\mu}$ up to a normalisation,
\begin{equation}
  \ket{\mu}= e^{|\mu|^2/2} \mbf{D}^K(z) \ket{K_0}~.
\end{equation}

\subsection{Reflection, translation operators and the Wigner function} 
For our purposes in Section \ref{sec_op_phase_sp}, we now briefly review an alternative and very well-known representation of the expressions for the Wigner function and its characteristic function.\footnote{From here onwards, we shall use the variables $\bm{\xi}=(\xi_q, \xi_p)^T$ to denote the Fourier conjugate variables of $q$ and $p$. } First, consider the \textit{reflection and translation operators} defined on the phase space, in terms of the position and the momentum basis, respectively, as (see e.g., \cite{Royer,de1998weyl})
\begin{eqnarray}
    \mbf{R}_{\bmx}=\mbf{R}_{q,p}:= \int \td y~ \ket{q+y/2} \bra{q-y/2}~e^{i p y}~=\int \td k~ \ket{p+k/2} \bra{p-k/2}~e^{-i q k}~,\label{reflection_op}\\
    \mbf{T}_{\bm{\xi}}=\mbf{T}_{\xi_q,\xi_p}:= \int \td p~ \ket{p+\xi_p/2} \bra{p-\xi_p/2}~e^{-i \xi_q p}~=\int \td q~ \ket{q+\xi_q/2} \bra{q-\xi_q/2}~e^{i \xi_p q}~.\label{translation_op}
\end{eqnarray}
It is easy to check that these two operators are related to one another by a symplectic Fourier transform.
The close connection of these two operators and the phase space description of quantum mechanics can be seen from the fact that the Wigner function and its characteristic function are given in terms of these operators as (with $\rho=\ket{\psi}\bra{\psi}$)
\begin{equation}\label{Wigner_refle_tra}
    W(\bmx)= \frac{1}{2\pi}\braket{\psi|\mbf{R}_{\bmx}|\psi}~=\frac{1}{2\pi}\text{Tr}\Big[\mbf{R}_{\bmx}\rho\Big]~,~~\text{and}~~~\tiW(\bm{\xi})=\braket{\psi|\mbf{T}_{\bm{\xi}}^\dagger|\psi}=\text{Tr}\Big[\mbf{T}_{\bm{\xi}}^\dagger\rho\Big]~.
\end{equation}
Comparing the second expression above with the one derived previously in terms of the displacement operator (see, e.g., \eqref{W_FT}), we see that $\mbf{T}_{\bm{\xi}}$ provides the momentum (or position) space representation of the displacement operator $\mbf{D}_{\bm{\xi}}$ - a statement which can also be directly verified by using the fact that $\mbf{T}_{\bm{\xi}}^\dagger=\mbf{T}_{\bm{-\xi}}$ (the role of both these operators essentially is to translate phase space variables).\footnote{In fact, some authors use the definition of $\mbf{D}_{q,p}$ as the definition of the translation operator, see, e.g.,  \cite{de1998weyl}. The name displacement operator is more commonly used in the context of quantum optics and coherent states. We will use the terms displacement operator and translation operator interchangeably in the following.} 

For a generic operator $\mbf{O}$, its Weyl transform, and the corresponding characteristic function can be expressed in terms of the reflection and the translation operators as, 
\begin{eqnarray}\label{Weyl_refle_tra}
   \mb{O}(\bmx) = \text{Tr}\Big[\mbf{R}_{\bmx}\mbf{O}\Big]~,~~~\text{and}~~\tilde{\mb{O}}(\bm{\xi}) =\int \td \bmx ~ \mb{O}(\bmx) ~\mathbb{D}^\dagger_{\bm{\xi}}(\bmx) =2 \pi ~\text{Tr}\Big[\mbf{D}_{\bm{\xi}}^\dagger \mbf{O}\Big]~=2 \pi ~\text{Tr}\Big[\mbf{T}_{\bm{\xi}}^\dagger \mbf{O}\Big]~.
\end{eqnarray}
These are also known as, respectively,  the \textit{centre representation} and the \textit{chord symbol} of the operator $\mbf{O}$ (see e.g., \cite{de1998weyl}). 

Since the reflection operator is a Hermitian operator which is associated with each point of the phase space, some authors also call it the \textit{ phase-point operators}. The matrix elements of this operator in the position basis are given by  \cite{Groenewold, wootters1987wigner}
\begin{equation}
	\bra{q^{\prime}}\mbf{R}_{q,p}\ket{q^{\prime\prime}} = \delta \Big(q-\frac{q^\prime+q^{\prime\prime}}{2}\Big)~ e^{i p (q^\prime-q^{\prime\prime})}~.
\end{equation}
Furthermore, using the definition of these operators, it can be checked that reflection and translation operators satisfy the following relations 
\begin{equation}
    \text{Tr}\bitd{\mbf{R}_{\bmx}\mbf{R}_{\bm{y}}} = 2 \pi \delta(\bmx-\bm{y})~,~~\text{Tr}\bitd{\mbf{T}^\dagger_{\bm{\xi}}\mbf{T}_{\bm{\eta}}}  = 2 \pi \delta(\bm{\xi}-\bm{\eta})~,~~\text{and}~~~\text{Tr}[\mbf{T}_{\xi}^\dagger \mbf{R}_{\bmx}]=\text{Tr}[\mbf{T}_{-\xi} \mbf{R}_{\bmx}]=\exp{\bitd{-i\braket{\bmx, \bm{\xi}}_s}}=\mathbb{D}^*_{\bm{\xi}}(\bmx)~.
\end{equation}
Thus, these operators are orthogonal, with the inner product being given by the Hilbert-Schmidt inner product. Moreover, they also form a complete set, so that any Hermitian operator can be expanded in the basis formed by them. As an example, the coefficient of the expansion of the density matrix in the basis of the reflection operators is just the Wigner functions,
\begin{equation}\label{rhoWA}
	\rho = \int W(\bmx) ~\mbf{R}_{\bmx} ~\td \bmx~.
\end{equation} 
For a generic operator $\mbf{O}$, these coefficients would be proportional to the Weyl symbol of the operator. A compact way for writing the relation between a quantum mechanical operator and its Weyl symbol, which can be derived easily from \eqref{Weyl_to_op}, is 
\begin{equation}
    \mbf{O} = \mb{O}(-i \partial_{\bmxi_{i}}) \exp{\bitd{i \bmxi_{i} \mbf{X}^i}}\Big|_{\bmxi=0}~,
\end{equation}
where $\bmxi_i$ denotes the $i$th component of the vector $\bmxi$, and so on. 

\paragraph{\textbf{Parity operators.}}
We can also consider the parity operator about the origin of the phase space, defined as 
\begin{equation}\label{parity_op_0}
    \mbf{\Pi} = \int \td q ~\ket{-q}\bra{q} = \int \td p ~\ket{p}\bra{-p} =  \frac{1}{4\pi} \int \td \bmx ~ \mbf{D}_{\bmx}~,
\end{equation}
where $\mbf{D}_{\bmx}=\mbf{D}_{q,p}=\exp\Big[i( p \mbf{Q} - q\mbf{P})\Big]$ is the standard form of the displacement operator defined in terms of the phase space variables.  Note that, unlike the one in eq. \eqref{displace_Fourier}, $\mbf{D}_{\bmx}$ is not defined in terms of the `Fourier space' variables $\bm{\xi}$ (or $\bm{\alpha}$). However, as we shall see later in Section \ref{sec_op_phase_sp}, these ($\bm{\xi}$) can be thought of as the coordinates of an extended or double phase space associated with canonical transformations  (see also \cite{Chountasis} for related discussion in the context of Wigner function).  

Using the definition of the parity operator about the origin of the phase space, it can be checked that $\mbf{\Pi}_{x}=  \mbf{D}_{\bmx}\mbf{\Pi}\mbf{D}_{\bmx}^\dagger =\mbf{R}_{\bmx}/2$
is the operator which reflects about the phase space point $\bmx$, and hence is the parity operator about this point\footnote{This operator is sometimes also referred to as the displaced parity operator. However, in the following, we shall mostly refer to the parity operator. Also note that this operator is essentially the same as the reflection operator, apart from a factor of two. Since different authors use either of these two operators, we have distinguished them to avoid confusion.} \cite{grossmann1976parity, Royer, Mollow}. The operator $\mbf{\Pi}_{x}$ (along with $\mbf{\Pi}$) is a Hermitian operator; however, it is not positive definite, since it has eigenvalues $\pm 1$. Alternative useful expressions for this operator can be derived in terms of the phase space displacement operator and are given by
\begin{equation}\label{parity_op}
\begin{aligned}
    \mbf{\Pi}_{x} = \mbf{\Pi}_{q,p} &= \frac{1}{4 \pi} \int \td k ~\td y ~\mbf{D}_{y,k}~e^{-i (kq - y p)}~=\frac{1}{4 \pi} \int \td k ~\td y ~\exp{\Big[{i \Big(k (\mbf{Q}-q) - y (\mbf{P}-p)}\Big)\Big]}~\\
    & \equiv \frac{1}{4 \pi} \int \td \xi_q ~\td \xi_p ~\mbf{D}_{\xi_q,\xi_p}~e^{-i (\xi_p q - \xi_q p)}~=\frac{1}{4 \pi} \int \td \xi_q ~\td \xi_p ~\exp{\Big[{i \Big(\xi_p (\mbf{Q}-q) - \xi_q (\mbf{P}-p)}\Big)\Big]}~.
\end{aligned} 
\end{equation}
This relation is just the statement that the parity operator (and hence the reflection operator) and the translation (displacement) operator are related by a symplectic Fourier transform, a fact which will be very useful for purposes in Section \ref{sec_op_phase_sp}. 
Here, we have explicitly written down two trivially equivalent forms in terms of the Fourier-transformed variables and the usual phase space coordinates to again emphasise the connection between them, namely, the Fourier-transformed variables can be thought of as the two coordinates of an extended phase space. Apart from their obvious physical significances, and their relation with the Wigner function and the Weyl transform (e.g., the Wigner function of a quantum state is just the expectation value of the operator $\mbf{\Pi}_{q,p}$  with respect to that state \cite{Royer}), these operators defined so far in this section will be of central importance later in our discussion in Section \ref{sec_op_phase_sp} on Weyl transform of superoperators.

Now we note a few useful relations satisfied by the parity (or reflection) and translation operators. The action of the parity operator on the position and the momentum eigenstates, and the unitary transformation actions on the canonical position and momentum operators are given by, 
\begin{align}
    \mbf{\Pi}_{q,p}\ket{y}&=\ket{2q-y}e^{2ip(q-y)}~,~~\mbf{\Pi}_{q,p}\ket{k}=\ket{2p-k}e^{-2iq(p-k)}~\label{parity_action}\\
    \mbf{\Pi}_{q,p}(\mbf{Q}-q)\mbf{\Pi}^{-1}_{q,p} &= -(\mbf{Q}-q),~~\mbf{\Pi}_{q,p}(\mbf{P}-p)\mbf{\Pi}^{-1}_{q,p} = - (\mbf{P}-p)~, \text{Tr}\bitd{\mbf{\Pi}_{\bmx}\mbf{\Pi}_{\bm{y}}} =  \frac{\pi}{2} \delta(\bmx-\bm{y})~.
\end{align}
On the other hand, the actions of the translation operator on the position and the momentum eigenkets are given by,
\begin{equation}\label{transl_action}
    \op{T}{\xi_q, \xi_p}\ket{y} = \ket{y+\xi_q}~ e^{i\xi_p (y+\frac{\xi_q}{2})}~,~~\op{T}{\xi_q, \xi_p} \ket{k} = \ketc{k+\xi_p}~e^{-i \xi_q(k+\frac{\xi_p}{2})}~. 
\end{equation}
At this point, it is interesting to note the following point regarding the definition of the reflection (or parity operator) and the translation operators, and the different roles played by the coordinates $q,p$ and $\xi_q, \xi_p$. Comparing the two definitions in eqs.  \eqref{reflection_op} and \eqref{translation_op} we see that in the definition of $\suop{T}{\xi}$, the Fourier and the phase space variables appear to have mixed. Of course, since $p$ (or $q$) being integrated over it does not affect the values of the final expressions for, say, Wigner characteristic function, its symplectic (inverse) Fourier transform, the Wigner function. In principle, one can define the translation operator in the following manner, with $\bm{\eta}=(\eta_q, \eta_p)^T$,  
\begin{equation*}
    \mbf{T}_{\bm{\xi}}=\mbf{T}_{\xi_q,\xi_p}:= \int \td \eta_p~ \ket{\eta_p+\xi_p/2} \bra{\eta_p-\xi_p/2}~e^{-i \xi_q \eta_p}~=\int \td \eta_q~ \ket{\eta_q+\xi_q/2} \bra{\eta_q-\xi_q/2}~e^{i \xi_p \eta_q}~.
\end{equation*}
However, even though the numerical values of the corresponding Wigner characteristic function (computed from $\text{Tr}\Big[\mbf{T}_{\bm{\xi}}^\dagger\rho\Big]$) and Wigner function do not change, the formal expressions now depend on the `mixed' variable combinations, such as on the wavefunction of the form $\psi(q+\xi_q)$ and so on, where the variable $\xi_q$ is integrated over. Again, this is due to the symmetry played by these two sets of variables, and we will discuss it further in the context of the canonical superoperators and double phase space formalism in Section \ref{sec_op_phase_sp}. 

In an analogous manner to the classical parity (and hence, reflection) and translations in phase space, the reflection and translation operators form a group having the following composition relations \cite{de1998weyl, gosson2017wigner},
\begin{align}\label{affine_group}
    \suop{\Pi}{x}\suop{\Pi}{y} &= \exp\bitd{2 i \braket{\bmx, y}_s} \suop{T}{2(x-y)}~,~~\suop{T}{\xi} \suop{T}{\eta} = \exp\bitd{- \frac{i}{2} \braket{\bm{\xi}, \bm{\eta}}_s} \suop{T}{\xi+\eta}~,\\
    \suop{\Pi}{x}\suop{T}{\xi}&=\exp\bitd{ i \braket{\bmx, \xi}_s} \suop{\Pi}{x-\frac{1}{2}\xi}~,~~\suop{T}{\xi}\suop{\Pi}{x}=\exp\bitd{ i \braket{\bmx, \xi}_s} \suop{\Pi}{x+\frac{1}{2}\xi}~.
\end{align}

\paragraph{\textbf{Invariance of the Wigner function under Galilean transformations.}}
Before proceeding, it will also be useful to discuss an important point: the invariance of the Wigner function under the Galilean change of reference frame \cite{Hillery:1983ms, bertrand1987tomographic, kruger1976quantum}, which will be employed later on when discussing the Weyl transformation of superoperators in the next section. Consider a Galilean change of reference frame, $q^\prime=q-v t$, and $t^\prime=t$, where $v$ is the (relative) velocity of the ``moving'' frame with respect to the ``fixed'' frame. If we denote the density matrix of a particle of mass $m$ in some inertial frame by $\rho$, then with respect to a frame moving uniformly with velocity $v$ with respect to the first frame, the state of the particle is given by the unitarily transformed density matrix, 
\begin{equation}\label{Galilean_on_rho}
    \rho^\prime= \mbf{D}^\dagger_{vt, mv}~ \rho ~\mbf{D}_{vt, mv} ~,~~\text{where}~~~\mbf{D}_{q,p}=\exp\Big[i( p \mbf{Q} - q\mbf{P})\Big]~.
\end{equation}
Now, using the expression in \eqref{Wigner_refle_tra}, in terms of the operator
$\mbf{\Pi}_{q,p}$, we have the following relation between the Wigner functions of the two frames (here $W^\prime(q^\prime, p^\prime)$ is the Wigner function of the particle according to an observer in the moving frame who is using the coordinate $q^\prime$),
\begin{equation}\label{WF_Glilean}
    W^\prime(q^\prime, p^\prime) = \frac{1}{\pi}\text{Tr}\Big[\mbf{\Pi}_{q^\prime p^\prime}\rho^\prime\Big] = \frac{1}{\pi}\text{Tr}\Big[\mbf{\Pi}_{q p}\rho\Big] = W(q,p)= W(q^\prime+vt,p^\prime+mv)~.
\end{equation}
This indicates that the Wigner function remains invariant under a Galilean change of reference frames (unlike the wavefunction).

\section{Double Weyl transformation, double Wigner function, and the Krylov operator basis}\label{sec_op_phase_sp}

Having discussed the phase space representation of Hilbert space states and quantum mechanical operators (in particular, those associated with the Krylov basis), in this section, we shall develop the phase space representation of superoperators and the supervectors, defined on the Liouville space. As in the previous section, our goal here is to find out the phase space representation of the Krylov basis - here the Krylov basis of operators associated with the Heisenberg time evolution of a quantum mechanical observable, which was recently utilised in the \textit{universal operator growth hypothesis} of \cite{Parker:2018yvk} to quantify the complexity of growth of an initial Hermitian operator under Heisenberg time evolution. After briefly reviewing the  Lanczos algorithm for operators in Section \ref{sub_Lanczos_op}, we first describe in detail the arguments behind the choice of the canonical superoperators and supervectors bases of the Liouville space in 
Section \ref{sub_canonical_super}. We then use these canonical supervector bases to define an extension of the Weyl transform for superoperators in Sections \ref{sec_DWT} and \ref{sub_spce_DWT}, as well as for the product of superoperators in Section \ref{sub_DWT_prod}. Finally, we use all these ingredients to construct the phase space representation of the operator Krylov basis elements, and as one of the main results of this section, elucidate their interpretation by writing these double phase space functions in terms of the $\star$-product between the Weyl symbol and the chord function of the Krylov basis operators (see Section  \ref{sub_Op_Kry_phase}).

Before moving on, here we very briefly mention some aspects of dynamics when viewed through the Wigner-Weyl phase space approach \cite{prosser1983correspondence, Osborn:1994sf, berezin2012schrodinger}. As discussed previously, in the Weyl quantisation, an observable in the Hilbert space is mapped to a function on the phase space through the Weyl transform. Now consider a Schrodinger picture observable $\mbf{O}$ which is transformed to a Heisenberg picture observable through the linear mapping (which is the Heisenberg picture evolution operator)
\begin{equation}
    \mbf{G}(s,t) : \mbf{O} \mapsto \mbf{U}^\dagger (t,s)\mbf{O} \mbf{U}(t,s)\equiv \mbf{O}(t,s)~,
\end{equation}
where $\mbf{U}(t,s)$ denotes the time evolution operator, $s$ being an initial arbitrary instant of time. As is well-known, since the Heisenberg picture evolution operator $\mbf{G}(s,t)$ has a simpler semiclassical form compared to its Schrodinger picture counterpart $\mbf{U}(t,s)$, in the phase space approach, this is the operator which is commonly studied \cite{Osborn:1994sf}. The Wigner-Weyl phase space representation of the operator $\mbf{G}(s,t)$ is a linear mapping ($\mb{G}_{s,t}$) between phase space functions that depends on the times $(t,s)$; i.e., it governs the time evolution of the Weyl transforms associated with Heisenberg picture observables. Symbolically,
\begin{equation}
    \mb{G}_{s,t}: \mb{O} \mapsto \mb{O}(t,s)~.
\end{equation}
In this sense, in the phase space approach, the time evolution dynamics in both classical and quantum mechanics can be understood as unitary flows in the same Hilbert space $L^2(\mb{R}^2)$. 

\subsection{ Liouvillian recursion: Lanczos algorithm in the Liouville space}\label{sub_Lanczos_op}

We start by considering the time evolution of an initial operator generated by a generic time-independent Hamiltonian in the Heisenberg picture, 
\begin{equation}
\mathbf{O}(t)=e^{i\mbf{H}t}\mathbf{O}(t=0)e^{-i\mbf{H}t},
    \label{operatorevolution}
\end{equation}
which, in terms of the \textit{Liouvillian superoperator}\footnote{\textbf{Notation.} We shall use calligraphic letters, with an overcheck to denote a superoperator - usually the same letter as the original operator from which the superoperator is constructed, if that is the case.} ($\calch{L}$), can be written as 
\begin{equation}
\mathbf{O}(t)=\sum_{n=0}^{\infty}\frac{(it)^n}{n!}\calch{L}^{n}\mathbf{O}_{0}~.
\end{equation}
The Liouvillian superoperator is defined as a linear map acting on the operators as $\calch{L}\mathbf{O}=[\mathbf{H},\mathbf{O}]$ and we have indicated the initial Hermitian operator $\mathbf{O}(t=0)$ as $\mathbf{O}_{0}$.  We shall use the notation $\ketc{\mathbf{O}}$ to denote a state in the Hilbert space of operators,\footnote{The operator Hilbert space spanned by these `supervectors' $\ketc{\mathbf{A}}$, for generic operator $\mbf{A}$  in the Hilbert space is sometimes called the \textit{Liouville space}, since in terms of these the von Neumann equation for the evolution of the density matrix reduces to the (quantum) Liouville equation. See \cite{Gyamfi:2020stx} for a recent review on properties of these spaces. We shall often use the terms `operator Hilbert space' and `Liouville space' interchangeably in the following.} and use the normalised version of the \textit{Hilbert-Schmidt norm} as the inner product (the infinite-temperature inner product),
\begin{equation}\label{Hilbert-schmidt}
\avgc{\mathbf{O_{1}}|\mathbf{O_{2}}}=\text{Tr}[\mathbf{O_{1}}^{\dagger}\mathbf{O_{2}}]/\text{Tr}[\mbf{I}]~,
\end{equation}
where $\text{Tr}[\mbf{I}]=D$ is the dimension of the Hilbert space.\footnote{For infinite-dimensional Hilbert spaces, we usually need to introduce a cut-off to compute this. Instead of the inner product defined with respect to an infinite temperature thermal state, we can consider a finite temperature state as well. In that case, the thermal density matrix will act as a regulator.} For the systems we are considering in this paper, the initial operators $\mbf{O}_0$ can be taken to be a function of the position and the momentum operators, which do not commute with the Hamiltonian.

The repeated action of the Liouvillian on the initial operator $\mathbf{O}_{0}$ generates a set of non-orthonormal operator sequences $\{\calch{L}^{n}\mathbf{O}_{0}\}$, with respect to the inner product on the space of operators chosen. From this set, we can successively use the Gram-Schmidt algorithm and generate an orthonormal set of operators, starting from $\mathbf{O}_{0}$ as the first operator, which is assumed to be normalised, $\ketc{\mc{O}_0}:=\ketc{\mbf{O_0}}$. For a Hermitian Hamiltonian generating the evolution, this procedure reduces to what is known as the Lanczos algorithm for the operator space \cite{viswanath1994recursion}, and the resulting orthonormal basis is called the Krylov basis.  The Lanczos algorithm is inductively defined by the following steps
\begin{equation}\label{Lanczos_alg}
\begin{aligned}
\ketc{\mathbf{A}_{n}}&:=\calch{L}\ketc{\mathbf{O}_{n-1}}-b_{n-1}\ketc{\mathbf{O}_{n-2}}~,\\
    b_{n}&:=\avgc{\mathbf{A}_{n}|\mathbf{A}_{n}}^{1/2}~,\\
    \ketc{\mathbf{O}_{n}}&:=b_{n}^{-1}\ketc{\mathbf{A}_{n}}~.        
\end{aligned}
\end{equation}
Note that one usually fixes $\ketc{\mathbf{O}_{-1}}=0$ and $b_0=0$, and the sequence of positive numbers $\{b_{n}\}$, which is generated as a product of the algorithm, is called the Lanczos coefficients.
Here, due to the nature of the inner product used and since the initial operator is Hermitian, the Liouvillian is tridiagonal with the diagonal entries all zero. 


In the rest of this section, our primary goal is to rewrite these expressions in terms of phase space quantities and understand their interpretation in the phase space language. To begin with, we note that the inner product \eqref{Hilbert-schmidt}, for Hermitian operators, can be straightforwardly written as an integral over phase space in terms of the corresponding Weyl symbols using the relation \eqref{transitions} (with an overall normalisation constant). Furthermore, the action of the Liouvillian on a generic operator, $\calch{L}\mathbf{O}=[\mathbf{H},\mathbf{O}]$, transforms to, in terms of phase space quantities, a star-commutator, i.e., $\calch{L}\mathbf{O}:=[\mathbf{H},\mathbf{O}] \rightarrow [\mathbb{H}(\bm{x}),\mathbb{O}(\bm{x})]_\star$.  One can now use this mapping iteratively for each step of the Lanczos algorithm in \eqref{Lanczos_alg}, and obtain a phase space version of the Lanczos algorithm in terms of the Weyl symbols of the Hamiltonian and the operator $\mbf{O}_0$. Note that, to obtain the Lanczos coefficients $b_n$ at each step, one also needs to find out the Weyl symbols of the unnormalized operators ($\mbf{A}_n$). In this approach, even though we do not need to introduce any additional structure beyond what we have already discussed in the previous section, it does not utilise the Liouville space formalism of supervectors and superoperators. In a second approach, which we elaborate below, we introduce a generalisation of the Weyl transforms defined for superoperators by employing some additional structure from the so-called double phase space formalism \cite{abraham2008foundations, littlejohn1986semiclassical}, which is very natural to use in this context, since it allows us to phrase the Lanczos algorithm in a form similar to its state-space counterpart we have discussed so far in this paper.

\subsection{Determining the canonical superoperators and bases in the Liouville space} 
\label{sub_canonical_super}
To define the Weyl transform of superoperators, we first need to find a canonical basis, with respect to which the transformation will be defined (much like position and momentum operators, and their eigenstates, which are used for defining the usual Weyl transform eq. \eqref{Weyl-tr}). The approach we adopt in the following to determine these involves first finding a set of canonical superoperators and subsequently identifying the corresponding canonical supervector bases as the common eigenbasis of these superoperators. To this end, we impose two physical conditions on different possible `phase space representations': first, it must incorporate a Galilean change of reference frame naturally, and secondly, it should also transform `naturally' under linear canonical transformations of the phase space coordinates \cite{Royer2}. We shall define these conditions more precisely later, but before doing that, we recall here that it is usually demanded that any well-defined phase space distribution function (including the Wigner function itself, see eq. \eqref{WF_Glilean}), should satisfy a set of conditions, which includes, among others, the requirement that the acceptable phase space distribution function should be invariant under a Galilean change of reference frame \cite{Hillery:1983ms, royer1989measurement}. Therefore, it is natural to impose the same restriction for defining the Weyl transforms for superoperators. 

We also note that, due to the natural structure of the operator basis of the Liouville space, the (operator) eigenbasis for the canonical superoperators that we shall construct will be defined from two operators. For example, one can associate four orthonormal bases with $\ket{q}, \ket{p}$, respectively, $\ketc{\ket{q}\bra{q^\prime}}$,  $\ketc{\ket{p}\bra{p^\prime}}$, $\ketc{\ket{q}\bra{p}}$, and $\ketc{\ket{p}\bra{q}}$, all of which are possible continuous bases in the Liouville space. Furthermore, in terms of the inner product in the Liouville space, the Wigner function and characteristic function  can be rewritten as 
\begin{equation}
    W(\bmx)=\frac{D}{\pi} \avgc{\mbf{\Pi}_{\bmx} |\rho}~,~~\text{and}~~\tiW(\bm{\xi}) = D \avgc{\mbf{T}_{\bm{\xi} }|\rho}=D \avgc{\mbf{S}^\dagger_{\bm{\xi} }|\rho}~.
\end{equation}

In the following, we first find out this canonical set of superoperators and the canonical bases associated with them, along the lines of \cite{Royer2}, and then provide an interpretation of them later on, following the more recent work of \cite{Saraceno:2015cek}. Finally, with the canonical superoperator and bases thus defined, we shall define the Weyl transformation of superoperators, which will help us formulate the phase space description of the operator Krylov space vectors and the associated Lanczos algorithm.

\paragraph{\textbf{Implementing the Galilean invariance.}} Let us now investigate the consequences of imposing Galilean invariance on the phase space representation, i.e., we look for a set of superoperators and a basis in the Liouville space which would naturally capture the effect of a Galilean change of reference frame from which the quantum system is viewed. First, note that the effect of a Galilean transformation on the state, given in eq. \eqref{Galilean_on_rho} can be written, by introducing a superoperator, $\check{\mc{D}}_{q,p}$, as\footnote{In terms of the notation used in \cite{Royer2}, this superoperator is written as, $\check{\mc{D}}_{qp}=\calch{D}_{qp}^{\leftarrow}\calch{D}_{qp}^{\dagger\rightarrow}$, where $\calch{A}^{\leftarrow}\mbf{O}=\mbf{A}\mbf{O}$, and $\calch{A}^{\rightarrow}\mbf{O}=\mbf{O}\mbf{A}$. Note that any superoperator can be written in terms of the set of operators $\{\calch{Q}^{\leftarrow}, \calch{P}^{\leftarrow}, \calch{Q}^{\rightarrow},\calch{P}^{\rightarrow}\}$. Although this set of notations is very clear, it is somewhat cumbersome to use in practice. Therefore, we shall not use it unless necessary.} 
\begin{equation}\label{rho_super}
    \rho^\prime = \calch{D}^{-1}_{vt,mv}~ \rho~, ~~~\text{where}~~\check{\mc{D}}_{qp} = \mbf{D}_{qp} \bullet \mbf{D}_{qp} ^\dagger~=\exp{\Big[i \Big(p \calch{Q}^{-}-q \calch{P}^{-}\Big)\Big]}~,
\end{equation}
where $\bullet $ is the place holder for the operator on which the superoperator acts, in this case, the density matrix of the `fixed frame', and the notation $\calch{A}^{-}$ denotes the superoperator $[\mbf{A, \bullet}]$, an example being the Liouvillian superoperator, which is $\calch{L}=\calch{H}^{-}$. Considering a basis in the Liouville space $\ketc{\mbf{C}_{qp};\varphi}$,\footnote{Here $\mbf{C}_{\bmx}$ is an operator to be determined. There can be additional labels to this, which we have denoted by $\varphi$ collectively.} we shall call this a phase space basis if its inner product with $\ketc{\rho}$ remains invariant under a Galilean transformation, i.e., 
\begin{equation}\label{qp_displacement}
    \avgc{\mbf{C}_{q^\prime p^\prime}; \varphi|\rho^\prime} = \avgc{\mbf{C}_{qp}, \varphi|\rho}~,~~\text{i.e.,}~~\ketc{\mbf{C}_{\bmx}; \varphi}\equiv\ketc{\mbf{C}_{qp}; \varphi} = \calch{D}_{q,p} \ketc{\mbf{C}_{q=0,p=0}; \varphi}~,
\end{equation}
where the second relation can be easily obtained from the first one using the Hilbert-Schmidt norm. It is important to notice that, as we have already indicated within the notation  $\ketc{\mbf{C}_{qp};\varphi}$, this state is associated with a Hilbert space operator $\mbf{C}$ which depends on the parameters $(p,q)$. We shall determine this operator shortly, but we mention here that the manner in which the condition in \eqref{qp_displacement} is imposed naturally implies that $q$ and $p$ are the coordinates on the phase space. Furthermore, the law of transformation of the Wigner function under a Galilean change of frame, written in terms of superoperators, $\avgc{\mbf{\Pi}_{q^\prime p^\prime}|\rho^\prime}=\avgc{\mbf{\Pi}_{q p}|\rho}$, already indicates that the parity operator about the phase space point $(q,p)$ (i.e., also the reflection operator) can form one such phase space basis. 

Since any orthonormal and continuous Liouville space basis is a canonical eigenbasis for a set of superoperators, a natural question we can now ask is: What is the canonical set of superoperators associated with the basis $\ketc{\mbf{C}_{qp};\varphi}$? As the notation used anticipates, let us denote, for the moment, the two commuting superoperators, whose eigenstates are $\ketc{\mbf{C}_{qp};\varphi}$, by $\calch{Q}_\varphi$ and $\calch{P}_\varphi$. Furthermore,  the relation \eqref{qp_displacement} which states that the superoperator $\calch{D}_{q,p}$ in eq. \eqref{rho_super}, which generates the basis $\ketc{\mbf{C}_{qp};\varphi}$ from a fixed or `standard' ket, is analogous to the relation $\ket{q} \propto \exp[{-i q \mbf{P]}\ket{q=0}}$ and a similar relation for the $\ket{p}$ states. Therefore, it can be seen from the structure of $\calch{D}_{q,p}$, the superoperators $\calch{Q}^{-}$ and $\calch{P}^{-}$ play the roles of the position (or momentum) operator on the Liouville space. Hence, it is natural to take $\calch{Q}_\varphi$ and $\calch{P}_\varphi$, along with $\calch{Q}^{-}$ and $\calch{P}^{-}$ as the 
canonical set of superoperators for our purposes.

The superoperators in this set satisfy the following commutation relations 
\begin{eqnarray}
    \comt{\calch{Q}_\varphi, \calch{P}_\varphi} =0~, ~~ \comt{\calch{Q}^{-}, \calch{P}^{-}}=0~,\label{can_commt1}\\
\comt{\calch{Q}_\varphi,\calch{Q}^{-}}=0~,~~\comt{\calch{P}_\varphi, \calch{P}^{-}}=0~,\label{can_commt2}\\ 
\comt{\calch{Q}_\varphi,\calch{P}^{-}}=i~,~~\comt{\calch{Q}^-,\calch{P}_\varphi}=i~\label{can_commt3}.
\end{eqnarray}
The first commutator in the first line is true by assumption, while the second one in this line can be easily checked to be true from the definitions of these superoperators.\footnote{This can alternatively be seen to be true according to the relation $\comt{\calch{A}^{-}, \calch{B}^{-}}=\comt{\mbf{A}, \mbf{B}}^{-}$ valid for any two superoperators, $\calch{A}^-, \calch{B}^-$. Here we also note down an additional expression, which states that $\comt{\calch{A}^{+}, \calch{B}^{+}}=\comt{\mbf{A}, \mbf{B}}^{+}$, where $\calch{A}^{+}$ stands for the superoperator associated with the anticommutator, i.e., $\calch{A}^{+}:=\frac{1}{2} \comt{\mbf{A}, \bullet}_{+}$. We refer to \cite{Royer2} for additional such useful formulas.} The commutation relations in the second line indicate that it is possible to find out two other additional continuous and orthonormal bases in the Liouville space, which are simultaneous eigenkets of, respectively,   $\calch{Q}_\varphi$ and $\calch{Q}^{-}$, and $\calch{P}_\varphi$ and $\calch{P}^{-}$ (this statement is also true, of course, for $\calch{Q}^{-}$ and $\calch{P}^{-}$ as well).  Finally, the relations in the third line are analogous to the canonical commutation relation between the position and the momentum operators.

\paragraph{\textbf{Implementing linear canonical transformations.}}
Given the requirement of natural transformation of a phase space basis under a Galilean transformation only, there are many possible choices of the two unknown canonical super operators $\calch{Q}_\varphi$ and $\calch{P}_\varphi$ which satisfy the above commutation relations (see \cite{Royer2} for discussion on some of these choices). We now impose a further restriction that the phase space representation should also naturally incorporate a linear canonical transformation of the phase space coordinates as well. We investigate the consequences of imposing this restriction in the context of Liouville space below; however, before proceeding, here we mention that canonical transformation in quantum mechanics is a widely studied topic, a complete discussion of which is beyond the scope of the present paper. We refer to \cite{garcia1980wigner, blaszak2013canonical, Curtright:2000ux} for the role of canonical transformations in the context of the Wigner-Weyl phase space version of quantum mechanics (see also \cite{marino2021advanced}), and, specifically, for further references of early works on the subject. It is well known that under a general canonical transformation, the Wigner functions are related by an integral kernel \cite{Curtright:2000ux}, whereas, for a linear canonical transformation, the Wigner functions are related by a phase space coordinate change \cite{marino2021advanced}.

We shall now proceed to determine the consequence of a linear canonical transformation for determining the canonical superoperators.  A variation of the following discussion can be found in \cite{Royer2}; however, it is also worth noting that our approach differs somewhat from that of \cite{Royer2}. 
Consider a linear canonical transformation of the phase space coordinates, $\bmx^\prime=V \bmx$, which by definition, preserves the Poisson brackets among the components of $\bmx$ or between those of $\bmx^\prime$. Thus, the matrix $V$ with real elements satisfies the symplectic condition 
\begin{equation}
    V J V^T=J~,~~~\text{where}~~J =\left(
		\begin{array}{ccc}
			0 & 1\\
			-1  & 0
		\end{array}
		\right)~,
\end{equation}
is the symplectic matrix, and $\det V=1$ (i.e., $V$ is an element of $\text{SL}(2, \mb{R}))$. As is usual, in quantum mechanics, we identify the canonical transformation with a unitary transformation defined on the Hilbert space \cite{dirac1981principles, leaf1969canonical}. Thus, denoting the unitary operator associated with the linear canonical transformation by $\mbf{V}$, we have the following transformation relations,
\begin{equation}\label{X_change_QCT}
    \bm{X}^\prime = \mbf{V} \bm{X} \mbf{V}^\dagger \equiv \calch{V} \bm{X}~,~~\text{where}~~\calch{V}= \mbf{V} \bullet \mbf{V}^\dagger ~,
\end{equation}
and can be thought of as the \textit{superoperator associated with a canonical transformation}.
Given a canonical transformation, the new phase space operators can be written in terms of the old ones, such that\footnote{As mentioned earlier, here we are considering only linear canonical transformations of classical phase space coordinates. Additionally, in general, the new coordinates can depend on time; we have assumed this is not the case here for simplicity.} $\mbf{Q}^\prime=Q(\mbf{Q}, \mbf{P)}$ and $\mbf{P}^\prime=P(\mbf{Q}, \mbf{P)}$, and moreover, the eigenvalue relations for the position and momentum operators transform to $\mbf{Q}^\prime \ket{q^\prime}=q \ket{q^\prime}$ and $\mbf{P}^\prime \ket{p^\prime}=p \ket{p^\prime}$ (or compactly, $\mbf{X}^\prime \ket{\bmx^\prime}=\bmx \ket{\bmx^\prime}$). Similar to the canonical position and momentum operators, we can now ask, how the two unknown canonical superoperators $\calch{Q}_\varphi$ and  $\calch{P}_\varphi$ transform under this canonical transformation? Denoting $\calch{X}_{\varphi}=(\calch{Q}_\varphi, \calch{P}_\varphi)^T$, we are interested in the transformed superoperators,   $\calch{X}_{\varphi}^\prime = \calch{V}\calch{X}_{\varphi}\calch{V}^{-1}$, specifically, how they are related to initial ones $\calch{X}_{\varphi}$ for a linear canonical transformation of the phase space coordinates.

To this end, we first determine how $\mbf{X}$ changes under linear canonical transformations.  Start from $\mbf{X}^\prime \ket{\bmx^\prime}=\bmx \ket{\bmx^\prime}$, which is equivalent to the relation
$\mbf{X}^\prime \ket{V \bmx}=\bmx \ket{V\bmx}$. Now changing the label $\bmx \rightarrow V^{-1} \bmx$, we see that the resulting relation $\mbf{X}^\prime \ket{ \bmx}=V^{-1} \mbf{X} \ket{\bmx}$ implies that the transformed canonical operators are given by $\mbf{X}^\prime= \mbf{V} \bm{X} \mbf{V}^\dagger =V^{-1} \mbf{X}$. 

Though not essential for our purposes in the following, one can derive the same relation by considering an explicit representation of the unitary operator $\mbf{V}$ for the linear canonical transformations being considered here.  As is usual, rewrite the unitary operator $\mbf{V}$ as the exponential of an anti-Hermitian operator 
\begin{equation}
    \mbf{V}=\exp{\mbf{F}(\mbf{X})}=\exp{\Big[-\frac{i}{2} \mbf{X}\cdot M\mbf{X}\Big]}~,
\end{equation}
with $M$ being a real symmetric $2\times2$ matrix. Then the superoperator  $\calch{V}$ is given by $\calch{V}=\exp{\calch{F}^{-}}$, with $\calch{F}^{-}=[\mbf{F}, \bullet]$. For the above expression of $\mbf{F}(\mbf{X})$, it can be checked that, $\calch{F}^{-} \mbf{X}=[\mbf{F},\mbf{X}]= - J M \mbf{X} $. Therefore, from the relation in \eqref{X_change_QCT}, we see that transformed operators are related to the old ones by the simple relation, $\mbf{X}^\prime= \mbf{V} \bm{X} \mbf{V}^\dagger =V^{-1} \mbf{X}$ where $V=\exp{[J M]}$. Here we have also denoted the transformation matrix between $\mbf{X}^\prime$ and $\mbf{X}$ as $V$, since, as can be checked easily, $V=\exp{[J M]}$ satisfies the condition $V J V^T=J$ (note also that $\det V=1$). 

From the preceding discussion, it is clear that the condition we need to impose on the to be determined superoperators $\calch{X}_{\varphi}$ is that they should satisfy the relation (see also the discussion in Appendix \ref{sec_CT_and_bases})
\begin{equation}\label{Xsu_tra_CT}
    \calch{X}_{\varphi}^\prime = \calch{V}\calch{X}_{\varphi}\calch{V}^{-1} = V^{-1}\calch{X}_{\varphi}~,~~\text{with}~~V=e^{J M}~. 
\end{equation}
This relation can be utilised to determine $\calch{X}_{\varphi}$ as follows. First, we recall that any superoperator $\calch{A}$ can be written as a function of the left and right multiplication superoperators, $\calch{A}^{\leftarrow}$ and  $\calch{A}^{\rightarrow}$, respectively. Therefore, keeping \eqref{Xsu_tra_CT} in mind, we determine how the superoperator  $\calch{V}$ is associated with the linear canonical transformation superoperators $\calch{X}^{\leftrightarrows}$. Using the transformation relation $\mbf{V} \bm{X} \mbf{V}^\dagger =V^{-1} \mbf{X}$, it can be checked that the required transformation rule is 
$\calch{V}\calch{X}^{\leftrightarrows}\calch{V}^{-1} = V^{-1}\calch{X}^{\leftrightarrows}$. This indicates that $\calch{X}_{\varphi}$ must be a linear combination of $\calch{X}^{\leftrightarrows}$, which we take to be of the form $\calch{X}_{\varphi}= c_1\calch{X}^{\leftarrow} + c_2 \calch{X}^{\rightarrow}$, $c_1$ and $c_2$ being two arbitrary complex numbers. 
These coefficients can be determined by making use of the commutation laws in eqs. \eqref{can_commt1}-\eqref{can_commt3}. The vanishing of the commutator between $\calch{Q}_{\varphi}$ and $\calch{P}_{\varphi}$ implies that $c_2=\pm c_1$ (the evolution of the commutator can be simplified by noticing the fact that $[\calch{A}^{\leftarrow}, \calch{B}^\rightarrow]=0$ for any two superoperators). If $c_2=-c_1$, then the superoperator $\calch{X}_{\varphi}$ would become proportional to $\calch{X}^{-}$, so that the canonical superoperator commutation relations in \eqref{can_commt3} would not be satisfied. Therefore, we must choose $c_1=c_2$, and using \eqref{can_commt3} it can be determined that $c_1=c_2=1/2$. Thus, finally, the canonical superoperators we are after are given by \cite{Royer2}, 
\begin{equation}
   \calch{Q}_{\varphi}=\frac{1}{2} (\calch{Q}^{\leftarrow}+\calch{Q}^{\rightarrow})=\calch{Q}^{+}~,~~\text{and}~~\calch{P}_{\varphi}=\frac{1}{2} (\calch{P}^{\leftarrow}+\calch{P}^{\rightarrow})=\calch{P}^{+}~.
\end{equation}
These will be collectively denoted as $\calch{X}^+=(\calch{Q}^{+},\calch{P}^{+})^T$, and are referred to as the Wigner-Weyl canonical superoperators.  In terms of these canonical superoperators, the 
canonical commutation relations in eq. \eqref{can_commt3} can be rewritten as $[\calch{Q}^+,\calch{P}^-]=i=[\calch{Q}^-,\calch{P}^+]$, or more compactly all these commutation relation are given by
\begin{equation}\label{can_commtf}
  [\calch{X}^+,\calch{X}^-]=i J~,
\end{equation}
which is the analogue of the usual canonical commutation relations 
$[\mbf{X,\mbf{X}}]=i J$ in quantum mechanics. A generic superoperator (i.e., a Liouville space operator $\calch{A}$) can be 
expressed as a function of these canonical superoperators.

Before proceeding to determine the eigenbases of this canonical set of superoperators, we mention here that some additional useful discussions on these superoperators and the implications of the linear canonical transformations are presented in the Appendix \ref{sec_CT_and_bases}.

\paragraph{\textbf{Determining the canonical supervector bases.}}
Since the Wigner-Weyl superoperator basis is the canonical one for our purposes, which naturally incorporates transformation under a Galilean change of reference frame and linear canonical transformation, we drop the label for other indices $\varphi$ from now on. To identify the eigenbases of the superoperators $\calch{X}^{\pm}$, we proceed as follows.  

To begin with, we notice that the eigenstates of $\calch{X}^{+}$ are proportional to $\ketc{\mbf{\Pi}_{\bmx}}$, i.e., $\ketc{\mbf{C}_{\bmx}}=\mc{N} \ketc{\mbf{\Pi}_{\bmx}}=\mc{N} \ketc{\mbf{R}_{\bmx}}/2$. It is easy to check this statement explicitly by using the expression for the operator $\mbf{\Pi}_{\bmx}$
given in eq. \eqref{parity_op}. Here $\mc{N}$ is a normalisation constant, which we determine from the normalisation condition. 
These states satisfy the additional conditions of completeness and orthonormality,
\begin{equation}
    \int \td \bmx ~ \ketc{\mbf{C}_{\bmx}}\brac{\mbf{C}_{\bmx}} = \mc{N}^2  \int \td \bmx ~ \ketc{\mbf{\Pi}_{\bmx}}\brac{\mbf{\Pi}_{\bmx}}=\check{1}~,~~\text{and}~~\avgc{\mbf{C}_{\bm{y}}|\mbf{C}_{\bmx}}=\mc{N}^2 \avgc{\mbf{\Pi}_{\bm{y}}|\mbf{\Pi}_{\bmx}}= \delta(\bm{y}-\bmx)~,
\end{equation}
where $\check{1}$ denotes the unit superoperator (e.g., $\mbf{1}^+=\check{1}$). The second condition determines the normalisation constant to be $\mc{N}=\sqrt{\frac{2 D}{\pi}}$. Note that, here keeping in mind the fact that we shall use these basis states for elements of the Liouville space Krylov basis elements, we have used the inner product relation \eqref{Hilbert-schmidt} which scales the basis elements by an additional factor of $\sqrt{D}$. 
This completes the characterisation of one class of canonical states in the Liouville space. 

Next, we define a second class of Liouville space basis states, which is related to the states $\ketc{\mbf{C}_{\bmx}}$ through a (symplectic) 
Fourier transformation (this is analogous to how the position and momentum eigenstates are related in ordinary quantum mechanics)
\begin{equation}\label{Ctilde_def}
     \ketc{\tilde{\mbf{C}}_{\bm{\xi}}}= \frac{1}{2\pi} \int ~ \td \bmx~\ketc{\mbf{C}_{\bmx}} ~e^{i \braket{\bmx,\bm{\xi}}_s}~.
\end{equation}
We need to determine the unknown operator, which we have denoted by $\tilde{\mbf{C}}_{\bm{\xi}}$ above. To this end, we notice that the state $\ketc{\mbf{T}_{\bm{\xi}}}$(i.e., $\ketc{\mbf{D}_{\bm{\xi}}}$) is an eigenstate of the superoperator $\calch{X}^{-}$, where $\tilde{\mc{N}}$ is a normalisation constant to be determined. Therefore, we can make the identification $\ketc{\tilde{\mbf{C}}_{\bm{\xi}}}=\tilde{\mc{N}}\ketc{\mbf{T}_{\bm{\xi}}}=\tilde{\mc{N}}\ketc{\mbf{D}_{\bm{\xi}}}$, where the normalisation constant can be determined now to be equal to $\tilde{\mc{N}}=\sqrt{\frac{D}{2\pi}}$. 
The completeness and orthonormality relations for these basis states are given by 
\begin{equation}
    \int \td \bm{\xi} ~ \ketc{\suopf{C}{\xi}}\brac{\suopf{C}{\xi}} = \tilde{\mc{N}}^2  \int \td \bmx ~ \ketc{\suop{T}{\xi}}\brac{\suop{T}{\xi}}=\check{1}~,~~\text{and}~~\avgc{\suopf{C}{\xi}|\suopf{C}{\eta}}=\tilde{\mc{N}}^2 \avgc{\suop{T}{\xi}|\suop{T}{\eta}}= \delta(\bm{\xi}-\bm{\eta})~,
\end{equation}
and the overlap of $\ketc{\suop{C}{\bmx}}$ and $\ketc{\suopf{C}{\bm{\xi}}}$ is 
\begin{equation}
    \avgc{\suop{C}{\bmx}|\suopf{C}{\bm{\xi}}} = \frac{1}{2 \pi} e^{i \braket{\bmx,\bm{\xi}}_s}~,
\end{equation}
which is a direct generalisation of the standard overlap formula between the position and the momentum eigenstates.
A general superoperator, $\calch{A}$, can be expanded in terms of these bases as 
\begin{equation}
    \calch{A} = \int \td \bmx ~\td \bm{y}~ \avgc{\suop{C}{x}|\calch{A}|\suop{C}{y}} \ketc{\suop{C}{x}}\brac{\suop{C}{y}}~=\int \td \bm{\xi} ~\td \bm{\eta}~ \avgc{\suopf{C}{\xi}|\calch{A}|\suopf{C}{\eta}} \ketc{\suopf{C}{\xi}}\brac{\suopf{C}{\eta}}~,
\end{equation}
and its trace is given by
\begin{equation}
    \check{\text{Tr}} \calch{A} = \int \td \bmx ~ \avgc{\suop{C}{x}|\calch{A}|\suop{C}{x}} = \int \td \bm{\xi} ~\avgc{\suopf{C}{\xi}|\calch{A}|\suopf{C}{\xi}}~.
\end{equation}

Before moving on to define the Weyl transform of a superoperator using these canonical Liouville space bases, it is important to note that these bases do not represent the density matrix of some quantum mechanical system, since, as previously mentioned, the parity operator is not positive definite \cite{Royer2}.

\subsection{ Double Weyl transformation of superoperators }\label{sec_DWT}
Having determined the canonical set of bases for Liouville space operators, our goal is now to understand the phase space representation of superoperators. To this end, we proceed in an analogous manner to that of the case of Hilbert space operators, and define the Weyl transformation of superoperators by introducing the parity and translation superoperators in terms of the canonical Wigner-Weyl supervector bases $\ketc{\mbf{C}_{\bmx}}$ and $\ketc{\tilde{\mbf{C}}_{\bm{\xi}}}$. This is also called the double Weyl transformation (DWT) - a terminology we shall use in the following to avoid any confusion with the usual Weyl transformations of Hilbert space operators, discussed in Section \ref{Weyl_Wigner}.\footnote{For different aspects of the phase space representation of the superoperators and supervectors, we refer to \cite{Royer2, Ghosh, Wilkie1, Saraceno:2015cek}.}   

We begin by defining the \textit{parity superoperator as}\footnote{Note that the corresponding reflection superoperator would be given by the expression \cite{Saraceno:2015cek}, 
\begin{align}
    \calch{R}_{{\bmx}, \bm{\xi}} = \int \td \bm{y} ~\ketc{\mbf{C}_{\bmx+\frac{\bm{y}}{2}}}\brac{\mbf{C}_{\bmx-\frac{\bm{y}}{2}}} ~e^{i \braket{\bm{y},\bm{\xi}}_s} =  \frac{\mc{N}^2}{4} \int \td \bm{y} ~\ketc{\mbf{R}_{\bmx+\frac{\bm{y}}{2}}}\brac{\mbf{R}_{\bmx-\frac{\bm{y}}{2}}} ~e^{i \braket{\bm{y},\bm{\xi}}_s}=4 \check{\Pi}_{{\bmx}, \bm{\xi}}~.
\end{align}
} 
\begin{align}\label{parity_supop}
    \check{\Pi}_{{\bmx}, \bm{\xi}} &:= \frac{1}{4} \int \td \bm{y} ~\ketc{\mbf{C}_{\bmx+\frac{\bm{y}}{2}}}\brac{\mbf{C}_{\bmx-\frac{\bm{y}}{2}}} ~e^{i \braket{\bm{y},\bm{\xi}}_s} = \frac{\mc{N}^2}{4} \int \td \bm{y} ~\ketc{\mbf{\Pi}_{\bmx+\frac{\bm{y}}{2}}}\brac{\mbf{\Pi}_{\bmx-\frac{\bm{y}}{2}}} ~e^{i \braket{\bm{y},\bm{\xi}}_s}~\\
    &= \frac{1}{4}\int \td \bm{\eta} ~\ketc{\tilde{\mbf{C}}_{\bm{\xi}+\frac{\bm{\eta}}{2}}}\brac{\tilde{\mbf{C}}_{\bm{\xi}-\frac{\bm{\eta}}{2}}} ~e^{-i \braket{\bm{x},\bm{\eta}}_s}~,
\end{align}
where the alternative expression in the second line is derived by using the relation in \eqref{Ctilde_def}. 
The action of this superoperator on the canonical Liouville space bases can be evaluated to be  (compare with the action of the parity operator on the position and the momentum eigenkets, equations \eqref{parity_action}), 
\begin{align}\label{parity_suop_action}
    \check{\Pi}_{{\bmx}, \bm{\xi}} \ketc{\mbf{C}_{\bm{x}^\prime}}= \exp{\bitd{{2i\braket{\bmx-\bm{x}^\prime, \bm{\xi}}_s}}}\ketc{\mbf{C_{2\bm{x}-\bm{x}^\prime}}}~,~~\text{and}~~~\check{\Pi}_{{\bmx}, \bm{\xi}} \ketc{\tilde{\mbf{C}}_{\bm{\xi}^\prime}} = \exp{\bitd{{-2i\braket{\bmx, \bm{\xi}-\bm{\xi}^\prime}_s}}}\ketc{\tilde{\mbf{C}}_{2 \bm{\xi}-\bm{\xi}^\prime}} ~.
\end{align}

Next, we define the \textit{translation superoperator} with the following expression, 
\begin{align}\label{transl_supop}
    \calch{T}_{{\bmx}, \bm{\xi}} &:= \int \td \bm{\eta} ~\ketc{\tilde{\mbf{C}}_{\bm{\eta}+\frac{\bm{\xi}}{2}}}\brac{\tilde{\mbf{C}}_{\bm{\eta}-\frac{\bm{\xi}}{2}}} ~e^{-i \braket{\bm{x},\bm{\eta}}_s} = \tilde{\mc{N}}^2 \int \td \bm{\eta} ~\ketc{\mbf{T}_{\bm{\eta}+\frac{\bm{\xi}}{2}}}\brac{\mbf{T}_{\bm{\eta}-\frac{\bm{\xi}}{2}}} ~e^{-i \braket{\bm{x},\bm{\eta}}_s} ~\\
    &= \int \td \bm{y} ~\ketc{\mbf{C}_{\bm{y}+\frac{\bm{x}}{2}}}\brac{\mbf{C}_{\bm{y}-\frac{\bm{x}}{2}}} ~e^{i \braket{\bm{y},\bm{\xi}}_s}~,
\end{align}
where the alternative expression in the second line can be obtained by using the relation in \eqref{Ctilde_def}. 
Its action on the canonical bases of the Liouville space is given by (compare these with the action of the translation operator on the position and the momentum eigenkets, written in \eqref{transl_action}),
\begin{align}\label{tran_suop_action}
    \calch{T}_{{\bmx}, \bm{\xi}} \ketc{\mbf{C}_{\bm{x}^\prime}}= \exp{\bitd{{i\braketb{\frac{\bmx}{2}+\bm{x}^\prime, \bm{\xi}}_s}}}\ketc{\mbf{C_{\bm{x}+\bm{x}^\prime}}}~,~~\text{and}~~~\calch{T}_{{\bmx}, \bm{\xi}} \ketc{\tilde{\mbf{C}}_{\bm{\xi}^\prime}} = \exp{\bitd{{-i\braketb{\bmx, \frac{\bm{\xi}}{2}+\bm{\xi}^\prime}_s}}}\ketc{\tilde{\mbf{C}}_{\bm{\xi}+\bm{\xi}^\prime}} ~.
\end{align}
The effect of the parity superoperator on the parity and translation operators is to produce new parity and translation operators, defined on a new point on the phase space, which are, respectively, the reflection of the initial point with respect to a plane through $\bmx$ (and constant $\bm{\xi}$) and a plane through $\bm{\xi}$ (with constant $\bmx$). Similarly, the action of the translation superoperator on the parity and translation operators is to produce new parity and translation operators, defined on a new point on the phase space, which are, respectively,  translations by either $\bmx$ and $\bm{\xi}$. It is important to note that, the action of these superoperators on $\mbf{\Pi}_{\bmx}$ and $\suop{T}{\xi}$ do not mix the coordinates $\bmx$ and $\bm{\xi}$ in the argument of these operators, they only give rise to phase factors.

By direct calculation using the definitions of the parity and the translation superoperators, it can be verified that they satisfy the following composition relations
\begin{align}
     \check{\Pi}_{{\bmx}, \bm{\xi}}  \check{\Pi}_{{\bm{y}}, \bm{\eta}} &= \exp\bitd{2 i \bift{\braket{\bmx, \bm{\eta}}_s-\braket{\bm{y}, \bm{\xi}}_s}} \calchs{T}{2(x-y)} {2(\xi-\eta)}~,~~\calchs{T}{x} {\xi}\calchs{T}{y} {\eta} = \exp\bitd{-\frac{i}{2} \bift{\braket{\bmx, \bm{\eta}}_s-\braket{\bm{y}, \bm{\xi}}_s}}\calchs{T}{x+y} {\xi+\eta}~,\\
      \check{\Pi}_{{\bmx}, \bm{\xi}}  \calchs{T}{y} {\eta} &= \exp\bitd{ i \bift{\braket{\bmx, \bm{\eta}}_s-\braket{\bm{y}, \bm{\xi}}_s}}  \check{\Pi}_{{\bmx-\frac{1}{2}\bm{y}}, \bm{\xi}-\frac{1}{2}\bm{\eta}} ~,~ \calchs{T}{x} {\xi}\check{\Pi}_{{\bm{y}}, \bm{\eta}}  = \exp\bitd{ i \bift{\braket{\bmx, \bm{\eta}}_s-\braket{\bm{y}, \bm{\xi}}_s}}  \check{\Pi}_{{\bmx+\frac{1}{2}\bm{y}}, \bm{\xi}+\frac{1}{2}\bm{\eta}}~.
\end{align}
These relations are direct analogues of those in \eqref{affine_group} for parity and translation operators to the corresponding superoperators.

The \textit{double Weyl transform of a superoperator} $\calch{A}$ is defined using the parity superoperator in an analogous manner to the case of usual operators, as,
\begin{equation}\label{DWT}
    \bbch{A} (\bmx, \bm{\xi}) := \int \td \bm{y}~\avgc{\suop{C}{x-\frac{y}{2}}|\calch{A}|\suop{C}{x+\frac{y}{2}}}~\exp{\bitd{i\braket{\bm{y},\bm{\xi}}_s}}~=4 \texch{Tr}\bitd{\check{\Pi}_{{\bmx}, \bm{\xi}} \calch{A}}=\texch{Tr} \bitd{\calch{R}_{{\bmx}, \bm{\xi}} \calch{A}}~.
\end{equation}
Similarly, the \textit{double chord function} of a superoperator is given by \cite{Saraceno:2015cek}
\begin{equation}
    \tilde{\bbch{A}}(\bmx, \bm{\xi}) := \int \td \bm{\eta} ~ \avgc{\suopf{C}{\eta + \frac{\xi}{2}}|\calch{A}|\suopf{C}{\eta -\frac{\xi}{2}}}~\exp{\bitd{i\braket{\bm{x},\bm{\eta}}_s}}= \texch{Tr} \bitd{\calch{T}^\dagger_{{\bmx}, \bm{\xi}} \calch{A}}~,
\end{equation}
which is analogous to the relation in \eqref{Weyl_refle_tra} for the characteristic function of the Weyl transformation of an operator. For further discussion of different properties of the reflection and translation superoperators, we refer to \cite{Saraceno:2015cek}.

The expectation value of a superoperator $\calch{A}$ in the state $\ketc{\mbf{O}}$ can be written as the following phase space average in terms of their respective DWT 
\begin{equation}\label{su_mat_el}
     \avgc{\mbf{O}|\calch{A}|\mbf{O}} = \int \td \bmx ~\td \bm{\xi}~\bbch{A}(\bmx, \bm{\xi})~\bbch{W}_{\mbf{O}}(\bmx, \bm{\xi})~,
\end{equation}
where $\bbch{W}_{\mbf{O}}(\bmx, \bm{\xi})$ denotes the DWT of the superoperator $\ketc{\mbf{O}}\brac{\mbf{O}}$ normalised by a factor of $(2\pi)^2$, hence the \textit{double Wigner function} (DWF) of this superoperator.\footnote{When the symbol $\bbch{W}$ is used to denote the DWT of a superoperator, an extra factor of $(2\pi)^2$ is assumed to be multiplied in the denominator, as in the case of usual Wigner functions and Weyl transforms of density matrices.} This has the following expression in terms of the Weyl symbol of the operator $\mbf{O}$ (which is essentially the wavefunctions of the operator in the canonical supervector basis) 
\begin{equation}\label{DWF_OO}
\begin{aligned}
    \bbch{W}_\mbf{O}(\bmx, \bm{\xi}) &= \frac{1}{8\pi^3 D} \int ~\td \bm{y} ~\mb{O} \bift{\bmx -\frac{\bm{y}}{2}} \mb{O}^* \bift{\bmx+\frac{\bm{y}}{2}} ~\exp{\bitd{i\braket{\bm{y},\bm{\xi}}_s}}~\\
    &=\frac{1}{\pi^2 D} \int \td u \td v~ 
    \braket{q_+-u|\mbf{O}|q_- + v}\braket{q_--v|\mbf{O}^\dagger|q_+ +u}~ \exp \bitd{2i (up_{+}+v p_{-})}~,
\end{aligned}
\end{equation}
where in the final expression, we have rewritten the arguments of $\mb{H}$ in terms of a new set of coordinates, defined by, $\bmx_{\pm}=\bmx \pm \frac{1}{2}\bm{\xi}$. In classical dynamics, the canonical transformation $\bmx_- \rightarrow \bmx_+$ is generated by the \textit{center generating function} $\mc{S}(\bmx)$ \cite{de1998weyl}, from which the chord $\bm{\xi}$ is recovered using the relation $\bm{\xi}=J \frac{\partial \mc{S}(\bmx)}{\partial \bmx}$. Also, in the above equation, $u,v$ are two dummy variables, which are defined as a combination of position coordinates. The first expression for $ \bbch{W}_\mbf{O}(\bmx, \bm{\xi}) $ shows that it is a convolution of the Weyl symbol of the operator $\mbf{O}$ with shifted arguments, or equivalently, it is a symplectic Fourier transform of a correlation between the Weyl symbol of the operator.

\paragraph{Examples of DWF: Weyl-ordered operators.}
Let us now discuss some simple examples, where the DWF in \eqref{DWF_OO} can be explicitly evaluated. First consider the case when $\mbf{O}=\mbf{Q}$, i.e., the position operator.  Then a straightforward calculation shows that the corresponding $\bbch{W}_\mbf{Q}(\bmx, \bm{\xi})$ has the following expression\footnote{We became aware that a similar computation was reported in a different context in a recent work \cite{Wang:2025ywg}, where the authors have advocated an approach to the eigenstate thermalisation hypothesis based on the Wigner-Weyl phase space version of quantum mechanics.}
\begin{equation}
	\bbch{W}_\mbf{Q}(\bmx, \bm{\xi})  = \frac{1}{2 \pi D} \Bitd{q^2 \delta(\xi_q)\delta(\xi_p)+\frac{1}{4}\delta(\xi_q) \frac{d^2 \delta(\xi_p)}{d \xi_p^2}}~.
\end{equation}
On the other hand, when $\mbf{O}=\mbf{P}$, the momentum operator, the expression for the DWF reduces to 
\begin{equation}
	\bbch{W}_\mbf{P}(\bmx, \bm{\xi})  = \frac{1}{2 \pi D} \Bitd{p^2 \delta(\xi_q)\delta(\xi_p)+\frac{1}{4}\delta(\xi_p) \frac{d^2 \delta(\xi_q)}{d \xi_q^2}}~.
\end{equation}
From these expressions, we see that the first one has support only over the $\xi_q=0$ plane of the double phase space, while the second one has support only over the $\xi_p=0$ plane. 

We now consider the more complicated case, namely the DWF associated with generic Weyl-ordered operators. As is well known, the Weyl transform of a given phase space function of $(q,p)$ would give a Weyl-ordered (or, also known as the symmetric ordering) \cite{Hillery:1983ms} operator, which will contain an arithmetic average products of $\mbf{Q}$ and $\mbf{P}$ appearing in all possible permutations. For our purposes here, we consider the following Weyl-ordered operator (which  has two equivalent forms)  \cite{mccoy1932function,  Hillery:1983ms},
\begin{equation}\label{Weyl_ornm}
    \mbf{O}_{n,m}^W(\mbf{Q},\mbf{P}) = \frac{1}{2^n} \sum_{j=0}^n ~\binom{n}{j}~\mbf{Q}^{n-j} \mbf{P}^{m}\mbf{Q}^j=\frac{1}{2^m} \sum_{i=0}^n ~\binom{m}{i}~\mbf{P}^{m-i} \mbf{Q}^{n}\mbf{P}^i~.
\end{equation}
Here, the subscripts $n,m$ indicate the total number of the position and momentum operators, respectively, in the Weyl-ordered operator - a property which is made explicit with the superscript.
We can, of course, consider a more general form of the Weyl-ordered operator, such as, 
\begin{equation}
    \mbf{O}^W(\mbf{Q},\mbf{P}) = \sum_{n=0}^{\infty} \frac{1}{2^n} \sum_{j=0}^n ~\binom{n}{j}~\mbf{Q}^{n-j} f_n(\mbf{P})~\mbf{Q}^j~,
\end{equation}
with $f_n(\mbf{P})~$ being a function of the momentum operator. However, for our purpose of explicitly evaluating the corresponding DWF, the operator $\mbf{O}_{n,m}^W(\mbf{Q},\mbf{P})$ is easier to deal with. 

As mentioned previously, the Weyl transform of the operator $\mbf{O}_{n,m}^W(\mbf{Q},\mbf{P}) $ is just $\mb{O}_{n,m}^W(q,p)=q^np^m$, and therefore, the DWF associated with this Weyl-ordered operator can be evaluated explicitly to be given by
\begin{equation}
    \begin{aligned}
          \bbch{W}_{\mbf{O}^W_{n,m}}(\bmx, \bm{\xi}) &= \frac{1}{8\pi^3 D} \int ~\td y ~\td k ~\exp\bitd{i\bift{y \xi_p-k\xi_q}} \bift{q^2-\frac{1}{4}y^2}^n\bift{p^2-\frac{1}{4}k^2}^m\\
          &=  \frac{1}{2\pi D} \sum_{i=0}^{ m}\sum_{j=0}^{ n} \binom{n}{j}\binom{m}{i} \frac{q^{2(n-j)}p^{2(m-i)}}{2^{2(i+j)}}~\frac{d^{2j} \delta(\xi_p)}{d \xi_p^{2j}} \frac{d^{2i} \delta(\xi_q)}{d \xi_q^{2i}} ~.
    \end{aligned}
\end{equation}

\paragraph{\textbf{Relating double Wigner function with Weyl transform of operators.}}
As mentioned above, from the first expression of \eqref{DWF_OO}, we see that the function $\bbch{W}_\mbf{O}(\bmx, \bm{\xi})$ represents the symplectic Fourier transform of a correlation between the Weyl symbol $\mb{O}(\bmx)$ of the operator $\mbf{O}$ at different phase space coordinates. As elaborated in the following, one can in fact proceed further, and derive a more direct connection between DWF of a generic superoperator of the form $\calch{B}_{\mbf{AB}}=\ketc{\mbf{A}}\brac{\mbf{B}}$, and the Weyl symbol of a certain operator related to $\mbf{A}$ and $\mbf{B}$. 

By following a similar set of steps in driving the final expression in \eqref{DWF_OO}, we obtain the expression for the DWF  
\begin{equation}\label{DWF_AB}
\begin{aligned}
    \bbch{W}_\mbf{AB}(\bmx, \bm{\xi}) &= \frac{1}{8\pi^3 D} \int ~\td \bm{y} ~\mb{A} \bift{\bmx -\frac{\bm{y}}{2}} \mb{B}^* \bift{\bmx+\frac{\bm{y}}{2}} ~\exp{\bitd{i\braket{\bm{y},\bm{\xi}}_s}}~\\
    &=\frac{1}{\pi^2 D} \int \td u \td v~ 
    \braket{q_+-u|\mbf{A}|q_- + v}\braket{q_--v|\mbf{B}^\dagger|q_+ +u}~ \exp \bitd{2i (up_{+} +v p_{-})}~,
\end{aligned}
\end{equation}
which, in turn, can be rewritten as the following two alternative expressions, 
\begin{align}\label{DWF_AB1}
    \bbch{W}_\mbf{AB}(\bmx, \bm{\xi}) =\bbch{W}_\mbf{AB}(\bmx_{+}, \bmx_{-})&= \frac{1}{\pi^2 D} \int \td u~ 
    \braket{q_+-u|\mbf{A}\mbf{\Pi}_{\bmx_{-}}\mbf{B}^\dagger|q_+ +u}~ \exp \bitd{2i up_{+}}~\\
    &=\frac{1}{\pi^2 D} \int \td v~ 
    \braket{q_--v|\mbf{B}^\dagger\mbf{\Pi}_{\bmx_{+}}\mbf{A}|q_- +v}~ \exp \bitd{2i vp_{-}}~,\label{DWF_AB2}
\end{align}
where $\mbf{\Pi}_{\bmx_{\pm}}$ denote parity operators defined on the points $\bmx_{\pm}$, respectively (see eq. \eqref{reflection_op}). These two expressions now make the interpretation of the DWF $\bbch{W}_\mbf{AB}(\bmx, \bm{\xi}) $ in terms of the Weyl symbols of the operators $\mbf{A}$ and $\mbf{B}$ clear:
\begin{center}
    \textit{The DWF ($\bbch{W}_\mbf{AB}(\bmx, \bm{\xi}) $) of superoperators (which, by definition, is proportional to the DWT of generic superoperators of the form $\calch{B}_{\mbf{AB}}=\ketc{\mbf{A}}\brac{\mbf{B}}$) is (apart from an overall constant) the Weyl transform of the operator $\mbf{A}\mbf{\Pi}_{\bmx_{-}}\mbf{B}^\dagger$ (with respect to the coordinates $\bmx_{+}=(q_+,p_+)$), or alternatively, it is the Weyl transform of the operator $\mbf{B}^\dagger\mbf{\Pi}_{\bmx_{+}}\mbf{A}$ (when the coordinates $\bmx_{-}=(q_-,p_-)$ are used)}.
\end{center}
This is one of the main results of this section and makes the roles of the coordinates $\bmx_{\pm}$ in the double phase space to the DWF manifest (this is why in the above equation we have written $\bbch{W}_\mbf{AB}$ as a function of $\bmx_{\pm}$). We shall discuss more on why the DWTs (and hence DWFs) are always functions of $\bmx_{\pm}$, and the connection between the two expressions \eqref{DWF_AB1} and \eqref{DWF_AB2} in the Appendix. \ref{Why_xpm}.

Writing the last statement explicitly in terms of the $\star$-product, we have the following alternative expressions for the DWF in terms of the Weyl symbol and the chord function of the operators
\begin{equation}\label{DWF_Weyl}
\begin{aligned}
   \bbch{W}_\mbf{AB}(\bmx, \bmxi) &= \frac{1}{2 \pi^3 D} \mb{A}\bift{\bmx+\frac{1}{2} \bmxi} \star \bitd{\mb{D}^*_{2\bmxi}(\bmx)\bbf{B}^*(-2\bmxi)} = \frac{1}{2 \pi^3 D}\bitd{\mb{D}_{2\bmxi}(\bmx)\bbf{A}(2\bmxi)} \star \mb{B}^*\bift{\bmx +\frac{1}{2} \bmxi}\\
    &= \frac{1}{2 \pi^3 D} \mb{B}^*\bift{\bmx -\frac{1}{2} \bmxi} \star \bitd{\mb{D}_{2\bmxi}(\bmx)\bbf{A}(2\bmxi)} =\frac{1}{2 \pi^3 D} \bitd{\mb{D}^*_{2\bmxi}(\bmx)\bbf{B}^*(-2\bmxi)} \star ~\mb{A}\bift{\bmx-\frac{1}{2} \bmxi},
\end{aligned}
\end{equation}
where $\mb{D}_{\bmxi}(\bmx)$ is the Weyl transform of the displacement operator.  

We discuss here two important special cases, where the last relation simplifies. 
\begin{itemize}
    \item First consider the case where either of the two operators $\mbf{A}$ or $\mbf{B}$ is an identity operator. When $\mbf{B}=\mbf{I}$, the second expression in the first line of \eqref{DWF_Weyl} gives $\bbch{W}_\mbf{AI}(\bmx, \bmxi)=\frac{1}{2 \pi^3 D}\bitd{\mb{D}_{2\bmxi}(\bmx)\bbf{A}(2\bmxi)}$, whereas, when $\mbf{A}=\mbf{I}$, the second expression in the second line of \eqref{DWF_Weyl} gives $\bbch{W}_\mbf{IB}(\bmx, \bmxi) = \frac{1}{2 \pi^3 D} \bitd{\mb{D}^*_{2\bmxi}(\bmx)\bbf{B}^*(-2\bmxi)}$. That these relations are correct can be directly checked using the expression in the first line of \eqref{DWF_AB}. 
    \item  Next consider the situation where the superoperator $\calch{B}_{\bm{z}_1\bm{z}_2}=\ketc{\suop{C}{z_1}}\brac{\suop{C}{z_2}}$, i.e., it is constructed from the canonical supervectors. Then from the formula in \eqref{DWF_Weyl}, or directly from \eqref{DWF_AB}, it can be checked that the corresponding DWF is just the product of the Weyl symbol of the displacement operator at two different points, i.e.,  $\bbch{W}_{\bm{z}_1\bm{z}_2}(\bmx, \bmxi)=\frac{1}{\pi^2}\exp \bitd{-2i \braket{\bmx-\bm{z}_2, \bmxi}_s} \delta(\bm{z}_1-\bm{z}_2)=\frac{1}{\pi^2} \mb{D}^*_{2\bmxi}(\bmx)\mb{D}_{2\bmxi}(\bm{z}_2 )\delta(\bm{z}_1-\bm{z}_2)$.
\end{itemize}


\paragraph{\textbf{Example.}} For completeness, here we also work out an example of the evaluation of the DWF of a superoperator of the form $\calch{B}_{\mbf{AB}}=\ketc{\mbf{A}}\brac{\mbf{B}}$ with $\mbf{A} \neq \mbf{B}$. Specifically, for the purpose of illustration, we consider the scenario where $\mbf{A}=\mbf{O}_{n,m}^W(\mbf{Q},\mbf{P})$ is the Weyl-ordered operator in \eqref{Weyl_ornm}, and $\mbf{B}=\mbf{I}$ is the identity operator. An explicit computation starting from the first expression of \eqref{DWF_AB}, shows that the corresponding DWF is given by
\begin{equation}
        \begin{aligned}
          \bbch{W}_{\mbf{O}^W_{n,m}\mbf{I}}(\bmx, \bm{\xi} ) &= \frac{1}{8\pi^3 D} \int ~\td y ~\td k ~\exp\bitd{i\bift{y \xi_p-k\xi_q}} \bift{q-\frac{1}{2}y}^n\bift{p-\frac{1}{2}k}^m\\
          &=  \frac{1}{2\pi D} \sum_{i=0}^{ m}\sum_{j=0}^{ n} \binom{n}{j}\binom{m}{i} \frac{(-1)^jq^{(n-j)}p^{(m-i)}}{2^{i+j}i^{i+j}}~\frac{d^{j} \delta(\xi_p)}{d \xi_p^{j}} \frac{d^{i} \delta(\xi_q)}{d \xi_q^{i}} ~.
    \end{aligned}
\end{equation}

\paragraph{\textbf{Matrix elements of an operator and the double Wigner functions.}}
We now show that the modulus square of the matrix elements of a generic operator $\mbf{O}$ (which we assume to be Hermitian for convenience) in a generic complete orthonormal basis $\ket{B_n}$, are related to the DWF of the superoperator $\ketc{\mbf{O}}\brac{\mbf{O}}$. To find the sought-after connection, we perform a direct calculation to show that the following relation is valid:
\begin{align}\label{matrxiel_DWF}
|\braket{B_m|\mbf{O}|B_n}|^2 &= \frac{1}{(2\pi)^2} \int \td \bmx_1 \td \bmx_2 ~ \mc{J} (\bmx_1, \bmx_2)~ W^B_m \bift{\bmx_1 -\frac{1}{2}\bmx_2}  W^B_n \bift{\bmx_1 +\frac{1}{2}\bmx_2} \nonumber\\
&=\frac{1}{(2\pi)^2} \int \td \bmx_+ \td \bmx_- ~ \mc{J} (\bmx_+, \bmx_-)~ W^B_m \bift{\bmx_-}  W^B_n \bift{\bmx_+}~,
\end{align}
where $W^B_m(\bmx)$ denotes the Wigner function associated with $\ket{B_m}$, and we have defined (with a slight abuse of notation compared to those used previously) $\bmx_{\pm}=\bmx_1\pm \frac{1}{2} \bmx_2$.
From the explicit expression for the function $\mc{J} (\bmx_1, \bmx_2)$, it can be seen that this is just the DWF of the superoperator $\ketc{\mbf{O}}\brac{\mbf{O}}$ (with a proportionality constant), i.e., $\mc{J} (\bmx_1, \bmx_2) = (2 \pi)^2 D \bbch{W}_\mbf{O}(\bmx_1, \bmx_2) $ (see eq. \eqref{DWF_OO}). 
One important aspect of the relation \eqref{matrxiel_DWF} is that it expresses the matrix elements in terms of the Wigner functions of the projectors $\ket{B_n}\bra{B_n}$, and does not involve those of $\ket{B_n}\bra{B_m}$, with $n \neq m$ (this would be the case if we directly use a relation like the one in \eqref{matrix_krylov}).

\subsection{Double Weyl transforms of some specific superoperators}\label{sub_spce_DWT}

We have shown above that the DWF, which is proportional to the DWT of superoperators of the form  $\calch{B}_{\mbf{AB}}=\ketc{\mbf{A}}\brac{\mbf{B}}$, is directly related to the Weyl symbol of an operator related to $\mbf{A}$ and $\mbf{B}$. However, for the DWT of a generic superoperator, it is not possible to provide such an interpretation, unless the explicit form of the superoperator under consideration is known in terms of  Hilbert space operators.\footnote{We also note here that, as we shall show later in the next subsection, for the DWT of the product of two superoperators, one of which is say of the form $\bbch{A}^{-}$ or $\bbch{A}^{+}$, and the other has the form  $\calch{B}_{\mbf{AB}}=\ketc{\mbf{A}}\brac{\mbf{B}}$, it is possible to given a similar interpretation of the DWT as a Weyl transform of a specific operator related to $\mbf{A}$ and $\mbf{B}$.}

We now explicitly evaluate DWT for some specific superoperators we have encountered so far. These will help us to connect the DWT of these superoperators with the usual Weyl transforms of the operators they are constructed from, and will be very useful later on. We start by considering the Liouvillian superoperator. 
Using the definition of this superoperator,  $\calch{L}\mathbf{O}=[\mathbf{H},\mathbf{O}]$ and utilising the composition relations for the parity operator (see eq. \eqref{affine_group}), we derive the following expression for $\bbch{L}(\bmx, \bm{\xi})$ in terms of the Weyl symbol of the Hamiltonian, 
\begin{equation}\label{Liouvillian_DWT}
    \bbch{L}(\bmx, \bm{\xi}) = \mb{H}(\bmx +\bm{\xi}/2) - \mb{H}(\bmx -\bm{\xi}/2)  = \mb{H}( \bmx_+) - \mb{H}(\bmx_-)~.
\end{equation}
As a simple example, let us consider the case where the Hamiltonian is quadratic of the form, $\bm{H}(\bmX)=\bmX \cdot H_0 \bmX$, where $H_0$ is a real symmetric matrix with constant coefficients. In this case the Hamiltonian kernel is $\mb{H}(\bmx)=\bmx \cdot H_0 \bmx$. Therefore, the DWT of the associated Liouvillian superoperator is given by, $\bbch{L}(\bmx, \bm{\xi}) =2 \bmx \cdot \bm{\xi} = (\bmx_{+}+\bmx_{-})\cdot (\bmx_{+}-\bmx_{-}) $.

The relation between the DWT of a specific superoperator which is constructed from a given operator and the Weyl transform of the operator, as in in \eqref{Liouvillian_DWT} is not specific to the Liouvillian superoperator; in fact, for any superoperator defined as $\calch{A}^{-}:=[\mbf{A}, \bullet]$, one can establish the following relation between its DWT, and the Weyl transform of the operator $\mbf{A}$, 
\begin{equation}\label{DWT_A_minus}
    \bbch{A}^-(\bmx, \bm{\xi}) = \mb{A}(\bmx +\bm{\xi}/2) - \mb{A}(\bmx -\bm{\xi}/2)  = \mb{A}( \bmx_+) - \mb{A}(\bmx_-)~.
\end{equation}
Similarly, for the superoperator $\calch{A}^{+}=\frac{1}{2} \comt{\mbf{A}, \bullet}_{+}$, it is possible to derive the following expression for its DWT
\begin{equation}\label{DWT_A_plus}
    2\bbch{A}^+(\bmx, \bm{\xi})  = 2\bbch{A}^+(\bmx_{+}, \bmx_{-}) = \mb{A}(\bmx +\bm{\xi}/2) + \mb{A}(\bmx -\bm{\xi}/2)  = \mb{A}(\bmx_+) + \mb{A}(\bmx_-)~.
\end{equation}
Before moving on, we note the following points regarding the DWT of the superoperators derived above
\begin{itemize}
    \item The relations in \eqref{DWT_A_plus} and \eqref{DWT_A_minus}  are direct consequences of the composition relation for the parity operators in \eqref{affine_group}.
    \item  The DWT (defined in \eqref{DWT}) of a superoperator constructed from an operator can be written as a combination of the Weyl symbol of that operator, whose arguments are functions of the variables $\bmx_{\pm}$. It is possible to generically explain this observation by finding out the representation of the parity superoperator in terms of the parity operator, which is presented in the Appendix. \ref{Why_xpm}. 
    \item  Since under the transformation $\bmxi \rightarrow -\bmxi$, we have $\bmx_{\pm} \rightarrow \bmx_{\mp}$, it can be seen that $\bbch{A}^{-}(\bmx, \bm{\xi})$ is an odd function of $\bmxi$, while $\bbch{A}^{+}(\bmx, \bm{\xi})$ is an even function of $\bmxi$. 
  \item  The two relations in \eqref{DWT_A_plus} and \eqref{DWT_A_minus} provide a way of writing the Weyl symbol of an operator in terms of the DWT
of two superoperators constructed from it. Namely, we have 
\begin{equation}
    \mb{A}(\bmx)=\bbch{A}^+\bift{\bmx-\frac{1}{2} \bmxi, \bm{\xi}} + \frac{1}{2}\bbch{A}^-\bift{\bmx - \frac{1}{2} \bmxi, \bm{\xi}}=\bbch{A}^+\bift{\bmx + \frac{1}{2} \bmxi, \bm{\xi}} - \frac{1}{2}\bbch{A}^-\bift{\bmx + \frac{1}{2} \bmxi, \bm{\xi}}~.
\end{equation}
This relation very clearly illustrates the connection between the Wigner-Weyl phase space representation of an operator and the double phase space representation of two of the `canonical' superoperators constructed from it. Furthermore, the structure of the double phase space representation is such that the `extra' coordinates $\bmxi$ cancel from the RHS of the above relation. 
\end{itemize}

\subsection{Double Weyl transformation of product of superoperators}\label{sub_DWT_prod}
 We now define the DWT of the product of two superoperators and write a version of the star product for the DWT. The DWT of the product of two superoperators is defined as the star product  of the DWT of the individual superoperators:\footnote{To distinguish the star product between the DWT of superoperators, from the one in \eqref{star_def} between the Weyl transform of two operators, we have changed the symbol to $*$.}
\begin{equation}
    \bbch{A} (\bmx, \bm{\xi}) * \bbch{B} (\bmx, \bm{\xi}) := \int \td \bm{y}~~ \avgc{\suop{C}{x-\frac{y}{2}}|\calch{A}\calch{B}|\suop{C}{x+\frac{y}{2}}}~\exp{\bitd{i\braket{\bm{y},\bm{\xi}}_s}}~.
\label{Weyl_product_sopertaors}
\end{equation}

Analogous to eq. \eqref{star_to_pq}, the double phase space integral of $*$-product between two DWT is equal to the double phase space integral of the ordinary product of the DWT of the individual superoperators, which in turn is equal to the trace of the product of the superoperators, i.e., 
\begin{equation}
    (2 \pi)^2 \texch{Tr} \bitd{\calch{A}\calch{B}} = \int \td \bm{x} ~\td \bmxi ~\bbch{A} (\bm{x}, \bmxi) *\bbch{B} (\bm{x}, \bmxi) = \int \td \bm{x} ~\td \bmxi ~\bbch{A} (\bm{x}, \bmxi) \bbch{B} (\bm{x}, \bmxi) ~.
\end{equation}

The even and odd properties of the DWT, as mentioned above, along with the last relation, indicate that the trace of the product of a `minus' and `plus' superoperators vanishes,
\begin{equation}
     (2 \pi)^2 \texch{Tr} \bitd{\calch{A}^{-}\calch{B}^{+}} = \int \td \bm{x} ~\td \bmxi ~\bbch{A}^{-} (\bm{x}, \bmxi) \bbch{B}^{+} (\bm{x}, \bmxi) = 0~. 
\end{equation}

\paragraph{\textbf{Examples.}} We now derive the expressions for the DWT of the product of two superoperators, in terms of the Weyl transform of the underlying operators, for some specific examples which are relevant for the purpose of work. 
First consider a product of two superoperators, $\calch{A}^{-}=[\mbf{A}, \bullet]$ and $\calch{B}^{-}=[\mbf{B}, \bullet]$. The DWT of their product can be written in terms of the Weyl transforms of $\mbf{A}$ and $\mbf{B}$ as 
\begin{equation}\label{DWT_AB_suop}
    \bbch{A}^{-}(\bmx, \bm{\xi}) * \bbch{B}^{-}(\bmx, \bm{\xi}) = \mb{A}(\bmx_{+}) \star \mb{B}(\bmx_{+}) + \mb{A}(\bmx_{-}) \star \mb{B}(\bmx_{-}) - \mb{A}(\bmx_{+}) \mb{B}(\bmx_{-}) - \mb{A}(\bmx_{-})\mb{B}(\bmx_{+})~,
\end{equation}
where the phase space coordinates $\bmx_{\pm}$ has been defined after eq. \eqref{Liouvillian_DWT}, and given by $\bmx_{\pm}=\bmx \pm \frac{1}{2}\bm{\xi}$. In particular, for the product of two Liouvillian superoperators, the corresponding DWT reduces the following relation in terms of the Weyl symbol of the Hamiltonian, 
\begin{equation}
    \bbch{L}(\bmx, \bm{\xi}) * \bbch{L}(\bmx, \bm{\xi}) = \mb{H}(\bmx_{+}) \star \mb{H}(\bmx_{+}) + \mb{H}(\bmx_{-}) \star \mb{H}(\bmx_{-}) -2 \mb{H}(\bmx_{+}) \mb{H}(\bmx_{-})~.
\end{equation}
Note that when one of the superoperator, say $\calch{B}$, is $\calch{I}^{-1}=[\mbf{I}, \bullet]$, $\mbf{I}$ being the identity operator, the RHS in \eqref{DWT_AB_suop} vanishes, which is consistent with the fact that $\calch{I}^{-1}=0$.

Similarly, for the superoperator $\calch{A}^{+}=\frac{1}{2} \comt{\mbf{A}, \bullet}_{+}$ and $\calch{B}^{+}=\frac{1}{2} \comt{\mbf{B}, \bullet}_{+}$, we can derive the following expression for the DWT of their product,
\begin{equation}\label{DWT_AB_plus}
    4 \bbch{A}^{+}(\bmx, \bm{\xi}) * \bbch{B}^{+}(\bmx, \bm{\xi}) = \mb{A}(\bmx_{+}) \star \mb{B}(\bmx_{+}) + \mb{A}(\bmx_{-}) \star \mb{B}(\bmx_{-}) + \mb{A}(\bmx_{+}) \mb{B}(\bmx_{-}) + \mb{A}(\bmx_{-})\mb{B}(\bmx_{+})~.
\end{equation}
As an example, when $\calch{A}$ and $\calch{B}$ are both the so-called \textit{energy superoperator} \cite{Wilkie1},
$\calch{H}^+:=\frac{1}{2} \comt{\mbf{H}, \bullet}_{+}$, we have from the last expression 
\begin{equation}
    4 \bbch{H}^+(\bmx, \bm{\xi})* \bbch{H}^+(\bmx, \bm{\xi}) = \mb{H}(\bmx_{+}) \star \mb{H}(\bmx_{+}) + \mb{H}(\bmx_{-}) \star \mb{H}(\bmx_{-}) +2 \mb{H}(\bmx_{+}) \mb{H}(\bmx_{-})~,
\end{equation}
where $\bbch{H}^+(\bmx, \bm{\xi})$ denotes the DWT of the energy superoperator. Furthermore, it is to check that when $\calch{I}^{+}$ is one of the two superoperators, eq. \eqref{DWT_AB_plus} reduces to \eqref{DWT_A_plus}, which is again consistent with the fact that $\calch{I}^{+}$ is the identity superoperator. Note that, in all the relations above, the final expressions on the right-hand side depend on the variables $\bmx_{\pm}$, whose significance was discussed previously.  Also, since under the transformation $\bmxi \rightarrow -\bmxi$, we have $\bmx_{+} \rightarrow \bmx_{-}$, and vice versa, the $*$-product of $\bbch{A}^{-}$ and 
$\bbch{B}^{-}$ is seem to be (from \eqref{DWT_AB_suop}) an even function of $\bmxi$. This statement is also true for the $*$-product of $\bbch{A}^{+}$ and 
$\bbch{B}^{+}$, as can be seen from the relation in \eqref{DWT_AB_plus}.

So far, the two examples we have discussed above, DWT of the product of two superoperators, can be explicitly written as combinations of the Weyl transform of the operators from which the superoperator is constructed.  Next, we consider the DWT of the product of two superoperators: $\calch{A}^{-}$ and  $\calch{B}_{\mbf{B}_1\mbf{B}_2}=\ketc{\mbf{B}_1}\brac{\mbf{B}_2}$, where writing the final expression in terms of the Weyl transforms of the underlying operators is slightly more challenging. The DWT can be evaluated to be given by the following expressions,
\begin{equation}\label{DWT_A*WAB}
\begin{aligned}
    \bbch{A}^{-}(\bmx, \bm{\xi}) *\bbch{W}_{\mbf{B}_1\mbf{B}_2}(\bmx, \bm{\xi}) &=\frac{1}{(2 \pi)^3 D} \int \td \bm{y} ~\bitd{\mb{A}\bift{\bmx -\frac{\bm{y}}{2}}, \mb{B}_1\bift{\bmx -\frac{\bm{y}}{2}}}_{\star}\mb{B}_2\bift{\bmx +\frac{\bm{y}}{2}} ~\exp\bift{i \braket{\bm{y}, \bm{\xi}}_s}~\\ 
    &=  \frac{4}{D(2 \pi)^2} \int \td u \td v~ 
    \braket{q_+-u|[\mbf{A}, \mbf{B}_1]|q_- + v}\braket{q_--v|\mbf{B}_2^\dagger|q_+ +u}~ \exp \bitd{2i (up_{+} +v p_{-})}~,
\end{aligned}
\end{equation}
where $\bmx_{\pm}=(q_{\pm},p_{\pm})^T$. Once again, we notice that the structure of the last line is the same as \eqref{DWF_AB}, hence we can follow a strategy similar to the one described after this relation to see that $ \bbch{A}^{-}(\bmx, \bm{\xi}) *\bbch{W}_{\mbf{B}_1\mbf{B}_2}(\bmx, \bm{\xi}) $ can be rewritten as a $\star$-product between the Weyl symbol of the operator $\mbf{A}_{c}=[\mbf{A}, \mbf{B}_1]$ and the chord function of $\mbf{B}_2^\dagger$ (i.e., the symplectic Fourier transform of the Weyl symbol of $\mbf{B}_2^\dagger$) with appropriate arguments. 

Similarly the DWT of the product of $\calch{A}^{+}$ and  $\calch{B}_{\mbf{B}_1\mbf{B}_2}=\ketc{\mbf{B}_1}\brac{\mbf{B}_2}$ can be expressed as the $\star$-product between the Weyl symbol of the operator $\mbf{A}_{c}=[\mbf{A}, \mbf{B}_1]_{+}$ and the chord function of $\mbf{B}_2^\dagger$. 

\paragraph{\textbf{Double phase space representation of out-of-time order correlators.}}
As an application of the formulas derived above, we consider the case of a double commutator between two operators, one of which is time-evolved, 
\begin{equation}
    F(t)=D^{-1}\text{Tr}\bift{[\mbf{V}, \mbf{O}(t)]^\dagger [\mbf{V}, \mbf{O}(t)]} = \avgc{[\mbf{V}, \mbf{O}(t)]\big|[\mbf{V}, \mbf{O}(t)]}~,
\end{equation}
where we have assumed that, like the initial operator $\mbf{O}$, the operator $\mbf{V}$ is also Hermitian, and derive its phase space representation. This squared commutator contains an out-of-time order correlator \cite{larkin1969quasiclassical} of the form  $\text{Tr}\bift{\mbf{V}\mbf{O}(t)\mbf{V}\mbf{O}(t)}$, which can be a very convenient probe of the signature of chaos in quantum systems \cite{maldacena2016bound, Garcia-Mata:2022voo}.  This squared commutator also belongs to the family of Q-complexities introduced in \cite{Parker:2018yvk}, since $F(t)$ can be written as $F(t)=\avgc{\mbf{O}(t)\big|(\calch{V}^{-})^2\big| \mbf{O}(t)}$, and the superoperator  $(\calch{V}^{-})^2$ being positive semi-definite, satisfies the criterion for Q-complexity superoperator \cite{Parker:2018yvk, Rabinovici:2025otw}. Employing eqs. \eqref{su_mat_el} and \eqref{DWT_AB_suop}, we can derive the following expression for the squared commutator as an integral over the double phase space
\begin{equation}
    F(t) = \int \td \bmx ~\td \bm{\xi}~\bbch{W}_{\mbf{O}}(\bmx, \bm{\xi},t) \bitd{\mb{V}(\bmx_{+}) \star \mb{V}(\bmx_{+}) + \mb{V}(\bmx_{-}) \star \mb{V}(\bmx_{-}) -2 \mb{V}(\bmx_{+}) \mb{V}(\bmx_{-})}~,
\end{equation}
where $\bbch{W}_{\mbf{O}}(\bmx, \bm{\xi},t)$ denotes the DWF of the superoperator $\ketc{\mbf{O}(t)}\brac{\mbf{O}(t)}$, given by
\begin{equation}\label{DWT_O(t)}
    \bbch{W}_{\mbf{O}}(\bmx, \bm{\xi},t) = \frac{1}{8\pi^3 D} \int ~\td \bm{y} ~\mb{O}_{n} \bift{\bmx -\frac{\bm{y}}{2},t} \mb{O}_{m}^* \bift{\bmx+\frac{\bm{y}}{2},t} ~\exp{\bitd{i\braket{\bm{y},\bm{\xi}}_s}}~.
\end{equation}
$\mb{O}_{n} \bift{\bmx,t}$ formally denotes the Weyl symbol of the Heisenberg time-evolved operator $\mbf{O}(t)$ \cite{Osborn:1994sf, prosser1983correspondence}. In this context, we also mention one problematic aspect of using the  $\star$-product relation \eqref{star_def} to compute the Weyl symbol of a Heisenberg time-evolved operator $\mbf{O}(t)=e^{i \mbf{H} t/\hbar} \mbf{O}_0e^{-i \mbf{H} t/\hbar} $, by calculating the Weyl symbols of the initial operator $\mbf{O}_0$, and the time-evolution operator $\mbf{U}(\bmx,t)=e^{-i \mbf{H} t/\hbar}$: since  $\mb{U}(t)$ has essential singularity at $\hbar \rightarrow 0$ it does not have a regular Taylor series expansion in the classical limit. We refer to \cite{Osborn:1994sf} for a careful analysis of the problem and an alternative method for evaluating the Weyl symbols of time-evolved Heisenberg operators and determining their semiclassical limit.

\subsection{Operator Krylov basis and phase space functions}\label{sub_Op_Kry_phase}
With the above construction for the DWT and star product operation for the Liouville space operators and bases, we can now find out the phase space representation of the Krylov basis supervectors and the action of the Liouvillian on these. 
To begin with, consider the DWF of the superoperators  $\calch{B}_{nm}=\ketc{\mbf{O}_n}\brac{\mbf{O}_m}$ constructed from the Liouville space Krylov basis supervectors. This is given by (as before the superscripts $K$ refer to the fact that these are phase space functions corresponding to the Krylov basis elements, here the operator Krylov basis)\footnote{To be in accordance with the notation used in the previous subsections these should have been written as $\bbch{W}^K_{\mbf{O}_n\mbf{O}_m}(\bmx, \bm{\xi})$. However, to simplify the notation, we have omitted explicitly writing the Krylov basis operators and only indicated them by the indices $n$ and $m$.}
\begin{equation}\label{Wign_super}
    \bbch{W}^K_{nm}(\bmx, \bm{\xi})= \frac{1}{(2\pi)^2}\int \td \bm{y}~~ \avgc{\suop{C}{x-\frac{y}{2}}\ketc{\mbf{O}_n}\brac{\mbf{O}_m}\suop{C}{x+\frac{y}{2}}}~\exp{\bitd{i\braket{\bm{y},\bm{\xi}}_s}}~ = 4 \texch{Tr} \bitd{\check{\Pi}_{{\bmx}, \bm{\xi}} \calch{B}_{nm}}~.
\end{equation}
An expression for $\bbch{W}^K_{nm}(\bmx, \bm{\xi})$ can be derived in terms of the Weyl symbols of the operators $\mbf{O}_{n}$ and $\mbf{O}_m$ (these are the wavefunctions of the Krylov basis operator in the canonical supervector basis) as
\begin{equation}\label{Wnmch_2}
    \bbch{W}^K_{nm}(\bmx, \bm{\xi}) = \frac{1}{8\pi^3 D} \int ~\td \bm{y} ~\mb{O}_{n} \bift{\bmx -\frac{\bm{y}}{2}} \mb{O}_{m}^* \bift{\bmx+\frac{\bm{y}}{2}} ~\exp{\bitd{i\braket{\bm{y},\bm{\xi}}_s}}~.
\end{equation}
As before, the functions $\bbch{W}^K_{nm}(\bmx, \bm{\xi})$ therefore represent the symplectic Fourier transform of a correlation between the Weyl symbols $\mb{O}_{n}(\bmx)$s of the operators $\mbf{O}_n$s at different points on the phase space. Furthermore, these functions being a special case of the general DWF in \eqref{DWF_AB}, can similarly be interpreted as the $\star$-product of the Weyl symbol of and chord functions of the Krylov operator basis elements.

As should be clear, the functions $\bbch{W}^K_{nm}(\bmx, \bm{\xi})$ are the analogues of the functions  $\mcW_{nm}^K$ in eq. \eqref{W_nm_K}. We record a few useful properties of these functions, which are analogues to those of the functions $\mcW_{nm}^K$ given in section \ref{Ph_sp_Krylov}, and can be checked by straightforward calculation. 

1. The orthonormality and the completeness of the Krylov supervector basis imply, respectively,
\begin{equation}
    \int \td \bmx ~\td \bm{\xi} ~\bbch{W}^K_{nm}(\bmx, \bm{\xi}) = \delta_{nm}~,~~(2\pi)^2 \sum_{n} \bbch{W}^K_{nn}(\bmx, \bm{\xi})=1~.
\end{equation}

2. The matrix elements of a superoperator with respect to the Krylov supervectors can be written in terms of these functions as
\begin{equation}\label{suop_matel}
    \avgc{\mbf{O}_m|\calch{A}|\mbf{O}_n} = \int \td \bmx ~\td \bm{\xi}~\bbch{A}(\bmx, \bm{\xi})~\bbch{W}^K_{nm}(\bmx, \bm{\xi})~.
\end{equation}

3. Similarly, we can show that the following two relations are true,
\begin{equation}
    \int \td \bmx ~\td \bm{\xi}~\bbch{W}^K_{nm}(\bmx, \bm{\xi})~\bbch{W}^K_{ij}(\bmx, \bm{\xi}) = \frac{1}{(2\pi)^2} \delta_{ni}\delta_{mj}~,
\end{equation}
and 
\begin{equation}
   \sum_{nm} \bbch{W}^K_{nm}(\bmx, \bm{\xi})~\bbch{W}^K_{nm}(\bm{y}, \bm{\eta}) = \frac{1}{(2\pi)^2} \delta(\bmx-\bm{y})\delta(\bm{\xi}-\bm{\eta})~.
\end{equation}
Note that analogous relations as those of the above are true for any other complete orthonormal supervector basis $\{\ketc{\mbf{B}_n}\}$, with $\bbch{W}^K_{nm}(\bmx, \bm{\xi})$, replaced by the appropriate DWT of  $\ketc{\mbf{B}_n}\brac{\mbf{B}_m}$ (with appropriate factors of $2 \pi$ in the denominator, as in \eqref{Wign_super}). 

We now discuss the double phase space version of the action of the Liouvillian on the Krylov supervectors, i.e., eq. \eqref{Lanczos_alg}.  The action of the Liouvillian on $\calch{B}_{nm}=\ketc{\mbf{O}_n}\brac{\mbf{O}_m}$ transforms to the following relation, when written in terms of a function in the double phase space, 
\begin{equation}\label{Liou_Krylov}
    \bbch{L}(\bmx, \bm{\xi}) * \bbch{W}^K_{nm}(\bmx, \bm{\xi}) = b_n \bbch{W}^K_{(n-1)m}(\bmx, \bm{\xi}) + b_{n+1} \bbch{W}^K_{(n+1)m}(\bmx, \bm{\xi})~,
\end{equation}
and similarly, 
\begin{equation}
 \bbch{W}^K_{nm}(\bmx, \bm{\xi})  *   \bbch{L}(\bmx, \bm{\xi}) = b_m \bbch{W}^K_{n(m-1)}(\bmx, \bm{\xi}) + b_{m+1} \bbch{W}^K_{n(m+1)}(\bmx, \bm{\xi})~.
\end{equation}

As in case of the Wigner function associated with the state Krylov basis vectors considered in \ref{Krylov_and_Moyal}, starting from an initial distribution $\bbch{W}^K_{00}(\bmx, \bm{\xi})$, the above $*$-operation in \eqref{Liou_Krylov} (with $m=n$) can be successively performed to derive other $\bbch{W}^K_{nn}(\bmx, \bm{\xi})$s, which are the phase space representations of the operator Krylov basis elements. 

To understand the above relations more explicitly, let us consider the expression for the DWT of the Liouvillian, derived in \eqref{Liouvillian_DWT}. It is clear that the action of the Liouvillian on, say, the phase-space Krylov functions for operators, as in \eqref{Liou_Krylov}, is determined by a combination of the classical Hamiltonian kernel $\mb{H}$. The matrix elements of the Liouvillian in the Krylov basis, according to eq. \eqref{suop_matel} are given by,
\begin{equation}
    \avgc{\mbf{O}_m|\calch{L}|\mbf{O}_n} = \int \td \bmx ~\td \bm{\xi}~\bbch{L}(\bmx, \bm{\xi})~\bbch{W}^K_{nm}(\bmx, \bm{\xi}) = \int \td \bmx_{+} ~\td \bmx_{-}~\bbch{W}^K_{nm}(\bmx_{+}, \bmx_{-}) \bift{\mb{H}( \bmx_+) - \mb{H}(\bmx_-)}~. 
\end{equation}
Using the expression for $\bbch{W}^K_{nm}(\bmx, \bm{\xi}) $ from eq. \eqref{Wnmch_2} 
and the definition for the $\star$-product in \eqref{star_Fourier}, it can be seen that the RHS of the last expression is given by 
\begin{equation}
    \avgc{\mbf{O}_m|\calch{L}|\mbf{O}_n}=~\frac{1}{D}\int \td \bmx ~\mb{O}_m^*(\bmx) \bitd{\mb{H}(\bmx), \mb{O}_n(\bmx)}_\star = \frac{1}{D}\int \td \bmx ~\mb{H}(\bmx) \bitd{\mb{O}_n(\bmx), \mb{O}_m^*(\bmx)}_\star,
\end{equation}
as expected from the definition of the Liouvillian superoperator.

Finally, consider the DWT of the action of the Liouvillian on the superoperator, $\calch{B}_{nm}=\ketc{\mbf{O}_n}\brac{\mbf{O}_m}$ (which is the left-hand side of \eqref{Liou_Krylov}). Since this is a special case of the general relation in \eqref{DWT_A*WAB}, this can be evaluated to be given by,
\begin{equation}
\begin{aligned}
    \bbch{L}(\bmx, \bm{\xi}) *\bbch{W}_{nm}(\bmx, \bm{\xi}) &=\frac{1}{(2 \pi)^3 D} \int \td \bm{y} ~\bitd{\mb{H}\bift{\bmx -\frac{\bm{y}}{2}}, \mb{O}_{n}\bift{\bmx -\frac{\bm{y}}{2}}}_{\star}\mb{O}_{m}^*\bift{\bmx +\frac{\bm{y}}{2}} ~\exp\bift{i \braket{\bm{y}, \bm{\xi}}_s}~\\ 
    &=  \frac{4}{D(2 \pi)^2} \int \td u \td v~ 
    \braket{q_+-u|[\mbf{H}, \mbf{O}_n]|q_- + v}\braket{q_--v|\mbf{O}_m^\dagger|q_+ +u}~ \exp \bitd{2i (up_{+} +v p_{-})}~,
\end{aligned}
\end{equation}
and has a similar interpretation in terms $\star$-product of the Weyl symbol of $[\mbf{H}, \mbf{O}_n]$ and chord function of $\mbf{O}_m^\dagger$. Furthermore, since both the LHS and the RHS of the relation \eqref{Liou_Krylov} can be written in terms of the $\star$-product (see the discussion after eq. \eqref{Wnmch_2}), we conclude that the steps of the Lanczos algorithm for the Krylov basis operators can be performed in terms of the Weyl transform and the chord symbol for the Krylov basis operators.

\section{Measures of wavefunction and operator spreading in the phase space}\label{SC_phase}
In the previous two sections, we have obtained, respectively, the phase space functions associated with a complete basis of states in the Hilbert space and a complete basis of supervectors in the Liouville space, with the special emphasis on the so-called Krylov basis, generated by the Hamiltonian or the Liouvillian acting on an initial seed state or operator.  We have chosen to focus on the Krylov basis, since it has gained a lot of attention in the recent literature due to its usefulness in quantifying notions of complexity associated with the time evolution of the initial seed state or the operator \cite{Parker:2018yvk, Balasubramanian:2022tpr}. In this section, our goal is to understand the complexities associated with the spreading of an initial time-evolved state or a Heisenberg operator measured in the respective Krylov basis, in terms of the phase space Krylov functions we have constructed in the previous two sections. We first consider the case of the time evolution of wavefunctions 
in the Krylov basis and the notion of Krylov state complexity in the next two subsections, and deal with the Krylov operator complexity associated with a time-evolved Heisenberg operator in Section \ref{sub_Kry_op}.

\subsection{Decomposition of a state, transition probability and the Krylov basis }\label{sec_Krylov_decomposition}
Before directly proceeding with the notion of complexity of the spread of the time evolution of an initial state, here we first discuss a specific decomposition of the time evolution of the probability of obtaining an initial state and its phase space representation. This will help us to gain some intuition about the time evolution of an initial state in terms of the Krylov basis and its phase space version.

Consider a quantum state $\ket{\psi}$, which has been obtained from the action of an operator $f(\mbf{A)}$ on a fixed initial state $\ket{\psi_0}$ on the Hilbert space. Here $f(\mbf{A)}$ is a well-behaved and analytic function, and $\mbf{A}$ is a Hermitian operator, and we assume that the state, $\ket{\psi_0}$ is normalised and the final state is $\ket{\psi}_{N}=\frac{1}{\sqrt{N}}\ket{\psi}$, is normalised accordingly, where $N= \braket{f(\mbf{A})^2}_0$.  For our purposes later on, here we assume that  $f$ is a polynomial function,\footnote{In general, the following argument is valid as long as $f(\mbf{A})$ also represents a Hermitian operator, i.e., $f$ is a real valued function on the spectrum of $\mbf{A}$.} and the operator $\mbf{A}$ to be the Hamiltonian of the quantum system. A well-known expansion \cite{Aharonov} indicates that the 
$\ket{\psi}$ can be decomposed as (see the discussion in Appendix. \ref{sec_orthogonal_Krylov})
\begin{equation}\label{psi_decompo}
    \ket{\psi} = \braket{f(\mbf{A})}_0 \ket{\psi_0} + \Delta f(\mbf{A})_0 \ket{\psi_{0\bot}}~,
\end{equation}
where $\braket{f(\mbf{A})}_0$ and $ \Delta f(\mbf{A})_0 $ denote, respectively,  the expectation value and the variance of the operator $f(\mbf{A})$ in the initial state, while $\ket{\psi_{0\bot}}$ denotes a state which is orthogonal to the initial state. Note that, here,
\begin{equation}
    \braket{f(\mbf{A})}_0  = \braket{\psi_0|\psi}~, ~~\text{and}~~ (\Delta f(\mbf{A})_0)^2  = 1-|\braket{\psi_0|\psi}~|^2 = 1-|\braket{f(\mbf{A})}_0|^2~.
\end{equation}
Now consider the phase space representation of the operator $\mbf{W}=\ket{\psi}_{N}~ _{N}\bra{\psi}$, which has the expansion, 
\begin{equation}
    \mbf{W}=\frac{1}{N}\ket{\psi}\bra{\psi} = \frac{1}{N}\Big(|\braket{f(\mbf{A})}_0|^2 \ket{\psi_0}\bra{\psi_0}+(\Delta f(\mbf{A})_0 )^2 \ket{\psi_{0\bot}}\bra{\psi_{0\bot}} +\braket{f(\mbf{A})}_0(\Delta f(\mbf{A})_0 ) \bitd{\ket{\psi_{0}} \bra{\psi_{0\bot}}+\ket{\psi_{0\bot}}\bra{\psi_{0}}}\Big)~.
\end{equation}
The Weyl transform of this operator can be expressed as 
\begin{equation}\label{Wigner_expansion}
N W(\bmx)=     |\braket{f(\mbf{A})}_0|^2 W_{0}(\bmx) + (\Delta f(\mbf{A})_0 )^2  W_{0\bot}(\bmx) +  \braket{f(\mbf{A})}_0 (\Delta f(\mbf{A})_0) \bitd{\mb{P}^\bot(\bmx) +\mb{P}_\bot(\bmx) }~,
\end{equation}
where $W_{0}(\bmx) $ and $W_{0\bot}(\bmx) $ represent, respectively, the Wigner functions associated with the states $ \ket{\psi_0}$ and $ \ket{\psi_{0\bot}}$, while, $\mb{P}_\bot(\bmx) $ and $\mb{P}^\bot(\bmx) $ denote Wigner functions corresponding to the operators $\ket{\psi_{0\bot}}\bra{\psi_{0}}$ and $\ket{\psi_{0}}\bra{\psi_{0\bot}}$.  Note that the quantity within curly brackets in the third term is real since $\mb{P}_\bot(\bmx) $ is the complex conjugate of $\mb{P}^\bot(\bmx) $ (see Appendix \ref{sec_orthogonal_Krylov}).  The expression in \eqref{Wigner_expansion} is the general expansion of the Wigner function of the state $\ket{\psi}$ in terms of the Wigner functions of the initial state, the state orthogonal to the initial state and their outer products. 

Next, we use the expansion in \eqref{psi_decompo} to write the transition probability between a generic time-dependent state, $\ket{\phi(t)}$ and the state $\ket{\psi}$
as 
\begin{equation}
\begin{split}
	|\braket{\phi(t)|\psi}|^2 = |\braket{f(\mbf{A})}_0|^2 	|\braket{\phi(t)|\psi_0}|^2 + (\Delta f(\mbf{A})_0)^2 |\braket{\phi(t)|\psi_{0\bot}}|^2\hspace{3 cm}\\+ \braket{f(\mbf{A})}_0(\Delta f(\mbf{A})_0) \bitd{\braket{\phi(t)|\psi_0}\braket{\psi_{0\bot}|\phi(t)}+\braket{\phi(t)|\psi_{0\bot}}\braket{\psi_{0}|\phi(t)}}~.
\end{split}
\end{equation}

Let us now use the last result to the case where $\ket{\phi(t)}$ is the time-evolved initial state $\ket{\Psi(t)}$, and consider the expansion of the transition probabilities $|\braket{\Psi(t)|K_n}|^2 $, where $\ket{K_n}$ denotes the elements of the Krylov basis constructed starting from the initial state. In that case the function $f(\mbf{A})$ would be the orthogonal polynomials $\mc{P}_n(\mbf{H})$ for each value of $n$, and $\braket{\mc{P}_n(\mbf{H})}_0=\delta_{n0}$. Furthermore, the states $\ket{K_n}=\mc{P}_n(\mbf{H}) \ket{\psi_0}$ are already orthonormal. Therefore, the expression for the probability of obtaining the $n$-th  Krylov basis in the time-evolved state can  be written as 
\begin{equation}
	|\braket{\Psi(t)|K_n}|^2 = \delta_{n0}^2 |\braket{\Psi(t)|\psi_0}|^2 + (\Delta \mc{P}_n(\mbf{H})_0)^2 |\braket{\Psi(t)|\psi_{0\bot}}|^2 + \delta_{n0} \Delta \mc{P}_n(\mbf{H})_0 
	 \bitd{\braket{\Psi(t)|\psi_0}\braket{\psi_{0\bot}|\Psi(t)}+\braket{\Psi(t)|\psi_{0\bot}}\braket{\psi_{0}|\Psi(t)}} ~.
\end{equation}
The first term above contributes only for $n=0$, for which the contributions of the second and the third terms vanish. On the other hand, for $n\neq 0$, the first and the third terms vanish, and the only contribution to the transition probabilities comes from the overlap of the time-evolved initial state with a state orthogonal to it. 

\subsection{Krylov state complexity as a phase space average}\label{sub_spread_ph_spa}
The central idea behind the definition of a notion of complexity associated with a time-evolved state is to construct a suitable \textit{cost function},  which would provide a measure of how `far' the time-evolved initial state has propagated in the given orthonormal and complete set of basis vectors in the Hilbert space ($\ket{B_n}$). This is typically defined in the literature through a weighted average over probabilities of obtaining the time-evolved state in an element of the basis, such as $\sum_{n} c_n |\braket{\Psi(t)|B_n}|^2$, where $c_n$ is taken to be a real positive and increasing sequence of numbers to ensure that cost of the time-evolved state increases 
if it spreads `deeper' with in the orthonormal basis.  As we discuss below, this quantity can be written as a phase space average over the product of the Wigner function of the time-evolved state and another phase space function that has the information of the orthonormal basis under consideration.  
The cost function, constructed in this manner, can be shown to be minimised in the Krylov basis generated from the action of the Hamiltonian on the initial state over a finite amount of time after the start of the time evolution, and hence is the cost function evaluated in the Krylov basis is defined as the complexity of the time-evolved state \cite{Balasubramanian:2022tpr}. 
Even though in this section we mostly consider quantities evaluated in the  Krylov basis, we briefly discuss comparison with a general complete orthonormal basis towards the end of the next section.

To begin with, from  the relation in \eqref{transition_state}, we see that the probability of obtaining the $n$-th element of the Krylov basis in the time-evolved state can  be written as 
\begin{equation}\label{pn_Wign}
	p_n(t)= |\braket{K_n|\Psi(t)}|^2= 2\pi \int W^K_{nn}(q,p) W (q,p,t) ~\td p ~\td q~. 
\end{equation}
As a simple consistency check, we can substitute the expansion of the Wigner function from eq. \eqref{WFexpanK} in the right side of this expression, and after simplification
using the orthonomality of the functions $\psi^K_n(q)$ (eq. \eqref{orthpsi}), see that it is indeed equal to $|\phi_n(t)|^2$. 
As an example,  the survival probability of the initial state $\ket{\psi_0}$ can be expressed as 
\begin{equation}
	|\braket{\psi_0|\Psi(t)}|^2= |\braket{K_0|\Psi(t)}|^2= 2\pi \int W_{00}^K(q,p) ~ W (q,p,t) ~ \td p ~\td q~. 
\end{equation}
Using these expressions for the probabilities $p_n(t)$, the expression for the Krylov state complexity can be written as  
\begin{equation}\label{SCWF2}
	\mathcal{C}(t) = \sum_{n} n ~|\phi_n(t)|^2= 2\pi \sum_{n} n  \int W^K_{nn}(q,p) W (q,p,t) ~ \td p \td q~. 
\end{equation}

In order to obtain a better understanding of the above expression for $\mathcal{C}(t)$, we define the following function in the phase space as a sum over the diagonal phase space Krylov functions $W^K_{nn}(q,p)$ (weighted by $2 \pi n$), i.e., 
\begin{equation}\label{weylK}
	\mathbb{K} (q,p) :=  2 \pi \sum_{n} n ~W^K_{nn}(q,p) ~,
\end{equation}
so that the expression for the Krylov state complexity in eq. \eqref{SCWF2} can be rewritten as 
\begin{equation}\label{SCwigner}
	 \mathcal{C}(t) = \int  W (q,p,t)  ~ \mathbb{K}  (q,p) ~ \td p ~\td q~ ~. 
\end{equation}
This relation indicates that the Krylov state complexity can be thought of as the phase-space average of the function $\mathbb{K}(q,p)$, defined in terms of the canonical phase space coordinates $(q,p)$ with the probability density given by the Wigner function of the time-evolved state. Furthermore, using the relation in eq. \eqref{expecttaion}, it can be easily checked that the expression for the Krylov state complexity in \eqref{SCwigner} is consistent with the fact that $\text{Tr} [\rho (t) \mathbf{K}] = \mathcal{C}(t)$, where  $\rho(t)$ is the density matrix associated with the time evolved state $\ket{\Psi(t)}$. We also note that, for a given Hamiltonian $\mbf{H}$ and an initial state $\ket{\psi_0}$, the Wigner function for the time-evolved state is fixed. Therefore, the effect of the Krylov basis on the growth of the Krylov state complexity is solely contained in the function $\mathbb{K}(q,p)$.

The long-time average of the Krylov state complexity, using the formula for the long-time average of the Wigner function from eq. \eqref{WFavg}, is obtained to be 
 \begin{equation}
 \bar{\mathcal{C}} = \sum_{a} \rho_0 (E_a)   \sum_{n,m}  P_{mn} (E_a) \int  ~ \mathcal{W}^K_{nm} (p,q) ~\mathbb{K}  (q,p)  ~ \td p ~\td q~. 
 \end{equation}
We can simplify this by using the integrals given in the Appendix. \ref{Integrals}, and noting that only when $m=n$, does the above integral over the phase space have a non-zero value. 
Thus, we get back the usual expression \cite{Fu:2024fdm}
 \begin{equation}
	\bar{\mathcal{C}} = \sum_{a,n} n ~\rho_0 (E_a)  P_{n} (E_a)~,
\end{equation}
where we have denoted $P_{n} (E_a) = \braket{E_a|\mathbf{P}_{n}|E_a}$.

\paragraph{\textbf{Spreading operator.}}
We now provide an interpretation of the function $\mathbb{K} (q,p)$ in \eqref{weylK} as follows. Writing this function as 
\begin{equation}
	\mathbb{K} (q,p) =  2 \int e^{2i p y}  \bra{q - y} \mathbf{K} \ket{q+ y}  ~ \td y~,
\end{equation} 
we see that the operator $\mathbf{K}$ is nothing but the so-called \textit{spreading operator} in the Krylov basis, defined as
\begin{equation}
	\mathbf{K} := \sum_{n} n \ket{K_n} \bra{K_n}~.
\end{equation}
Furthermore, substituting the expression for the Wigner function in terms of the reflection operator (see the first relation in \eqref{Wigner_refle_tra}) on the right side of the eq. \eqref{SCwigner} (when written at an arbitrary time $t$), and comparing with the left side, one can derive the expected relation between the spreading operator, the function $\mathbb{K}(q,p)$, and the reflection operator 
\begin{equation}
 	\mathbf{K} = \int \mathbb{K}(\bmx) ~\suop{R}{x}  ~ \td \bmx~.
\end{equation}
This is a consequence of the completeness of the reflection operator, and analogous to the relation \eqref{rhoWA} connecting the density matrix, Wigner function and the reflection operator $\suop{R}{x}$. 

The discussion above suggests that the function $\mathbb{K} (\bmx)$ can be formally regarded as the Weyl transform of the spreading operator; hence, the Krylov state complexity can be understood as the phase-space average of the Weyl transform of the spreading operator. In other words, the Krylov state complexity is the phase space integral of the function $\mathbb{K} (\bmx)$, which is smeared by the time-dependent Wigner function. Wavefunction spreading in the Krylov basis in this picture, therefore, means that, after the initial time, the time-evolving Wigner function has increasing overlap with the phase space function $\mathbb{K} (\bmx)$, which is, in turn, a sum of phase space Krylov functions. Alternatively, it is straightforward to verify that we can also express the Krylov state complexity as an overlap of time-evolved phase space Krylov functions with an initial Wigner function, which is fixed by the initial state. As an example, the Wigner function of a Gaussian wave packet, which is localised around a point $\bmx$ in the phase space, would similarly be a localised Gaussian function \cite{de1998weyl,hudson1974wigner}. Hence, for this initial state, the Krylov state complexity is the smearing of this Gaussian function with $\mb{K}(\bmx,t)$. 

As we shall elaborate below, the notion of state complexity quantifies how much structure the Wigner function of the time-evolved state develops under evolution, and accordingly, $\mathcal{C}(t)$ is a measure of this in terms of complete orthonormal set of basis function in phase space, induced by the Krylov subspace generated by the Hamiltonian on the given initial state.

We also note that the action of $\mb{K}(\bm{x})$ on the phase-space Krylov functions 
$\mathcal{W}^K_{ij} (\bm{x})$, which can be evaluated to be 
\begin{equation}
    \mathbb{K}(\bm{x}) \star \mathcal{W}^K_{nm} (\bm{x})=n~ \mathcal{W}^K_{nm} (\bm{x})~,
\end{equation}
are exact analogues of the star-eigenvalue equation for the Hamiltonian \eqref{star_eigen}, where $n$ are the eigenvalues of the spreading operator $\mbf{K}$, and $W_{nn}^K(\bm{x})$ play the role of the Wigner functions of the eigenstates, since the Krylov basis elements are eigenstates of the spreading operator. Moreover, using this, the star-composition of $\mathbb{K}(\bm{x}) $ and the generating functions \eqref{gen_function} is given by 
\begin{equation}
    \mathbb{K}(\bm{x}) \star G^K(\bm{\mu};\bm{x}) = \mu_2 \partial_{\mu_2} G^K(\bm{\mu};\bm{x})~.
\end{equation}

\paragraph{\textbf{Statistics of the wavefunction spreading operator.}} Generalising the above discussion, one can consider the statistics of the spreading operator and write it in terms of the Wigner function. 
Consider the time evolution of an initial state $\ket{\psi_0}$ generated by the time-independent Hamiltonian of a particle in one dimension moving under the influence of a potential, and perform a measurement of the spreading operator at time $t$.  
The probability distribution of the result of the measurement, $P(j, t)$  \cite{Fu:2024fdm}, can be written in terms of the Wigner function (using the relation \eqref{pn_Wign}) as 
\begin{align}\label{SOsta}
		P(j, t)  &= \sum_n | \bra{ K_n} e^{-it H} \ket{\psi_0}| ^2 \delta (j-n)=  \sum_n   |\phi_n(t)|^2 ~ \delta (j-n)~\nonumber\\
        &= 2\pi \sum_n \delta (j-n)\int W^K_{nn}(\bmx) W (\bmx,t) ~\td \bmx.
\end{align}
The moments of the characteristic function of $P(j, t) $ (the Fourier transform of $P(j, t)$) are proportional to the so-called generalised Krylov state complexity \cite{Fu:2024fdm}. These moments contain finer information about the total distribution, compared to only the Krylov state complexity, which is the first moment of this distribution.\footnote{For discussions on time evolution of these generalised Krylov state complexity measures in generic quantum chaotic systems we refer to \cite{Camargo2025higher}, where it was argued that for specific initial states, these higher order quantities can be more sharper probes of quantum chaos than usual Krylov state complexity.}   

\paragraph{\textbf{In the Fourier representation.}}
We can write down the expression for the Krylov state complexity in \eqref{SCWF2} in terms of the Fourier transformations of the Wigner function and the phase space Krylov functions as well. Performing the transformations, we have the following expression for the Krylov state complexity 
\begin{align}
    \mathcal{C}(t) &= \sum _n n \int \frac{\td \bm{\xi}}{2 \pi}~ \braket{K_n|\mbf{D}_{-{\bm{\xi}}}|K_n}~\braket{\Psi(t)|\mbf{D}_{\bm{\xi}}|\Psi(t)}
    = (2 \pi)^{-1} \int \td \bm{\xi} ~ \text{Tr} (\mbf{D}_{-\bm{\xi}} \mathbf{K})~\text{Tr}(\rho(t) \mbf{D}_{\bm{\xi}} )~.
\end{align}
As before, this shows that the Krylov state complexity can be viewed as the characteristic function of the spreading operator, smeared over a region of phase space determined by the characteristic function of the Wigner function of the time-evolved state. In this interpretation, the region over which the distribution $\tilde{W}(\bm{\xi},t)=\text{Tr}(\rho(t) \mbf{D}_{\bm{\xi}} )$ has a support that changes with time, and Krylov state complexity therefore measures the overlap of this changing distribution with the constant one, namely, $\tilbb{K}(\bm{\xi})=\text{Tr} (\mbf{D}_{-\bm{\xi}} \mathbf{K})$.

\paragraph{\textbf{Complexity of evolution and phase space distributions.}}
For classical dynamical systems, it is well-known that the complexity of the dynamics and chaotic properties, usually quantified by the exponential divergence of nearby trajectories and the associated Lyapunov exponent, are closely related \cite{benatti2009dynamics}. Even though the standard notion of trajectories cannot be straightforwardly extended to quantum dynamical systems, one can adopt an approach based on phase-space distribution functions for this purpose - thereby quantifying the complexity of quantum evolution.  In fact, a notion of complexity of quantum dynamics based on classical and quantum phase space distribution functions has been studied extensively in the literature, which is somewhat different from the notions of complexity based on the Krylov basis we have considered here. For classical dynamical systems, the 2nd moment of the Fourier component of the classical phase space distribution function is known to grow linearly for an integrable system, while for a chaotic system it grows exponentially, thereby making this quantity a suitable probe for the complexity of quantum dynamics \cite{PhysRevE.56.5174, gong2003chaos,  gu1985group, Wang:2019juv}. The rate of exponential growth in the latter case is determined by the Lyapunov exponent, which characterises the local exponential instability of the nearby trajectories.  Analogously, for a single-particle quantum system, it has been shown that the 2nd moment of the harmonics of the Wigner function is a good measure of the complexity of quantum dynamics\footnote{For quantum systems with a small number of degrees of freedom, the exponential growth in the number of harmonics persists only up to Ehrenfest time scale $t_E \propto \ln \hbar$, after which it becomes, at most,  linear \cite{Sokolev}. Also note that an approach based on the number of harmonics of the Wigner function, to quantify the complexity of many-body quantum dynamics was later formulated in \cite{balachandran2010phase}.} since it can be used to detect the crossover from integrability to chaos \cite{Sokolev, Beneti, PhysRevE.56.5174}.

We now discuss the connection and differences between this notion of the complexity of quantum motion, as measured by the growth of the number of harmonics of the time-evolved Wigner function and the Krylov state complexity discussed above. First consider the Fourier expansion of the Wigner function expressed in terms of the action-angle variables $(I,\theta)$, 
\begin{equation}\label{Wign_Four_expa}
    W(I, \theta,t)= \frac{1}{\pi} \sum_{n=-\infty}^{\infty} W_{n}(I,t)~ e^{i n\theta}~,~~I\geq 0~, \theta \in [0, 2 \pi)~.
\end{equation}
Then, a good measure of the  complexity of the evolution or the complexity of the time-evolved Wigner function is the square root of the second moment ($\braket{n^2}_t$) of the harmonic distribution $W_n(I,t)$, where \cite{Sokolev, balachandran2010phase, Beneti}
\begin{equation}\label{2nd_momnet}
    \braket{n^2}_t = \frac{1}{\mc{N}_{2}} \sum_{n=-\infty}^{\infty} n^2 \int_{0}^{\infty} \td I ~ |  W_{n}(I,t)|^2~, ~~\text{with}~~\mc{N}_{2}= \sum_{m=-\infty}^{\infty} \int_{0}^{\infty} \td I ~ |  W_{n}(I,t)|^2~.
\end{equation}
As the definition suggests, this quantity is a weighted average over the expansion coefficients $W_{n}(I,t)$. To distinguish this from the notions of complexity based on Krylov subspace we have discussed so far, we refer to $\mcC_{h}(t)=\sqrt{\braket{n^2}_t }$ as the \textit{complexity of harmonics}.

Comparing the relations in \eqref{Wign_Four_expa} with the one in \eqref{WFexpanK}, we see that both represent an expansion of the Wigner function of the time-evolved state (albeit, expressed in different sets of canonical coordinates) in a complete basis. Of course, the crucial difference between them is that in \eqref{Wign_Four_expa}, the expansion is not in general with respect to a complete basis in the phase space,
whereas the expansion in \eqref{WFexpanK} is indeed with respect to a complete basis $\mcW^{K}(\bmx)$ for the phase space functions on $L^2(\mb{R}^2)$. Nevertheless, the 
Krylov state complexity defined through the expansion coefficients $\phi_{n}(t)$ as in \eqref{SCWF2},
or a generalisation of it with $n$ replaced by $n^k$ ($k =1,2,3,\cdots$) \cite{Fu:2024fdm},
\begin{equation}\label{Kryl_state}
    \mc{C}^{(k)}(t)=\sum_{n} n^k ~|\phi_n(t)|^2~,
\end{equation}
can be seen to have a similar status as that of the complexity of harmonics: they both are quantifiers of how the expansion coefficients of the Wigner function change with time, and measure the complexity of the time-evolved state.  In \eqref{2nd_momnet}, the presence of an extra integral over a phase space variable (the action variable $I$) is due to the fact mentioned above: the Fourier expansion of $W(I, \theta,t)$ does not represent an expansion with respect to a complete basis for the phase space functions. 

The previous discussion should make it clear that the Krylov state complexity and the complexity of harmonics (studied in \cite{Sokolev, balachandran2010phase, Beneti}, among others) can be thought to belong to the same class of complexity measures for the time evolution of a quantum state, when viewed through its phase space representation, namely the Wigner function. This also suggests that one can define a general class of complexity measures from the expansion coefficients of the Wigner function $W(\bmx,t)$ in a complete orthonormal basis for the phase space functions on $L^2(\mb{R}^2)$. Note that, however, if this phase space basis is constructed from the Weyl transforms of the elements of a complete basis in the Hilbert space, then a theorem established in \cite{Balasubramanian:2022tpr} implies that a general `cost function' of the form \eqref{Kryl_state} will be minimised for a finite amount of time after the start of the evolution, when computed with respect to the Krylov basis generated by the Hamiltonian starting from the initial state (see also the discussion in the next section).  

It is also well-known that the complexity of harmonics is closely related to the fidelity between the initial state and a suitably perturbed and subsequently reversed time-evolved version of the initial state \cite{benatti2009dynamics, Sokolev}. In the Appendix. \ref{sec_fidelity} we discuss a natural double phase space representation of the relevant fidelity, and show how it is related to the Wigner function of the time-evolved initial state.

\subsection{Krylov operator complexity as a  double phase space average}\label{sub_Kry_op}
Similar to the Krylov state complexity, the Krylov operator complexity can also be expressed in terms of the double phase space quantities defined in Section \ref{sub_Op_Kry_phase}. From \eqref{suop_matel}, we first note that the transition probability of a time-evolved operator $\mbf{O}(t)$ in terms of the Krylov supervector element $\mbf{O}_n$ can be written as 
\begin{equation}
   | \avgc{\mbf{O}_n|\mbf{O}(t)}|^2= (2 \pi)^2\int \td \bmx ~\td \bm{\xi}~\bbch{W}_\mbf{O}(\bmx, \bm{\xi},t)~\bbch{W}^K_{nn}(\bmx, \bm{\xi})~,
\end{equation}
where $\bbch{W}_\mbf{O}(\bmx, \bm{\xi},t)$ is the DWT of the superoperator $\ketc{\mbf{O}(t)}\brac{\mbf{O}(t)}$, given in \eqref{DWT_O(t)}. 

Since the Krylov operator complexity is defined as a weighted average of these transition probabilities \cite{Parker:2018yvk}, it is given by the following expression in terms of the phase space quantities,
\begin{align}\label{KC_phase}
    \mc{C}_{K}(t)= \sum_{n=0}^{D_K-1} n ~| \avgc{\mbf{O}_n|\mbf{O}(t)}|^2 &=   (2 \pi)^2
    \sum_{n=0}^{D_K-1} n ~\int \td \bmx ~\td \bm{\xi}~\bbch{W}_\mbf{O}(\bmx, \bm{\xi},t)~\bbch{W}^K_{nn}(\bmx, \bm{\xi}) \nonumber\\
    &= \int \td \bmx ~\td \bm{\xi}~\bbch{W}_\mbf{O}(\bmx, \bm{\xi},t)~\bbch{K}(\bmx, \bm{\xi})~,
\end{align}
where $\bbch{K}(\bmx, \bm{\xi})$ is defined as the weighted sum of the $\bbch{W}^K_{nn}(\bmx, \bm{\xi})$s, and as in the case of the wavefunction spreading operator discussed in Section \ref{sub_spread_ph_spa}, it can be thought of as the DWT of the Krylov spreading superoperator,
\begin{equation}
    \calch{K}:=\sum_{n=0}^{D_K-1} n  ~\ketc{\mbf{O}_{n}}\brac{\mbf{O}_n}~.
\end{equation}
Analogous to the Krylov state complexity, the relation in \eqref{KC_phase} can be intuitively understood as the overlap between two double-phase space distributions, one of which is time-dependent. Hence, the growth of an initial operator in the Krylov subspace corresponds to the fact that the distribution $\bbch{W}_\mbf{O}(\bmx, \bm{\xi},t)$
has access to (or overlaps with) an increasing number of phase space Krylov functions for operators. 

Even though, the expression \eqref{KC_phase} is analogous to expression in \eqref{SCwigner} for the Krylov state complexity, it is further possible to rewrite the expression for Krylov operator complexity in \eqref{KC_phase} in terms of the centre variables $\bmx$ and Weyl symbols of the operators $\mbf{O}(t)$ and $\mbf{O}_n$ only without the integral over the chord variables $\bm{\xi}$. The corresponding expression can be checked to be given by
\begin{equation}
    \mc{C}_{K}(t) =\frac{1}{4\pi^2 D^2}\sum_{n=0}^{D_K-1} n \int \td \bmx ~\td \bm{y}~\mb{O}_n^*(\bmx) \mb{O}^*_n(\bm{y}) \mb{O}(\bmx,t) \mb{O}(\bm{y},t) ~.
\end{equation}

The Krylov operator complexity provides the upper bound on the growth of a class of observables, called $Q$-complexities in \cite{Parker:2018yvk}. Each member of the set of $Q$-complexities is related to a hermitian superoperator $\calch{Q}$ satisfying certain generic properties (see \cite{Parker:2018yvk} for detail of these properties), and is defined by the expectation value of this superoperator in the superstate $\ketc{\mbf{O}(t)}$ generated by the Liouvillian dynamics, i.e., $C_{Q}(t)=\avgc{\mbf{O}(t)|\calch{Q}|\mbf{O}(t)}$.
In terms of the double phase space average, this can be rewritten as 
\begin{equation}
    C_{Q}(t) =  \int \td \bmx ~\td \bm{\xi}~\bbch{W}_{\mbf{O}}(\bmx, \bm{\xi},t) \bbch{Q}(\bmx, \bm{\xi})~.
\end{equation}
Then the so-called $Q$-complexity theorem reduces to the following inequality, 
\begin{equation}
    \int \td \bmx ~\td \bm{\xi}~\bbch{W}_{\mbf{O}}(\bmx, \bm{\xi},t) \bift{\bbch{Q}(\bmx, \bm{\xi})-c \bbch{K}(\bmx, \bm{\xi})} \leq 0~,
\end{equation}
where $c$ is a constant, which quantifies how `complex' the initial operator is and the action of the Liouvillian on the eigenstates of the superoperator $\calch{Q}$ \cite{Parker:2018yvk}.

\section{Dynamics of the Wigner function and wavefunction spreading }\label{sec_WF_dot}
We now take into account the time evolution equation of the Wigner function for the systems we are considering and use it to study the time evolution
equations of the Krylov state complexity.  
First, we use the well-known fact that the evolution equation for the Wigner function can be thought of as a quantum Liouville equation, which reduces to the classical Liouville equation in the limit of vanishing $\hbar$, to find out the classical and quantum contributions of the time evolution of the Wigner function towards the time evolution of the Krylov state complexity itself. In the Appendix. \ref{sec_W_dot_2}, we derive another equation for the time derivative of the Wigner function in terms of the phase space Krylov functions. 

\subsection{Classical and quantum contributions to the time rate of change of  the Krylov state complexity}\label{sub_quan_class}
We derive an expression for the derivative of the Krylov state complexity for the particular forms of the Hamiltonians under consideration, i.e., a single particle moving in one dimension under the effect of a well-behaved potential, from the time evolution equation for the Wigner function. To this end, we note that when the  total 
Hamiltonian can be separated into kinetic and potential terms (with the potential energy function denoted by $V(q)$), it is well-known that the time evolution equation for the Wigner function $ W (q,p,t) $ can also be separated  into two parts - contributions from the 
kinetic and potential terms of the Hamiltonian, respectively. The simplified form for the evolution equation, assuming that the potential function $V(q)$ can be expanded in a Taylor series (as well as setting the boundary terms coming from partial integration over the derivative of the wavefunction to zero), is given by \cite{Hillery:1983ms, Case, wigner1932quantum}
\begin{eqnarray}
\label{WigFdot2}
    \frac{\partial  W (q,p,t) }{\partial t} =  \frac{\partial  W_{T} (q,p,t) }{\partial t} + \frac{\partial  W_V (q,p,t)}{\partial t} \hspace{3.7cm} \\
    =-\frac{p}{m} \frac{\partial  W (q,p,t) }{\partial q}  + \sum_{\lambda} \frac{1}{\lambda !} \bigg(\frac{1}{2i}\bigg)^{\lambda-1}
		\frac{\partial^\lambda  V(q)}{\partial q^\lambda} 	\frac{\partial^\lambda  W (q,p,t) }{\partial p^\lambda}~.
\end{eqnarray}
In the above expression, $m$ is the mass of the particle, $V(q)$ denotes the potential, while the sum in the second term is over odd integers only.\footnote{This expression can be written in a more compact form in terms of the star product as mentioned in the following \cite{moyal1949quantum}, but for our purposes below, this is more suitable. Also, this formula for the time evolution of the Wigner function is an asymptotic series and may not be convergent in general. Of course, for a potential which is a finite order polynomial in $q$, the series gets truncated and is exact.} Importantly, note that (when appropriate powers of $\hbar$ are restored) each of the terms in the summation in the second line of \eqref{WigFdot2} is proportional to powers of $\hbar^{\lambda-1}$, whereas the first term does not have any $\hbar$ dependence, and hence, this term is entirely classical.
Therefore, the time evolution equation for the Wigner function, also called the \textit{quantum Liouville equation}, which
governs quantum dynamics in phase space, consists of the classical Liouville equation (terms which are independent of $\hbar$) and a series of correction terms with increasing powers of $\hbar$. Since these correction terms depend on the derivatives of the potential, it can be seen that all of them vanish when the potential is such that
its third or higher order derivatives vanish, and therefore, in such cases, the dynamical equation of motion of the Wigner function in phase space is the same as classical dynamics under $V(q)$.\footnote{However, one should keep in mind that any distribution $W(q,p)$ in the phase space in not allowed, since the positivity of the density matrix restricts the permissible distributions \cite{Hillery:1983ms}. This restriction on the possible initial condition is the key difference between the classical and quantum dynamics in such systems.} We also note that the expression for the time evolution equation for the 
Wigner function can be written in a more compact form in terms of the star product \cite{moyal1949quantum}, as (with the appropriate factors of $\hbar$ restored)
\begin{equation}\label{Moyal_eq}
    i \hbar \frac{\partial W (\bm{x,t})}{\partial t} = \mb{H}(\bm{x}) \star W (\bm{x,t}) - W (\bm{x,t}) \star \mb{H}(\bm{x}) = [\mb{H}(\bm{x}) , W (\bm{x,t}) ]_\star~.
\end{equation}
However, for our purposes below, the explicit expression in \eqref{WigFdot2} in terms of the derivatives of the potential function is the one that will be primarily used. 

We can now consider the time derivative of the Krylov state complexity from the expression given in eq. \eqref{SCwigner} or \eqref{SCWF2},  and use the above equation to obtain the classical and 
quantum contributions of the Wigner function evolution equation to the time evolution of $\mcC(t)$. The contribution of the classical part of the equation \eqref{WigFdot2} to the time derivative of the Krylov state complexity is given by (an over-dot denotes a derivative with respect to time)
\begin{equation}\label{SCclaiical}
	\dot{\mathcal{C}}_{c} (t)= \int   \mathbb{K}  (q,p)  \Bigg[ - \frac{p}{m}  \frac{\partial  W (q,p,t) }{\partial q} 
	+\frac{\partial  V (q) }{\partial q} \frac{\partial  W (q,p,t) }{\partial p}  \Bigg] ~ \td p \td q ~. 
\end{equation}
Here the subscript $c$ denotes the classical, i.e., $\mathcal{O}(\hbar^0)$ contribution of the quantum Liouville equation. Therefore, for systems such as the free particle, a particle moving in a constant force, or the harmonic oscillator,  the time rate of change of the Krylov state complexity is solely determined by the classical Liouville equation for the Wigner function.

On the other hand, the contribution of the lowest order quantum correction (of order of $\hbar^2$, denoted by the subscript $qu(2)$ in the following) of the Wigner function evolution equation to the time derivative of the Krylov state complexity is given by\footnote{We emphasise here that this may not be the lowest order quantum correction term in an expansion $\mathcal{C}(t)$ in $\hbar$. This term only represents the contribution of the lowest-order quantum correction to the classical Liouville equation to the time derivative of the Krylov state complexity. A similar statement is true for the expression in \eqref{SCclaiical} as well.}
\begin{equation}\label{Scqu2}
		\dot{\mathcal{C}}_{qu(2)} (t)= - \frac{1}{24} \int \mathbb{K} (q,p)~ \frac{\partial^3  V (q) }{\partial q^3} \frac{\partial ^3 W (q,p,t) }{\partial p^3} ~ \td p ~\td q ~.
\end{equation}
Even though, this is beyond the scope of the present work, it will be very much desirable to explicitly evaluate these quantum contributions of the Wigner function evolution to the time evolution 
of Krylov state complexity in a systematic manner for different orders of the Planck constant in some concrete examples, specifically Hamiltonians whose classical counterpart is chaotic\footnote{As we have mentioned previously, even though we have written most of the relations in this paper for an isolated system describing a particle moving in one dimension governed by a time-independent Hamiltonian, where there can not be any chaotic motion, most of the expressions presented in this paper can be extended to higher-dimensional systems.} and study the effect of these quantum corrections to the spreading of an initial wavefunction in the Krylov basis. We hope to report results along these lines in future work.

Let us now compare the expressions we have obtained so far in the Krylov basis and the analogues of the Krylov state complexity with those in other complete orthonormal bases in the subspace. When computed  in terms of another complete orthonormal basis, say, $\{\ket{\mathcal{B}_n}\}$, the expression for the cost (which, by definition, is the Krylov state complexity when expressed in terms of the Krylov basis) can be  similarly written as (compare with the expression in eq. \eqref{SCWF2})
\begin{equation}
	\mathcal{C}_{\mathcal{B}}(t) = 2 \pi \sum_{n} n   \int W^{\mathcal{B}}_{nn}(q,p) W (q,p,t)  ~ \td p ~\td q~. 
\label{cost}
\end{equation}
Here, the functions $W^{\mathcal{B}}_{nn}(q,p)$ are the analogues of the functions $W_{nn}^K(q,p)$ in \eqref{Wndef}, and are nothing but the Wigner function associated with the elements' complete basis, $\mathcal{B}$, i.e., 
\begin{equation}
	\begin{split}
		W^{\mathcal{B}}_{nn} (q,p) = \frac{1}{\pi} \int e^{2i p y}  ~\braket{q - y| B_n}\braket{B_n|q+ y}  ~ \td y~\\
		=  \frac{1}{\pi} \int e^{2i p y} ~ \psi^{\mathcal{B}}_n(q-y) \psi^{\mathcal{B}*}_n(q+y) ~\td y~.
	\end{split}
\end{equation}
It was proven in \cite{Balasubramanian:2022tpr}, by considering a Taylor series expansion of the cost \eqref{cost}, that the cost is minimum up to a finite time when computed in the Krylov basis compared to 
any other orthonormal basis  $\mathcal{B}$.  From the expression for the Krylov state complexity given in \eqref{SCwigner}, this statement can be rewritten in terms of the phase space 
average of various powers of the derivative of the Wigner function and the function $\mathbb{K}(q,p)$ and $\mathbb{B}(q,p)$, where $\mathbb{B}(q,p)$ 
denotes the Weyl transform of the spreading operator corresponding to the orthonormal complete basis $\mathcal{B}$.  Denoting the derivative of the Krylov state complexity around
$t=0$ by  $C_{\mathcal{B}}^{(m)}= C_{\mathcal{B}}^{(m)} (0)=\frac{d^m C_{\mathcal{B}}(t)}{dt^m}\big|_{t=0}$, assuming that the cost function defined above has  
convergent Taylor series expansion over a finite duration of time when computed with respect to orthonormal basis under consideration, and also denoting the sequence 
of the derivative of the cost at $t=0$ as $S_\mathcal{\mcK}$ and $S_\mathcal{B}$, in the Krylov and the basis $\mathcal{B}$, respectively, it was shown in \cite{Balasubramanian:2022tpr} that these sequence satisfy the relation $S_\mathcal{K} \leq S_{\mathcal{B}}$.  
Looking at the expression for Krylov state complexity in \eqref{SCwigner}, it can be seen that when it is written in terms of a general complete orthonormal basis (which can be the Krylov basis as well), we can express these derivatives as 
\begin{equation}\label{derivative0}
C^{(m)}_{\mathcal{B}}= \int  W^{(m)}(q,p) ~ \mathbb{B}  (q,p)  ~ \td p ~\td q~,
\end{equation}
where $ W^{(m)}(q,p)  = \frac{\partial^m W (q,p,t)}{\partial t^m}\big|_{t=0}$ denotes the $m$th derivative of the Wigner function around $t=0$.  
The first-order derivative of the cost can be separated into contributions coming from the classical Liouville equation and quantum corrections of different orders, as in eqs. \eqref{SCclaiical} and \eqref{Scqu2}, with
$\mathbb{K}(q,p)$ replaced by $\mathbb{B}(q,p)$. Therefore, for potentials for which its third and higher order derivatives vanish (such as the harmonic oscillator potential or a linear potential), the statement that the spreading is minimised in the Krylov basis over a finite amount of time, at least in the first order, reduces to a statement about the contribution coming from the classical part of the quantum Liouville equation to the first-order time derivative of the Krylov state complexity. 

We also note that, when the time derivatives of the Wigner function $W(q,p,t)$ are square integrable,\footnote{For a generic quantum state, this is not true in general. However, if $W(q,p,t)$ belongs to the Schwartz space (which, in turn, would be the case if wavefunction $\Psi(q,t) \in \mc{S}(\mb{R})$ itself is a Schwartz function since the Wigner transform preserve Schwartz space), the quantum Liouville equation indicates that $W^{(m)}(\bmx)$ would be square integrable in $L^2(\mb{R}^2)$.} and the dimension of the Krylov space is finite, application of the Cauchy-Schwarz inequality in \eqref{derivative0} indicates that the derivatives of the Krylov state complexity satisfy the following upper bound,
\begin{equation}
    |C^{(m)}| \leq \sqrt{\frac{\pi D_K}{3}(D_K+1)(2D_K+1)}~~ \Bitd{\int\td \bmx~|W^{(m)}(\bmx)|^2}^{1/2}~.
\end{equation}
We have removed the subscript $\mc{B}$ from the derivatives of the complexity since the upper bound above is independent of the orthonormal basis used in the subspace.

\paragraph{\textbf{Early-time growth of the Krylov state complexity.}}
It is well-known that the Krylov state complexity for any generic initial state grows quadratically at early times after the initial time $t=0$ \cite{Balasubramanian:2022tpr, Gautam:2023bcm}. We now demonstrate that the early-time growth of the Krylov state complexity can be derived from the evolution equation for the Wigner function. Using the relation between the derivatives of the 
Wigner function and the Krylov state complexity (eq. \eqref{derivative0})
one can show that  $\mcC^{(1)}=0$, and $\mcC^{(2)}=2b_1^2 t^2$ in the Krylov basis. This confirms the known fact that the Krylov state complexity grows quadratically at early times, with the growth rate determined by the Lanczos coefficient $b_1$. The details of the computations are provided in the Appendix.  \ref{1storder}.

\section{Discussions and conclusions}\label{sec_dis_conc}
The phase space representation of quantum mechanics is an alternative  and very illuminating approach to the more standard 
Hilbert space-based formulation of the same, since, through appropriate mapping between Hilbert space operators and functions defined on the phase space, one can 
make a direct comparison of expectation values of different quantum mechanical observables and their classical counterpart in the scenario where a well-defined classical limit exists. The Wigner quasi-probability distribution function, which is the phase space image of the density matrix (or a general trace-class operator) obtained through the Weyl transform (which is defined for general Hilbert-Schmidt operators), is the canonical quantity in the phase space formulation of quantum mechanics, and plays an analogous role to the classical phase space distribution function, since the expectation value of a quantum mechanical observable in a quantum state can be written as an integral of the Weyl transform of that operator (also known as the symbol or the classical kernel of that operator) smeared with the Wigner function of the quantum state. Therefore, it is natural to expect that the concept of `complexity' of a time-evolved quantum state, viewed as a measurable physical quantity, can also be written as a phase space integral in the same spirit. For the recently introduced measure of Krylov state and operator complexities, we have shown this to be the case in this work. 

In contrast to classical mechanics, where progressively smaller scales in the phase space are explored by the dynamics (which can be exponential in time, when the system is classically chaotic), the finiteness of the Planck's constant imposes a fundamental limit to the resolution up to which the structures of the quantum phase space distribution function (e.g., the Wigner function) can be probed. This fact indicates a fundamental difference in the quantitative nature of the evolution of the concept of complexity in time evolution between classical and quantum dynamics. Nevertheless, focusing on the time evolution of a pure state, from the coefficients of the expansion of the Wigner function of the time-evolved state in a complete orthonormal basis for the phase space functions, one can construct a general class of quantities, which can be thought of as quantifiers of the complexity of the time evolution of the quantum state. As we have argued in this work, the measure of complexity obtained from the second moment of the harmonic distribution of the Wigner function components, which has been studied thoroughly in the literature (and dubbed as the complexity of harmonics in this paper), and the recently introduced notion of the Krylov state complexity can be thought to belong to this same general class of complexity measures.  

We began by constructing the phase space Krylov functions, which are a set of complete orthonormal functions induced by the Weyl transforms of the elements of the Krylov basis states, and provided a convenient basis for expansion of any well-behaved phase space distribution function. These functions were then used to rewrite the familiar expression for the Krylov state complexity as a phase space average, where a Weyl symbol is smeared with the time-dependent Wigner function (in the Schrodinger picture). Casting the Krylov state complexity as a phase space average helps us straightforwardly identify the contributions of the classical part of the quantum Liouville equation to the time rate of change of the complexity, as well as those of higher-order quantum corrections.

Next, we discussed in detail how one can extend the phase space picture to describe a complete basis of operators (supervectors in the Liouville space) - a topic that is relatively less explored compared to its state counterpart. After carefully identifying the set of canonical superoperators and their associated supervector basis following \cite{Royer2}, we defined a version of the Weyl transformation for superoperators and subsequently constructed the double phase space functions onto which the operator Krylov basis maps under this transformation. We established an important and generic relationship between the generalisation of the Wigner function for superoperator projections (the DWF) and the Weyl symbol for the underlying operators (the relations in \eqref{DWF_Weyl}) and the Weyl symbol of the displacement operator, and worked out explicit expressions for the DWT of specific important superoperators, as well as products of them. 

To summarise, in this work we have taken a step towards understanding the phase space structure of the Krylov state and operator basis in the Wigner-Weyl formulation of quantum mechanics, and utilised them to recast the measures of complexity of the time-evolved initial state and operators in terms of the phase space functions, thereby making the connection of these complexities with similar measures constructed from the phase space distribution function more concrete. 

We now briefly discuss some possible future directions along which the results presented in this work can be pursued further.

\textbf{Relation with other phase space distribution functions.} In this paper, in expressing the Krylov state complexity as a phase space average, we have utilised the relation   \eqref{expecttaion} in terms of the Wigner function. However, one can also utilise other phase-space quasiprobability distribution functions, such as the Husimi function \cite{husimi1940some} or the Glauber-Sudarshan function  \cite{sudarshan, Glauber} to write such expectation values. It will be interesting to utilise these functions for the purposes of the present paper, specifically by applying the time-evolution equation to these functions and understanding the structure of the Krylov basis. 

\textbf{Semi-classical approximations and the phase space Krylov functions for classical chaotic systems.}
For quantum systems whose classical counterpart is chaotic, Berry's conjecture \cite{Berry:1977wpp, voros1976semi, voros1977asymptotic} provides a form for the averaged Wigner function associated with an eigenstate, namely, it states that the expression for the Wigner function $W_{aa}(\bmx)$ (in \eqref{star_eigen}) over a narrow energy shell around $E_a$ is approximated by the associated classical energy surface in phase space, i.e.,\footnote{This formula is written for systems with $N$-dimensional configuration space with $N\geq2$, and it is assumed for simplicity that all the coordinates in the corresponding phase space takes values on the real line with out any restriction (otherwise, one needs to specify the ranges of the integrations). Therefore, $\bmx=(q_1, q_2,\cdots, q_N, p_1, p_2,\cdots, p_N)^T$. Analogously, the integral measure $\td \bmx$ represents an integral over $2N$ variables.} 
\begin{equation}\label{Berry_conj}
    \bar{W}_{aa}(\bmx) \simeq \frac{\delta(H(\bmx)-E_a)}{\mc{Z}(E_a)}~,~~\text{where}~~\mc{Z}(E_a) = \int \td\bmx ~\delta(H(\bmx)-E_a)~,
\end{equation}
is the `area' of the classical energy surface for the classical Hamiltonian function $H(\bmx)$. In the context of this work, one can use this conjecture to try to derive a semiclassical expression for the phase space Krylov functions, say using the formula \eqref{WKnm_star_def}. Specifically, combining Berry's conjecture with the expression for the star product of the three functions given in \eqref{star_three}, we have\footnote{Note that the approximation sign is used in the first line of the expression for $\mcW_{nm}^K(\bmx)$ to indicate that averaged Wigner functions for the energy eigenstates are used instead of `bare' ones. Of course, the exact Wigner function of the initial state $W_0(\bmx)$ (which is not an energy eigenstate, otherwise the Krylov basis construction would be trivial) appears without any averages, whose exact expression we assumed to be known. E.g., one can choose an initial state for which the Wigner function is a general Gaussian function of the position and momentum coordinates. On the other hand, the approximation in the second line is due to the application of Berry's conjecture.} 
\begin{align} \label{WKnm_semicla}
     \mcW_{nm}^K(\bmx) & \approx \frac{1}{\pi^2}\sum_{a,b} \mc{P}_{n}(E_{a})\mc{P}_{m}(E_{b})  \int \td \bmx_1 \td\bmx_2 ~ \bar{W}_{aa}(\bmx_1) W_{00}(\bmx_2) \bar{W}_{bb}(\bmx+\bmx_2-\bmx_1) \exp{\big[ 4 i \Delta(\bm{x}, \bm{x}_1,\bm{x_2})\big]}\\
     &\approx \frac{1}{\pi^2}\sum_{a,b} \frac{\mc{P}_{n}(E_{a})\mc{P}_{m}(E_b) }{\mc{Z}(E_a) \mc{Z}(E_b) }\int \td \bmx_1 \td\bmx_2 W_{00}(\bmx_2) \delta\bift{H(\bmx_1)-E_a}\delta\bift{H(\bmx+\bmx_2-\bmx_1)-E_b}\exp{\big[ 4 i \Delta(\bm{x}, \bm{x}_1,\bm{x_2})\big]}~ .
\end{align}
From this expression, it can be seen that the integral has a meaningful contribution only when the following conditions are satisfied simultaneously: $H(\bmx_1)=E_a$ and $H(\bmx+\bmx_2-\bmx_1)=E_b$. One can also directly use numerical approximations to evaluate the second line of \eqref{WKnm_semicla}, and use these to obtain an approximate expression for the Krylov state complexity in \eqref{SCWF2}. However, a detailed analysis of this problem is beyond the scope of the present paper and will be addressed in future work.

\textbf{Quantum systems with discrete phase space.}
In this work, we have only considered quantum systems that have a continuous (and semi-classically well-defined) phase space. However, these represent only a subclass of quantum systems, since, in general, they can have a discrete phase space. For quantum systems with discrete phase space, one can formulate a description based on a discrete version of the Wigner function \cite{wootters1987wigner}. 
We hope to return to the problem of understanding the associated phase space Krylov functions and description of the time evolution of complexity in future works \cite{Inprep}. 

\textbf{Growth of superoperators.} In Section \ref{sec_op_phase_sp}, we have discussed the phase space representation of operators, which is then used to describe the growth of an initial operator under Heisenberg time evolution. Continuing in a similar fashion, one can, in principle, construct the phase space representation of the supersuperoperators and use it to describe the evolution of an initial superoperator. 

\begin{center}
	\bf{Acknowledgments}
\end{center}
The work of Kunal Pal is  supported by the YST Program at the APCTP through the Science and Technology Promotion Fund and Lottery Fund of the Korean Government. This was also supported by the Korean Local Governments -
Gyeongsangbuk-do Province and Pohang City. Kunal Pal would like to thank Gwangju Institute of Science and Technology for the hospitality where a part of this work was carried out. This work was supported by the Basic Science Research Program through the National Research Foundation of Korea (NRF) funded by the Ministry of Science, ICT \& Future Planning (NRF-2021R1A2C1006791), the Korea government(MSIT)(RS-2025-02311201), and the framework of international cooperation program managed by the NRF of Korea (RS-2025-02307394), the Creation of the Quantum Information Science R\& D Ecosystem (Grant No. RS-2023-NR068116) through the National Research Foundation of Korea (NRF) funded by the Korean government (Ministry of Science and ICT). This research was also supported by the Regional Innovation System \& Education (RISE) program through the (Gwangju RISE Center), funded by the Ministry of Education (MOE) and the (Gwangju Metropolitan City), Republic of Korea (2025-RISE-05-001).

\appendix

\section{Phase space Krylov functions and orthogonal decomposition of the initial state}\label{sec_orthogonal_Krylov}
In this appendix, we discuss the Krylov phase space functions from the perspective of the decomposition of the action of a Hermitian operator on a given quantum state. This decomposition was used in Section \ref{sec_Krylov_decomposition} to write a decomposition of the transition probability $p_n(t)$.

We start from the following well-known expression, which states that the action of any Hermitian operator ($\mbf{A}$) on any quantum state (say, $\ket{\psi_0})$ can be decomposed in two parts, one of which is proportional to the state itself, and the other one is a term which is a contribution along a state orthogonal to that quantum state, i.e., 
\begin{equation}\label{operator_decom}
    \mbf{A}\ket{\psi_0} = \braket{\mbf{A}} \ket{\psi_0} + \Delta \mbf{A} \ket{\psi_{0\bot}}~,
\end{equation}
where $\braket{\psi_0|\psi_{0\bot}}=0$. As this decomposition indicates, the coefficients along the $\ket{\psi_0}$
and $\ket{\psi_{0\bot}}$ are nothing but the average and the standard deviation of the operator in the initial quantum state. Now consider applying this expansion to the case where the initial quantum state under consideration is an element of a set of complete and orthonormal basis, $\ket{B_{n}}$. We also fix the operator to be the Hamiltonian of the quantum system for future reference. As is clear, in general, the state orthogonal to $\ket{B_{n}}$ can be a linear combination of all other elements $\ket{B_j}, j \neq n$, i.e., $\ket{B_{n_\bot}}= \sum_{j(\neq n)} c_{j} \ket{B_j}$. 

From this perspective, the case when the complete orthonormal basis is the Krylov basis generated by the Hamiltonian starting from an initial state, the expression for the orthogonal state which is `spanned' after the action of the Hamiltonian is the `simplest' in some sense, since, for the $n$th element of the basis, $\ket{K_{n_\bot}}$ is a sum of the $(n-1)$th and $(n+1)$th state of the Krylov basis.\footnote{Note that, in making this statement, we assumed that the case of energy eigenstates is not applicable here, since its time evolution would be trivial.} Comparing with \eqref{H_Krylov}, we see that \footnote{We emphasise that, in general the state orthogonal to $\ket{K_n}$ has a form $\ket{K_{n}}_\bot= \sum_{j(\neq n)} c_{j} \ket{K_j}$. Only for the action of the Hamiltonian on $\ket{K_n}$, it is a combination of $(n+1)$th and $(n-1)$th elements. This is not true for the action of a generic Hermitian operator on the Krylov basis elements. }
\begin{equation}
    \ket{K_{n_\bot}} = (\Delta \mbf{H}_n)^{-1}\bitd{b_n \ket{K_{n-1}}+b_{n+1}\ket{K_{n+1}}}~,
\end{equation}
where $(\Delta \mbf{H}_n)^2$ is the variance of the Hamiltonian with respect to the $n$-th element of the Krylov basis, and can be checked to be given in terms of the Lanczos coefficients to be $(\Delta \mbf{H}_n)^2=b_n^2+b_{n+1}^2$, i.e., independent of the Lanczos coefficients $a_n$. Note that the preceding discussion is valid for any closed quantum system whose time evolution is generated by a Hermitian Hamiltonian.

In reference \cite{vaidman1992minimum}, the relation \eqref{operator_decom} was used to provide a very simple derivation of the minimum time needed for an initial state to evolve into its orthogonal state under the evolution generated by a time-independent Hamiltonian. Considering the state $\ket{K_{n}}$, the minimum time required to evolve to a state orthogonal to it is given in terms of the Lanczos coefficients as (restoring appropriate factor of $\hbar$) $\frac{\pi\hbar }{2 \Delta \mbf{H}_n}=\frac{ \pi\hbar }{\sqrt{b_n^2+b_{n+1}^2}}$.\footnote{We emphasise that this case is, in general, distinct from description of the time evolution starting from an initial state $\ket{\psi_0}$ in the Krylov basis. In that case, the time evolution of the $n$-th element of the Krylov basis is not very meaningful.}
For chaotic quantum mechanical Hamiltonians, starting from a random initial state, the $b_n$ Lanczos sequence typically has a pick for some value of $n$ (say $n_M$), usually for a value which is close to the start of the Krylov chain, rather than the end (see e.g., \cite{Balasubramanian:2022tpr, Gautam:2023bcm}). From the expression for the minimum time, it can then be seen that, among all the elements of the Krylov basis, the minimum time required to evolve to the orthogonal state has the smallest value for either of the three states,  $\ket{K_{n_M-1}}, \ket{K_{n_M}}$ or $\ket{K_{n_M+1}}$, depending on the numerical values of the Lanczos coefficients $b_{n_M-1}$ and $ b_{n_M+1}$.

We now work out how these relations can be expressed in terms of the phase space functions. Consider the operator (this is not Hermitian, in general)  $\mbf{P}^\bot=\ket{\psi_0}\bra{\psi_{0\bot}}$, and denote the Weyl transform of this operator (with an overall normalisation factor of $1/(2 \pi)$) as $\mb{P}^\bot(\bmx)$. Then the phase space description of the relation \eqref{operator_decom} is given by
\begin{equation}
    \mb{A}(\bmx) \star \mb{P}^\bot(\bmx) = \braket{\mbf{A}}  \mb{P}^\bot(\bmx) + \Delta \mbf{A} W_{0\bot}(\bmx)~,
\end{equation}
where $W_{0\bot}(\bmx)$ denotes the Wigner function of the state $\ket{\psi_{0\bot}}$. Alternatively, one also has,
\begin{equation}
    \mb{A}(\bmx) \star W_0(\bmx)= \braket{\mbf{A}}  W_0(\bmx) + \Delta \mbf{A} \mb{P}_{\bot} (\bmx)~,
\end{equation}
where $\mb{P}_{\bot} (\bmx)~$ denotes the Weyl transform 
of the operator $\mbf{P}_\bot=\ket{\psi_{0\bot}}\bra{\psi_0}$, and hence, $\mb{P}_{\bot} (\bmx)^*=\mb{P}^{\bot} (\bmx)$. 

The function $\mb{P}^\bot(\bmx) $, which has the following explicit expression in terms of the position space wavefunction,
\begin{equation}
    \mb{P}^\bot(\bmx)  = \frac{1}{2\pi} \int \td y~\psi_0 (q-y) \psi_{0\bot} (q+y)~ e^{2 i p y}~,
\end{equation}
satisfies the following relations 
\begin{equation}
    \mb{P}^\bot(\bmx)  \star W_0(\bmx)  = 0,~~\mb{P}^\bot(\bmx)  \star \mb{P}^\bot(\bmx)= 0~,~~\mb{P}^\bot(\bmx) \star \mb{P}^\bot(\bmx) = (2 \pi)^{-1}  W_0(\bmx)~.
\end{equation}
The last relation, along with the normalisation condition of the Wigner function, indicates that the phase space integral of the square of $\mb{P}^\bot(\bmx)$ is equal to $(2 \pi)^{-1}$.
Furthermore, it can be seen that, in terms of the phase space eigenfunctions $\mcW_{nm}(\bmx)$, the function $\mb{P}^\bot(\bmx)  $ has the  expansion 
\begin{equation}
    \mb{P}^\bot(\bmx)   = \sum_{nm} \mcW_{nm}(\bmx) \braket{E_{n}|\mbf{P}^\bot|E_{m}}~.
\end{equation}

\section{Canonical transformations and the Wigner-Weyl basis of superoperators}\label{sec_CT_and_bases}
In this Appendix, we present some additional arguments on the choice of the superoperators in the Wigner-Weyl canonical basis, and discuss a few consequences of canonical transformations on these Liouville space bases that we determined in Section \ref{sec_op_phase_sp}.

First, we derive the transformation relation in \eqref{Xsu_tra_CT} for the canonical superoperators under a linear canonical transformation from the transformation rule for the Liouville space basis states. Recall that, under a linear canonical transformation $\bmx^\prime=V \bmx$, the eigenbasis of the canonical position and momentum operators transform as \cite{dirac1981principles}, $\ket{\bmx^\prime}=\mbf{V}\ket{\bmx}=\ket{V\bmx}$, where, as in Section \ref{sec_op_phase_sp}, $\mbf{V}$ denotes the unitary transformation corresponding to the linear canonical transformation matrix $V$. This relation indicates that, the action of the superoperator $\calch{V}$ on the Liouville space states $\ketc{\mbf{C}_{\bmx}; \varphi}$ should be as follows 
\begin{equation}
    \ketc{\mbf{C}_{\bmx^\prime}, \varphi}=\calch{V} \ketc{\mbf{C}_{\bmx}, \varphi}=\ketc{\mbf{C}_{V\bmx}; \varphi}~,~~\text{with}~~\calch{X}_\varphi \ketc{\mbf{C}_{\bmx}, \varphi}= \bmx \ketc{\mbf{C}_{\bmx}, \varphi}~,~~\text{and}~~\calch{X}^\prime_\varphi \ketc{\mbf{C}_{\bmx^\prime}; \varphi}= \bmx \ketc{\mbf{C}_{\bmx^\prime}; \varphi}~.
\end{equation}
Now starting from the last relation above, i.e., eigenvalue equation for the transformed canonical superoperators, and changing the label $\bmx$ to $V^{-1} \bmx$, we have $\calch{X}^\prime_\varphi \ketc{\mbf{C}_{\bmx}; \varphi}= V^{-1} \calch{X}_\varphi  \ketc{\mbf{C}_{\bmx}; \varphi}$. This gives the desired relation \eqref{Xsu_tra_CT}, $\calch{X}^\prime_\varphi=V^{-1} \calch{X}_\varphi$. 

Next, we derive the transformation rules for the parity and translation operators under a linear canonical transformation. We expect that, like a state $\ket{\bmx}$, the effect of the canonical transformation on the operators would be to change their phase space coordinate labels as $\bmx \rightarrow V \bmx$.  To check this, consider the unitary transformed version of the parity operator in \eqref{parity_op}, which we can write as
\begin{equation}\label{Parity_cano_tr}
   \mbf{V} \mbf{\Pi}_{\bmx} \mbf{V}^\dagger= \frac{1}{4 \pi} \int \td \bm{y} ~e^{i \braket{\bmx, \bm{y}}_s}~ \mbf{V} \mbf{D}_{\bm{y}} \mbf{V}^\dagger = \frac{1}{4 \pi} \int \td \bm{y} ~e^{i \braket{\bmx, \bm{y}}_s}~\exp\bitd{i \braket{V^{-1}\mbf{X},\bm{y}}_s}~,
\end{equation}
where we have denoted $\bm{y}=(y,k)^T$ and used the relation, valid for linear canonical transformations, $\mbf{X}\mbf{V}\mbf{X}^\dagger=V^{-1}\mbf{X}$.  Now performing a change of coordinates, $\bm{y} \rightarrow V^{-1}\bm{y}$, subsequently using the facts that the Jacobian determinant associated with the transformation is unity, and the relation $VJV^T=J$ implies, $(V^{-1})^TJV^{-1}=-(V JV^T)^{-1}=J$, we have,
\begin{eqnarray}\label{Parity_cano_tr2}
    \mbf{V} \mbf{\Pi}_{\bmx} \mbf{V}^\dagger= \int \td \bm{y}~ e^{i \braket{\bmx, V^{-1}\bm{y}}_s}~\exp\bitd{i \braket{\mbf{X},\bm{y}}_s} = \int \td \bm{y}~ e^{i \braket{V\bmx, \bm{y}}_s}~\exp\bitd{i \braket{\mbf{X},\bm{y}}_s} = \mbf{\Pi}_{V \bmx}~,
\end{eqnarray}
as expected. Using the above transformation relation for the parity operator and the fact that parity and translation operators are related by a symplectic Fourier transform, one can derive the following transformation rule of the translation operator under linear canonical transformations, 
\begin{equation}
    \mbf{V} \mbf{T}_{\bm{\xi}} \mbf{V}^\dagger = \suop{T}{V\xi}~.
\end{equation}

All the continuous orthonormal bases of the Liouville space can be generated by the action of a unitary superoperator on other Liouville space basis vectors. E.g., the action of a unitary superoperator $\calch{U}$ on the bases $\ketc{\ket{q}\bra{q^\prime}}$ produces the new continuous bases 
\begin{equation}
    \ketc{\ket{q}\bra{q^\prime}, \theta} = \calch{U}\ketc{\ket{q}\bra{q^\prime}} ~,~~\text{with}~~\calch{U}=\int ~\td q~ \td q^\prime ~\ketc{\ket{q}\bra{q^\prime}, \theta} \brac{\ket{q}\bra{q^\prime}}~.
\end{equation}

\section{Representation of the parity and translation superoperators and the double Weyl transformations}\label{Why_xpm}
In this appendix, we first derive expressions for the parity and the translation superoperators in terms of the parity and the translation operators and provide an explanation why the DWT of a superoperator depends on the combination $\bmx \pm \frac{1}{2} \bm{\xi}$. A somewhat different set of arguments can be found in \cite{Saraceno:2015cek}, where the authors arrived at a similar conclusion.

We begin by addressing the question of how the parity and translation operators are related to the corresponding superoperators. 

\textbf{Proposition.} \textit{We propose that the sought-after relations are given by: $\check{\Pi}_{{\bmx}, \bm{\xi}} = \suop{\Pi}{\bm{a}(\bmx, \bmxi )} \bullet \suop{\Pi}{\bm{b}(\bmx, \bmxi )}$ and $\calch{T}_{{\bmx}, \bm{\xi}} = \suop{T}{\bm{c}(\bmx, \bmxi )} \bullet \suop{T}{\bm{d}(\bmx, \bmxi )}^\dagger$, where $\bm{a}$-$\bm{d}$ are functions of the coordinates $\bmx$ and $\bmxi$, which are determine by demanding that the parity and translation superoperators satisfy the relations \eqref{parity_suop_action} and \eqref{tran_suop_action}, when acted upon the canonical supervectors bases (i.e., equivalently on the parity and translation operators). The required expressions for the four unknown functions are $\bm{c}=\bm{a}=\bmx_{+}$ and $\bm{d}=\bm{b}=\bmx_{-}$.}

The choice of the above forms of $\check{\Pi}_{{\bmx}, \bm{\xi}}$ and $\calch{T}_{{\bmx}, \bm{\xi}}$, even though, are well-motivated, can nevertheless seem a bit ad hoc. To this end, we note that $\check{\Pi}_{{\bmx}, \bm{\xi}}$, defined as a superoperator of the form $\mbf{A} \bullet \mbf{B}$ can not have the translation operator in its expression (i.e., neither of $\mbf{A} \bullet \mbf{B}$ can be $\mbf{\Pi}$), and similarly, the expression for $\calch{T}_{{\bmx}, \bm{\xi}}$ can not contain the parity operator. This is because, otherwise, as can be easily checked using the composition rules in \eqref{parity_action}, it is not possible to satisfy the required relations in \eqref{parity_suop_action} and \eqref{tran_suop_action}.  Another argument why the parity and the transactional superoperators should be of the form $\mbf{A} \bullet \mbf{B}$, when written in terms of corresponding operators, is given below. For an alternative approach to reach the same conclusion, we refer to \cite{Saraceno:2015cek}.  

\textbf{Proof.} To determine the unknown functions $\bm{a}(\bmx, \bmxi )$ and $\bm{b}(\bmx, \bmxi)$, we consider the action of $\check{\Pi}_{{\bmx}, \bm{\xi}} = \suop{\Pi}{\bm{a}(\bmx, \bmxi )} \bullet \suop{\Pi}{\bm{b}(\bmx, \bmxi )}$ on the bases $\ketc{\suop{C}{\bmx}}$ and $\ketc{\suopf{C}{\xi}}$, and demanding that they match with \eqref{parity_suop_action}, finally obtain the following relations: $\bm{a}(\bmx, \bmxi )+\bm{b}(\bmx, \bmxi )= 2 \bmx$, and $\bm{a}(\bmx, \bmxi )-\bm{b}(\bmx, \bmxi )= \bmxi$. Thus, $\bm{a}=\bmx_{+}=\bmx +\frac{1}{2} \bm{\xi}$ and $\bm{b}=\bmx_{-}=\bmx -\frac{1}{2} \bm{\xi}$. By following a similar procedure, we determine $\bm{c}=\bm{a}=\bmx_{+}$ and $\bm{d}=\bm{b}=\bmx_{-}$. To summarise, the expressions for the parity and the translation superoperators in terms of the respective operators are given by
\begin{equation}\label{pari_tran_suop}
    \check{\Pi}_{{\bmx}, \bm{\xi}} = \suop{\Pi}{\bm{x}_{+}} \bullet \suop{\Pi}{\bm{x}_{-}}~,~~\text{and}~~\check{T}_{{\bmx}, \bm{\xi}} = \suop{T}{\bm{x}_{+}} \bullet \suop{T}{\bm{x}_{-}}^\dagger~,~~\text{with}~~\bmx_{\pm}=\bmx \pm \frac{1}{2}\bmxi~.
\end{equation}
This completes the proof of the above proposition.

To see that the expressions for the parity and translation superoperators can be written as in eq. \eqref{pari_tran_suop} from a different perspective, consider a generic superoperator $\calch{S}_{\bm{a}, \bm{b}}$, depending on two labels $\bm{a}, \bm{b}$, which, when expanded in the basis $\ketc{\suop{C}{\bmx} }\brac{C_{\bm{y}}}$ takes the form,
\begin{equation}
   \calch{S}_{\bm{a}, \bm{b}} = \mc{N}_0 \int \td \bmx 
   ~\td \bm{y}~ \Upsilon (\bm{a}, \bm{b};\bmx,\bm{y}) \ketc{\suop{C}{\bmx} }\brac{C_{\bm{y}}}~,~~ \Upsilon (\bm{a}, \bm{b};\bmx,\bm{y})= \avgc{\suop{C}{\bmx}|\calch{S}_{\bm{a}, \bm{b}}|\suop{C}{\bm{y}}}~,
\end{equation}
where $\mc{N}_0 $ is a constant to be determined. 
Assuming that the set of superoperators, $ \calch{S}_{\bm{a}, \bm{b}}$ with different labels actually forms an orthonormal basis in the space of superoperators, just like $\ketc{\suop{C}{\bmx} }\brac{C_{\bm{y}}}$, the inverse transformation of the previous relation can be written as 
\begin{equation}
   \ketc{\suop{C}{\bmx} }\brac{C_{\bm{y}}} = \mc{N}_0  \int \td \bm{a} 
   ~\td \bm{b}~ \Upsilon^{*} (\bm{a}, \bm{b};\bmx,\bm{y}) ~\calch{S}_{\bm{a}, \bm{b}}~,~~ \Upsilon (\bm{a}, \bm{b};\bmx,\bm{y})= \avgc{\suop{C}{\bmx}|\calch{S}_{\bm{a}, \bm{b}}|\suop{C}{\bm{y}}}~.
\end{equation}
Now returning to the case of the parity superoperator, we notice that, for a given choice of the superoperator basis $\calch{S}_{\bm{a}, \bm{b}}$, we need to determine $\bm{a}, \bm{b}$, such that (Jacobian determinant being unity)
\begin{equation}
   \calch{S}_{\bm{a}, \bm{b}} = \mc{N}_0  \int \td \bmx 
   ~\td \bm{y}~ \Upsilon (\bm{a}, \bm{b};\bmx,\bm{y}) \ketc{\suop{C}{\bmx+\frac{1}{2}\bm{y}} }\brac{C_{\bmx-\frac{1}{2}\bm{y}}}~\equiv \check{\Pi}_{{\bmx}, \bm{\xi}} ~,~~ \Upsilon (\bm{a}, \bm{b};\bmx,\bm{y})= \avgc{\suop{C}{\bmx+\frac{1}{2}\bm{y}}|\calch{S}_{\bm{a}, \bm{b}}|\suop{C}{\bmx-\frac{1}{2}\bm{y}}}~.
\end{equation}
Clearly, if we do not have any other information on the superoperator basis $ \calch{S}_{\bm{a}, \bm{b}} $, then it is not possible to proceed much further. However, for our purposes here, we notice from the structure of the parity superoperator in \eqref{parity_supop} and the well-known relation 
\begin{equation}
    \text{Tr}\bitd{\suop{\Pi}{\bmx_1}\suop{\Pi}{\bmx_2}\suop{\Pi}{\bmx_3}\suop{\Pi}{\bmx_4}} = \frac{\pi}{8} \delta\bift{\frac{1}{2}\bitd{(\bmx_1+\bmx_3)-(\bmx_2+\bmx_4)}} \exp 
    \bift{2i\braket{\bmx_1, \bmx_2}_s +2i\braket{\bmx_3, \bmx_4}_s }~,
\end{equation}
the parity superoperator should be of the form $\check{\Pi}_{{\bmx}, \bm{\xi}} = \suop{\Pi}{\bm{a}(\bmx, \bmxi )} \bullet \suop{\Pi}{\bm{b}(\bmx, \bmxi )}$, where the variables $\bm{a}(\bmx, \bmxi )$ and $\bm{b}(\bmx, \bmxi )$, satisfy the relation $\bm{a}(\bmx, \bmxi )+\bm{b}(\bmx, \bmxi )= 2 \bmx$, and $\bm{a}(\bmx, \bmxi )-\bm{b}(\bmx, \bmxi ) =\bmxi+\text{constant}$. Choosing the constant to be zero, we get back the relation in \eqref{pari_tran_suop}. The constant $\mc{N}_0$ can also be determined to be equal to $D^{-1}$. 
A similar set of arguments can be made to determine that the expression for the translation superoperator is given by the second relation in \eqref{pari_tran_suop}. We also note that the above construction implies that $\calch{S}_{\bm{a}, \bm{b}}=\suop{\Pi}{\bm{a}} \bullet \suop{\Pi}{\bm{b}}$ and $\calch{\tilde{S}}_{\bm{a}, \bm{b}}=\suop{T}{\bm{a}} \bullet \suop{T}{\bm{b}}$ are two sets of orthonormal superoperator basis, for generic values of $\bm{a}$ and $\bm{b}$; for specific values of these variables the elements of these bases coincide with the parity and translation superoperators. 

Having determined the expressions for the parity and the translation superoperators in terms of the corresponding operators,  it is now straightforward to see why the DWT (as well as the double chord function) are functions of the combinations $\bm{a}(\bmx, \bmxi )$ and $\bm{b}(\bmx, \bmxi )$. Consider expanding a generic superoperator $\calch{S}$ in the orthonormal superoperator basis $\calch{S}_{\bm{a}, \bm{b}}$ (we shall omit some proportionality constants in the following)
\begin{equation}
    \calch{S} \propto \int \td \bm{a} \td \bm{b} ~\Upsilon (\bm{a}, \bm{b}) ~\calch{S}_{\bm{a}, \bm{b}}~.
\end{equation}
Now taking $\bm{a}=\bmx_{+}=\bmx +\frac{1}{2} \bm{\xi}$ and $\bm{b}=\bmx_{-}=\bmx -\frac{1}{2} \bm{\xi}$, and subsequently making a change of variables $\bmx_{\pm} \rightarrow (\bmx, \bmxi)$ (which has unit Jacobian determinant) we see that the double phase space coordinate function $\Upsilon (\bmx_{+}, \bmx_{-})$ is nothing but the DWT of the superoperator, and is a function of the variables $\bmx \pm \frac{1}{2} \bmxi$, i.e., 
\begin{equation}
    \Upsilon (\bmx_{+}, \bmx_{-}) = \text{Tr} \bitd{\calch{S}\check{\Pi}_{\bmx, \bmxi}} = \bbch{S}(\bmx_{+}, \bmx_{-})~.
\end{equation}

\paragraph{\textbf{Linear canonical transformations and the double Wigner function.}} In Section \ref{sec_DWT} we derived two alternative expressions for the DWF of the superoperator $\calch{B}_{\mbf{AB}}=\ketc{\mbf{A}}\brac{\mbf{B}}$ (see eqs. \eqref{DWF_AB1} and \eqref{DWF_AB2}). To see the connection between these two alternative interpretations, let us now consider a special case, namely, a classical linear canonical transformation $\bmx_{-} \rightarrow \bmx_{+}$, which we write as $\bmx_{-}=V \bmx_{+}$ \cite{de1998weyl}. Denoting the unitary operator associated with this linear canonical transformation as $\mbf{V}$, we have $\mbf{V} \mbf{\Pi}_{\bmx_{-}} \mbf{V}^\dagger=\mbf{\Pi}_{\bmx_{+}}$ (see the relation \eqref{Parity_cano_tr2}). Now taking $\mbf{A}=\mbf{B}=\mbf{V}$, we see that 
$\bbch{W}_\mbf{VV}(\bmx, \bm{\xi})$ is (apart from an overall constant) just the Weyl symbol of the parity operator $\mbf{\Pi}_{\bmx_{+}}$ (with respect to the $\bmx_{+}$ coordinates) or alternatively, the Weyl symbol of $\mbf{\Pi}_{\bmx_{-}}$ (with respect to the $\bmx_{-}$ coordinates). Hence, $\bbch{W}_\mbf{VV}(\bmx, \bm{\xi})$ is just a constant $2/\pi^2$. In other words, the DWF of the superoperator, $\calch{B}_{\mbf{VV}}=\ketc{\mbf{V}}\brac{\mbf{V}}$, where $\mbf{V}$ is a unitary operator corresponding to the symplectic matrix $V$ representation the classical canonical transformation  $\bmx_{-}=V \bmx_{+}$, is just a constant.

\section{Fidelity and double phase space representation}\label{sec_fidelity}
In this Appendix, we discuss the double phase space representation of a fidelity that is 
related to the complexity of harmonics, and show how one can express this fidelity in terms of the Wigner function of the time-evolved initial state, thereby making its connection with the measures of complexity explicit.

Consider the time evolution of an initial density matrix, which in terms of the Liouvillian superoperators is written as 
\begin{equation}
    \rho(t) = e^{-i \mbf{H} t} \rho(0) e^{i \mbf{H} t} ~~\rightarrow ~~ \ketc{ \rho(t) }= e^{-i \calch{L}t} \ketc{\rho(0)}~.
\end{equation}
After a time $T$, from start of the evolution, a unitary perturbation is applied to the system, which we assume to be of the form $P(\theta)= \exp{\bift{- i \theta \mbf{M}} }$, where $\mbf{M}$ is a Hermitian operator, and $\theta$ is a real parameter. 
After the action of the perturbation operator, the perturbed state ($\tilde{\rho}(T, \theta)$) is evolved backward, without changing the Hamiltonian, for an amount of time $T$, so that  the final state of the system is given by
\begin{equation}
    \tilde{\rho}(0;T, \theta) = e^{i \mbf{H} T} \tilde{\rho}(T, \theta) e^{-i \mbf{H} T} ~ \rightarrow~~ \ketc{\tilde{\rho}(0;T, \theta)} = e^{i \calch{L}t} e^{i \calch{M}^{-}\theta} e^{-i \calch{L}t} \ketc{\rho(0)}~.
\end{equation}
Here we have defined the superoperator $\calch{M}^{-}=[\mbf{M,\bullet}]$. The fidelity between two states, namely, the initial time at $t=0$, and the reversed state at this time, can be written in terms of these superoperators as 
\begin{equation}
    F_\rho(\theta;t=T) = N^{-1} \tr \bitd{\tilde{\rho}(0;T, \theta) \rho(0)} =DN^{-1} \avgc{ \rho(0)|\tilde{\rho}(0;T, \theta)}~,
\end{equation}
where $N=\tr [\rho(0)^2]$. For a specific choice of the perturbation operator $\mbf{M}$, the fidelity has been shown to be related to the measure of number of harmonics of the Wigner function, in particular, fidelity was shown to be a decreasing function of the complexity of harmonics of the Wigner function associated with the state $\rho(T)$ \cite{benatti2009dynamics, Sokolev}.

The above expression for the fidelity can be rewritten  as follows, 
\begin{equation}
\begin{aligned}
    F_\rho(\theta;T) = DN^{-1} \avgc{\rho(0)|e^{i \calch{L}T} e^{i \calch{M}^{-}\theta} e^{-i \calch{L}T}|\rho(0)} &= DN^{-1} \avgc{\rho(T)|e^{i \calch{M}^{-}\theta}|\rho(T)}\\
    &=DN^{-1}\sum_{n} \frac{(i \theta)^n}{n!}\avgc{\rho(T)|(\calch{M}^{-})^n|\rho(T)}~.
    \end{aligned}
\end{equation}
The last expression shows that the fidelity can be thought of as the generating function 
for the set of quantities, $F_n(T)=\avgc{\rho(T)|(\calch{M}^{-})^n|\rho(T)}$. For even values of $n$, superoperators $(\calch{M}^{-})^n$ are positive-semidefinite, so that $F_n(T)$ (with $n=2,4, \cdots$) are positive for any value of $T$. Nevertheless, we can write these quantities in terms of the double phase space functions for any value of $n$ as 
\begin{equation}
    F_n(T) =  \int \td \bmx ~\td \bm{\xi}~\bbch{M}_n^{-}(\bmx, \bm{\xi})~\bbch{W}_{\rho(T)}(\bmx, \bm{\xi})~,
\end{equation}
where $\bbch{M}_n^{-}(\bmx, \bm{\xi})$ denotes the DWT of the superoperator $(\calch{M}^{-})^n$, and hence given by 
\begin{equation}
    \bbch{M}_n^{-}(\bmx, \bm{\xi}) = \underbrace{\bbch{M}^{-}(\bmx, \bm{\xi}) * \bbch{M}^{-}(\bmx, \bm{\xi}) * \bbch{M}^{-}(\bmx, \bm{\xi})*\cdots}_{\text{$n$ times}}~,
\end{equation}
i.e., $n$ times the $*$-products of DWT of $\calch{M}^{-}$, and $\bbch{W}_{\rho(T)}(\bmx, \bm{\xi})$ denotes the DWF of the superoperator, $\ketc{\rho(T)}\brac{\rho(T)}$, and hence can be rewritten in terms of the Wigner function of the time evolved state as (see eq. \eqref{DWF_OO})
\begin{equation}
    \bbch{W}_{\rho(T)}(\bmx, \bm{\xi}) = \frac{1}{2\pi D} \int ~\td \bm{y} ~W_{\rho} \bift{\bmx -\frac{\bm{y}}{2}, T} W_{\rho} \bift{\bmx+\frac{\bm{y}}{2}, T} ~\exp{\bitd{i\braket{\bm{y},\bm{\xi}}_s}}~.
\end{equation}

It is also to be noted that, if the superoperators $(\calch{M}^{-})^n$ are such that their eigenstates satisfy certain conditions (see \cite{Parker:2018yvk}) we can regard $F_n(T)$ (with $n=2,4, \cdots$) as a class of $Q$-complexities evaluated at time $T$.

\section{Some integrals involving phase space Krylov functions $\mathcal{W}^K_{ij} (q,p)$}\label{Integrals}

Here we provide expressions of some integrals involving the functions $\mathcal{W}^K_{ij}(q,p)$ introduced in the main text, which are very useful in computing the early time behaviour of the Wigner function and the Krylov state complexity. 

First, consider the integral over phase space of the symmetric and antisymmetric combination of $\mathcal{W}^K_{ij}(q,p)$ (with $i\neq j $) 
weighted with the function $\mathbb{K}(q,p)$, i.e.,
\begin{equation}
	\begin{split}
	I^{\pm}_{ij} = \int \Big(\mathcal{W}^K_{ij} (q,p) \pm \mathcal{W}^K_{ji}(q,p)  \Big)~  \mathbb{K}(q,p)  ~ \td p ~\td q~
	 = \frac{1}{\pi} \int  e^{2ipy}  ~  \mathbb{K}(q,p) ~ \big(\psi_i^K \chi_j^{K*} \pm \psi_j^K \chi_i^{K*}\big)~\td y  ~ \td p ~\td q~.
\end{split}
\end{equation}
Using the definition of the functions $\mathbb{K}(q,p)$, and introducing new variables, $u=q-y$ and $v=q+y$,\footnote{Recall that, according to our convention, $\psi^K_n$ are always 
functions of the only $u$, while, $\chi_n^K$ are solely functions of $v$.} we can rewrite this as 
\begin{equation}\label{Iij}
	I^{\pm}_{ij}  =  \sum_{n} n ~ \int  \chi_n  (v)~\psi^{K*}_n(u) ~ \Big(\psi_i^K  (u)\chi_j^{K*} (v) \pm \psi_j^K(u) \chi_i^{K*}(v)\Big)  ~ \td u ~\td v~=0~.
\end{equation}
To show these integrals vanish, we have used the orthogonality property of the functions $\psi^K_n(u)$ (and $\chi_n(v)$). Note that each term of the above integrals, i.e., integrals of the form, vanishes ($i \neq j$)
\begin{equation}
		I_{ij}  =  \int \mathcal{W}^K_{ij} (q,p)~ \mathbb{K}(q,p)  ~ \td p ~\td q~= 0 ~,~ \text{for}~i \neq j~.
\end{equation} 

Next, we evaluate the integral over the diagonal elements,  
\begin{equation}\label{Intjj}
		I_{jj} = \int \mathcal{W}^K_{jj} (q,p) ~\mathbb{K}(q,p) ~\td p ~\td q~
\end{equation}
Proceeding as before, we arrive at the following expression for these integrals,
\begin{equation}
	I_{jj}  = \sum_{n} n ~ \int  \chi_n^K  (v)\psi^{K*}_n(u)  ~ \psi_j^K (u)\chi_j^{K*} (v) ~ \td u ~\td v~= j~.
\end{equation}

Another way of deriving these expressions for the integrals is to use star product relations. Using eq. \eqref{star_to_pq}, we can rewrite the expression for $I_{ij}$ as 
\begin{equation}
 I_{ij}   =  \int \mathcal{W}^K_{ij} (q,p)~ \mathbb{K}(q,p)  ~ \td p ~\td q~= \int  \mathbb{K}(q,p) \star \mathcal{W}^K_{ij} (q,p)~ ~ \td p ~\td q~,
\end{equation}
where the star product expression can be evaluated to be $\mathbb{K}(q,p) \star \mathcal{W}^K_{ij} (q,p)=j \mathcal{W}^K_{ij} (q,p) $, so that one gets back the expression $I_{ij}=j ~\delta_{ij}$ by applying \eqref{orthpsi}. 

By using these integrals, it is possible to recover expressions of different quantities in terms of $\phi_n(t)$ and their derivatives from their phase space counterparts. 

\section{Time evolution of the Wigner function in the basis of phase space Krylov functions}\label{sec_W_dot_2}
In this appendix, we derive a formula for the time derivative of the time-evolved Wigner function in terms of the phase space Krylov functions by utilising its expansion in the Krylov basis, as given in eq. \eqref{Wignerfunction}, and the discrete Schrodinger equation for the time derivative of the expansion coefficients $\phi_n(t)$. This form will be useful later on when investigating the early time behaviour of the Wigner function in the Krylov basis, and hence relating it to the early time growth of the Krylov state complexity.
Taking a time derivative of the expansion of the Wigner function in eq. \eqref{WFexpanK}, and using the discrete Schrodinger equation in the Krylov basis \cite{Balasubramanian:2022tpr},
\begin{equation}\label{discreteseq}
	i \partial_t \phi_n(t)=  a_n \phi_n(t)+ b_{n+1}\phi_{n+1}(t)+b_n \phi_{n-1}(t)~,
\end{equation}
we have
\begin{equation}\label{WigFdot1}
	\begin{split}
		\frac{\partial W(q,p,t)}{\partial t} = i \sum_{n,m}  \mathcal{W}^K_{nm} (p,q) \Bigg[ - \phi^*_m(t)~\Big(a_n \phi_n(t)+ b_{n+1}\phi_{n+1}(t)
		+b_n \phi_{n-1}(t)\Big)+  \phi_n(t)~\Big(a_m \phi^*_m(t)+ b_{m+1}\phi^*_{m+1}(t)
		+b_m \phi^*_{m-1}(t)\Big)\Bigg]~.
	\end{split}
\end{equation}
Now introducing the functions $\mathcal{M}_{nm}(t)$ through the definition (these functions can be thought of as the elements of a matrix $\mathcal{M}(t)$ with time-dependent coefficients)
\begin{equation}
	\mathcal{M}_{nm}(t) := \phi_n(t)~\Big(a_m \phi^*_m(t)+ b_{m+1}\phi^*_{m+1}(t)
	+b_m \phi^*_{m-1}(t)\Big)~,
\end{equation}
the evolution equation for the Wigner function can be rewritten in a much simplified form 
\begin{equation}\label{Wigner_dot}
	\frac{\partial W(q,p,t)}{\partial t}  
	=i \sum_{n,m}  \mathcal{W}^K_{nm} (p,q) ~ \Big[	\mathcal{M}_{nm}(t)  - 	\mathcal{M}^{*}_{mn}(t)  \Big]~.
\end{equation}
This is essentially an expansion of the time derivative of the Wigner function in terms of the phase space Krylov functions, and can be thought of as the alternative form for the quantum Liouville equation for the Wigner function discussed in Section \ref{sub_quan_class}. Note also that, while the expressions in \eqref{WigFdot2} describe the time evolution equation for the Wigner function for specific class of Hamiltonians, \eqref{Wigner_dot} is valid for any generic Hamiltonian whose associated classical counterpart is described in the phase space by continuous coordinates (this statements is, of course, also true for \eqref{Moyal_eq}). 

From the initial conditions on the Krylov chain, $\phi_n(t=0) = \delta_{n0}$, we see that at the initial time $t=0$, among the matrix elements of $\mathcal{M}_{nm}(t)$,  the only non-zero 
elements are 
\begin{equation}
	\mathcal{M}_{00}(0) = a_0~,~~ \mathcal{M}_{10}(0)  = b_1~. 
\end{equation}
As we shall discuss in Appendix \ref{1storder}, these two elements determine the first-order derivative of the Wigner function at $t=0$, i.e., the early time behaviour of the Wigner function. 

Now, it is a straightforward task to obtain an equation for the time derivative of the Krylov state complexity, which we have rewritten in terms of the Wigner function (see eq. \eqref{SCwigner}), by substituting \eqref{Wigner_dot} into the time derivative of \eqref{SCwigner} as
\begin{equation}
	\begin{split}
			\dot{\mathcal{C}} (t) = i \int  \mathbb{K}  (q,p) \sum_{n,m}  \mathcal{W}^K_{nm} (p,q)~  \Big[\mathcal{M}_{nm}(t)  - 	\mathcal{M}^{*}_{mn}(t)  \Big] ~ ~ \td p ~\td q~\\
			=  i \sum_{n,m} \Big[\mathcal{M}_{nm}(t) - \mathcal{M}^{*}_{mn}(t)  \Big]    \int  \mathbb{K}  (q,p)  \mathcal{W}^K_{nm} (p,q)  ~ \td p ~\td q~. 
	\end{split}
\end{equation}
In the second line above, we have rearranged the factors in a way that separates out the phase space integral from the time-dependent terms. 
Using some useful integrals evaluated in the Appendix. \ref{Integrals}, we see that the above expression can be simplified to 
\begin{equation}
		\dot{\mathcal{C}} (t) = i \sum_{n} n ~ \Big[\mathcal{M}_{nn}(t)  - 	\mathcal{M}^{*}_{nn}(t)  \Big] ~,
\end{equation}
so that one can recover the usual definition of $\dot{\mathcal{C}} (t)$ in terms of $\dot{\phi}_n(t)$ from the  explicit expressions of $\mathcal{M}_{mn}(t)$.

\section{Behaviour of Krylov state complexity at early times from Wigner function evolution}	\label{1storder} 
In this appendix, we derive expressions for the first and second order time derivatives of the Wigner function $ W (q,p,t)$ around $t=0$, and use them to show that the first order derivative of the Krylov state complexity vanishes at $t=0$, while its second order derivative is $2b_1^2$. 

From the initial conditions on the Krylov chain at $t=0$ and eq. \eqref{discreteseq}, one has, up to second order, the following relations 
\begin{align}\label{intitial}
	\phi_n(0)&= \delta_{n0}~,~~\dot{\phi}_n(0) = -i (a_n \delta_{n0}+ b_n \delta_{n1} )~,\\
	~~\ddot{\phi}_n(0)&= - (a_n^2+b_{n+1}^2)\delta_{n0}-(a_{n-1} b_n+a_n b_n) \delta_{n1}-b_nb_{n-1} \delta_{n2}~.
\end{align}
Then, from the expansion in \eqref{WFexpanK} (or \eqref{WigFdot1}), one can derive the expressions for the Wigner function at the initial time, and its derivative at $t=0$ in terms of the phase space Krylov functions to be
\begin{align}
	W(q,p,0) &= \mathcal{W}^K_{00}(q,p)~,~~\text{and}\\
		\frac{\partial  W (q,p,t) }{\partial t} \Big|_{t=0}&=ib_1 \Big(\mathcal{W}^K_{01}(q,p)-\mathcal{W}^K_{10}(q,p)\Big)= 2ib_1\mathcal{W}^K_{[01]}(q,p)~. 
\end{align}
Using the property \eqref{antisW} of $\mathcal{W}^K_{nm}$, and recalling that $b_0=0$,  we can rewrite the  expression for the first order derivative of the Wigner function as 
\begin{equation}
	\frac{\partial  W (q,p,t) }{\partial t} \Big|_{t=0}= -\frac{i}{\pi} \int dy ~ e^{2i p y} \bra{q - y} [\mathbf{P}_0, H] \ket{q+y}~,
\end{equation}
where $\mathbf{P}_n= \ket{K_n}\bra{K_n}$ denotes the projector onto the $n$th Krylov basis state. 
Now, substituting this expression for the Wigner function and its derivative in eq. \eqref{derivative0}, and subsequently using eqs.  \eqref{weylK}, and \eqref{Wndef}, along with the orthogonality property of  $\psi^K_n(q)$, one can show that $C^{(1)}_{\mathcal{K}}=0$ after some straightforward calculation. Thus, unlike the Krylov state complexity, the first-order derivative of the Wigner function is non-zero at the initial time; only its phase space integral with the Weyl transformation of the spreading operator is the quantity that vanishes. 

Similarly, we can also obtain a simple expression for the second-order derivative of the Wigner function at $t=0$, and use it to show that the SC grows quadratically at the initial time.
Using the relations in eq. \eqref{intitial} valid for $t=0$, and the expansion of the Wigner function in \eqref{WFexpanK}, we obtain the 
second-order derivative at $t=0$ to be\footnote{In the following expression, we suppress dependence of $\mathcal{W}^K_{nm}(q,p)$ on $q$ and $p$ for convenience.} 
\begin{equation}
\begin{split}
    \frac{\partial^2  W (q,p,t) }{\partial t^2} \Big|_{t=0} =  -2b_1^2 \big(\mathcal{W}^K_{00}-\mathcal{W}^K_{11}\big) - 2(a_1b_1-a_0 b_1) \mathcal{W}^K_{(01)}-2 b_1 b_2 \mathcal{W}^K_{(02)}~.
\end{split}
\end{equation}
We can evaluate the contribution of each of these three terms in the second-order derivative of the Krylov state complexity at $t=0$ by using the integrals evaluated in the Appendix. \ref{Integrals}. 
From eq. \eqref{Iij}, we see that the contribution of the last two terms in the SC vanishes, while eq. \eqref{Intjj} indicates that the contribution of $\mathcal{W}^K_{00}(q,p)$
is also zero, and  the term proportional to $\mathcal{W}^K_{11}(q,p)$ is the only term which has a contribution equal to $2b_1^2$. Thus, we recover the quadratic growth of  SC,   $\mathcal{C}(t) \approx b_1^2 t^2$ at early times, with the growth rate being proportional to the square of the first $b_n$ coefficients. 
\bibliography{reference}
\bibliographystyle{JHEP}
\end{document}